\documentclass[printer]{aa}
\usepackage{txfonts} 
\usepackage{multicol}
\usepackage{graphicx}
\usepackage{url}
\usepackage{hyperref}
\usepackage{psfrag}
\DeclareGraphicsExtensions{.pdf,.png,.jpg,.ps,.eps}
\usepackage{natbib}
\usepackage{subfig}
\bibpunct{(}{)}{;}{a}{}{,} 

\begin{document}
\title{The first interferometric survey in the K-band of massive YSOs}
\subtitle{On the hot dust, ionised gas, and binarity at au scales}

\author{E.~Koumpia\inst{\ref{inst1},\ref{inst11}}, W.-J.~de Wit\inst{\ref{inst2}}, R.~D.~Oudmaijer\inst{\ref{inst1}}, A.~J.~Frost\inst{\ref{inst3}}, S.~Lumsden\inst{\ref{inst1}}, A.~Caratti o Garatti\inst{\ref{inst4},\ref{inst10}}, S.~P.~Goodwin\inst{\ref{inst7}}, B.~Stecklum\inst{\ref{inst5}}, I.~Mendigut{\'i}a\inst{\ref{inst6}}, J.~D.~Ilee\inst{\ref{inst1}}, M. Vioque\inst{\ref{inst8},\ref{inst9}}} 

\institute{School of Physics \& Astronomy, University of Leeds, Woodhouse Lane, LS2 9JT Leeds, United Kingdom\label{inst1}
\and
ESO Vitacura, Alonso de C{\'o}rdova 3107 Vitacura, Casilla 19001 Santiago de Chile, Chile\label{inst2}
\and 
Institute of Astronomy, KU Leuven, Celestijnenlaan 200D, 3001, Leuven, Belgium\label{inst3}
\and
Dublin Institute for Advanced Studies, Astronomy \& Astrophysics Section, 31 Fitzwilliam Place, Dublin 2, Ireland\label{inst4}
\and
Th{\"u}ringer Landessternwarte Tautenburg, Sternwarte 5, D-07778 Tautenburg, Germany\label{inst5}
\and
Centro de Astrobiolog{\'i}a (CSIC-INTA), Departamento de Astrof{\'i}sica, ESA-ESAC Campus, 28691 Madrid, Spain\label{inst6}
\and
Department of Physics and Astronomy, University of Sheffield, Hicks Building, Hounsfield Road, Sheffield, S3 7RH, United Kingdom\label{inst7}
\and
Joint ALMA Observatory, Alonso de C{\'o}rdova 3107, Vitacura, Santiago 763-0355, Chile\label{inst8}
\and
National Radio Astronomy Observatory, 520 Edgemont Road, Charlottesville, VA 22903, USA\label{inst9}
\and
INAF, Osservatorio Astronomico di Capodimonte, via Moiariello 16, 80131 Napoli, Italy\label{inst10}
\and \email{ev.koumpia@gmail.com}\label{inst11}
}
\date{Received date/Accepted date}

\abstract
{Circumstellar discs are essential for high mass star formation, while multiplicity, in particular binarity, appears to be an inevitable outcome since the vast majority of massive stars ($>$ 8~M$_{\sun}$) are found in binaries (up to 100\%).}{We spatially resolve and constrain the sizes of the dust and ionised gas emission of the innermost regions towards a sample of MYSOs for the first time, and provide high-mass binary statistics of young stars at 2-300~au scales.}{We observe six MYSOs with VLTI (GRAVITY, AMBER), to resolve and characterise the 2.2~$\mu$m hot dust emission originating from the inner rim of circumstellar discs around MYSOs, and the associated Br$\gamma$ emission from ionised gas. We fit simple geometrical models to the interferometric observables, and determine the inner radius of the dust emission. We place MYSOs with K-band measurements in a size-luminosity diagram for the first time, and compare our findings to T Tauris and Herbig AeBes. We also compare the observed K-band sizes to the sublimation radius predicted by three different disc scenarios. Lastly, we apply binary geometries to trace close binarity among MYSOs.}{When the inner sizes of MYSOs are compared to those of lower mass Herbig AeBe and T Tauri stars, they appear to follow a universal trend at which the sizes scale with the square-root of the stellar luminosity. The Br$\gamma$ emission originates from a similar or somewhat smaller and co-planar area compared to the 2.2~$\mu$m continuum emission. We discuss this new finding with respect to disc-wind or jet origin. Finally, we report an MYSO binary fraction of 17-25\% at milli-arcsecond separations (2-300~au).}{The size-luminosity diagram indicates that the inner regions of discs around young stars scale with luminosity independently of the stellar mass. At the targeted scales (2-300~au), the MYSO binary fraction is lower than what was previously reported for the more evolved main sequence massive stars, which, if further confirmed, could implicate the predictions from massive binary formation theories. Lastly, we spatially resolve the crucial star/disc interface in a sample of MYSOs, showing that au-scale discs are prominent in high-mass star formation and similar to their low-mass equivalents.}


\titlerunning{The first interferometric survey in the K-band of massive YSOs} 
\authorrunning{Koumpia et al.} 
 \maketitle

\vspace{0.2cm}

\section{Introduction}  
\label{intro}

Massive stars ($>$ 8~M$_{\sun}$) are among the most influential objects in galaxies. Their birth, evolution, and death as supernovae not only affect their immediate vicinity but highly contribute to the dynamical and chemical structure, and the evolution of their host galaxies. But what sets the conditions for the formation of a high-mass star, and how does the process of accretion/ejection manifest in the dynamic environments of massive star-formation? The main theoretical challenge has been to find a mechanism that can sustain the mass accretion towards the central protostar in the presence of a high radiation pressure, which halts the infalling matter \citep[e.g.,][]{Wolfire1987,Kuiper2018}. To explain this, the theory is converging towards the formation of massive stars via accretion discs, similar to their low mass counterparts \citep[e.g.,][]{Kuiper2010,Kuiper2011,Haemmerle2017}, or a multidirectional mass accretion \citep{Goddi2020}.

The search for circumstellar discs around massive young stellar objects (MYSOs) and the determination of the disc properties (i.e., size, mass, infall rates) are essential. Detecting and characterising such discs has been extremely challenging, mainly due to the scarcity, highly embedded nature of their host environments, and to the large distance (typically several kpc) to massive star forming regions. It is mostly because of high angular resolution and infrared interferometry that we are now finally in a position to start revealing and exploring these relatively rare objects in more detail. On the large scales (300-2000~au), millimeter line observations (ALMA, SMA, VLA) reveal Keplerian-like disc structures in cold (T$\sim$50~K) material \citep{Ilee2016,Ilee2018b,Johnston2015,Johnston2020}, while substructures down to $\sim$45~au were recently traced by ALMA \citep[e.g.,][]{Beuther2017,Maud2019}. At smaller scales (down to a few au), where the accretion onto the star takes place, the disc is traced using near and mid-IR (hot/warm) emission \citep[e.g.,][]{Boley2013}. Direct evidence of such hot discs is more scarce, while there is only one imaged case presented 10 years ago \citep{Kraus2010}, due to the extremely challenging nature of the required observations. Therefore, accessing and studying a sample of hot discs around MYSOs at au scales is of great importance. In addition, observations of the inner rim of such discs is of particular interest, as it is where interaction between the disc and the star manifests \citep[e.g., magnetospheric accretion, boundary layer accretion, disc-wind; e.g.,][]{Mendigutia2020}, and eventually where planet formation occurs. \citet{Beltran2016} provides a thorough review on discs around luminous young stellar objects and the processes that shape them.

Infrared interferometers trace the size scales and material temperatures of accretion discs around MYSOs uniquely, as well as binarity, highly collimated jets, and stellar winds. 

Determining the binary properties of high-mass young stars is now particularly relevant since it has become clear that the binarity fraction of OB-type populations is close to 100\% \citep{Chini2012}. Binarity is known to significantly affect the evolution and fate of massive stars \citep{Sana2012}. Multiple systems of massive stars with separations as large as several hundreds of au are mostly predicted by numerical simulations as a result of fragmentation processes during the collapse phase \citep{Myers2013}, while closer binaries ($<$ 100~au) may instead form during an accretion disc fragmentation \citep{Meyer2018} or via orbital decay during internal \citep[e.g., capture in competitive accretion, magnetic braking;][]{Bonnell2005,Lund2018} or external interactions \citep[e.g., with other stars;][]{Bate2002}. However, despite these theoretical findings, reproducible quantitative predictions of massive binary properties are currently lacking. Observational studies of massive binaries and multiple systems in the pre-main sequence (PMS) phase are necessary to inform and distinguish between different scenarios of their formation. 

To date, only a few studies have been dedicated to the multiplicity of MYSOs. \citet{Pomohaci2019} performed a search for wide ($>$ 1000 au) binary companions of 32 MYSOs and conclude that the total multiplicity fraction of MYSOs may be nearly 100\%, and reported mass ratios higher than 0.5, in agreement with studies on intermediate mass Herbig AeBe PMS stars \citep{Wheelwright2010}. Direct detections of high-mass binaries covering tighter separations ($<$ 500 au) were obtained serendipitously. In particular, only a handful of such protobinary systems are known, with ranging separations of several hundreds of au \citep[PDS~27: 30 au, PDS~37: 48 au, V921 Sco: 45 au, NGC~7538~IRS1: 430 au, IRAS 17216-3801: 170 au, IRAS 07299-1651: 180 au;][]{Koumpia2019,Kraus2012,Beuther2017,Kraus2017,Zhang2019}. 

Here, we present the first interferometric survey in K-band of six massive YSOs using the unique spatial capabilities of AMBER (2 MYSOs) and GRAVITY (4 MYSOs) on the Very Large Telescope Interferometer (VLTI). The VLTI can achieve angular resolutions down to 1.7 mas with AMBER and GRAVITY on the four 8.2-m Unit Telescopes (UTs) bringing the inner regions of embedded high mass stars within reach. We focus on the characteristic size and geometry of the hot dust traced via the 2.2~$\mu$m continuum emission, the ionised gas traced via the hydrogen recombination emission (Br$\gamma$), and finally the binarity of MYSOs at milli-arcsecond scales. In Sect.~\ref{observ} we describe our target selection and the interferometric observations (GRAVITY and AMBER) along with the data reduction process, and the interferometric observables. In Sect.~\ref{geom_model} we spatially resolve the hot innermost parts of the MYSOs, and we trace the dust and gas components down to a few au from the central star. In particular, we apply simple geometrical models to fit the interferometric observables and we constrain the size and the geometry of the 2.2~$\mu$m continuum emission. In the same section we constrain the size and the geometry of the Br$\gamma$ emission towards the MYSOs. In Sect.~\ref{broad}  we investigate the our results with respect to the size-luminosity relationship and we assess the statistics on the MYSO binarity at mas separations (2-300 au). Lastly, we discuss our findings and summarise the results in Sect.~\ref{discussion} and Sect.~\ref{conclusions} respectively.

\section{Observations and data reduction}

\label{observ} 

\subsection{Target selection}

Conducting interferometry in the near-IR with the VLTI and, in particular AMBER and GRAVITY, requires relatively bright targets. In the past, most MYSOs have been considered faint for VLTI observations on a single-field mode as the brightest typically have 2MASS K-band magnitudes between 7 and 8, close to the sensitivity limits of the instruments. 

To identify program stars, we exploit the largest and most complete sample of massive young stellar objects to date: the Red MSX Source (RMS\footnote{http://rms.leeds.ac.uk/}) survey \citep[see][]{Urquhart2011,Lumsden2013}. This is an unbiased survey of MYSOs throughout the Galaxy and is based on the Midcourse Space Experiment (MSX) which conducted a mid-IR survey of the Galactic Plane. The candidate MYSOs have been subject to an extensive multi-wavelength campaign to confirm their MYSO nature, and establish their kinematic distances and bolometric luminosities. The RMS survey also establishes kinematic distances and bolometric luminosities. The selection of targets was bound to near-IR bright MYSOs with 2MASS K-band magnitudes $<$ 8, with the stipulation that they must be observable from Paranal, resulting in a sample of 29 MYSOs.
Of the objects selected, approximately 50 percent are not located near a suitable optical guide star, which is a requirement for the wavefront correction by the adaptive optics system MACAO located in the Coud\'{e} focus of each UT. In this study, we present the analysis and results of six MYSOs (brightest in the K-band) comprising the largest sample of MYSOs with NIR interferometric observations to date. These objects are massive (8~M$_{\sun}$ $<$ M $<$ 15~M$_{\sun}$) and are representative of the unbiased MYSO survey they are drawn from in terms of typical luminosities (L$_{\rm bol}$ $\sim$ 10$^{4}$~L$_{\sun}$) at the average distances of 3.4 kpc. The program stars with successful observations are listed for the whole sample in Table~\ref{MYSOs_info}. The associated calibrators with their corresponding spectral type, K-magnitude, and size expressed as the diameter of a uniform disc are presented in Table~\ref{calibrators}. The calibrators were observed to calibrate the atmospheric transfer function, and were selected based on their single nature, their comparable brightness to the science target, size (small enough to be unresolved), and their proximity to their associated science target. They were also used during the telluric correction process, after removing the observed absorption/emission features around the emission lines of interest (e.g., Br$\gamma$).

\begin{table*}[ht!]
\caption{List of observed MYSOs with their coordinates taken from the RMS survey. Their stellar luminosity, mass, K-band magnitude, distance and presence of the CO bandheads are also listed.}
\resizebox{\textwidth}{!}{
\begin{tabular}{clccccccc}
\hline
Source & R.A. & Dec. & L$_{\rm *}$ & Mass & K-band & Distance & CO & Simbad name \\
 & (J2000) & (J2000) & (L$_{\sun}$) & (M$_{\sun}$) & (mag) & (kpc) &  \\
\hline
G231.7986$^{b,c}$ & 07:19:35.93 & -17:39:18.0 & 10,000 $^{+2,600}_{-2,800}$ & 11.7$^{+1.1}_{-1.6}$ & 6.4 & 2.53$^{+0.18}_{-0.15}$ & n & MSX6C G231.7986-01.9682 \\ 
G233.8306$^{\dagger}$ & 07:30:16.72 & -18:35:49.1  & 13,000 $\pm$ 4,400 & 11.0$^\star$ & 6.1 & 3.3 & y & RAFGL 5232\\
G282.2988$^{a,c}$ & 10:10:00.32 & -57:02:07.3 & 6,400 $^{+1,000}_{-800}$ & 9.0 $^{+0.7}_{-0.2}$ & 7.0 & 1.63$\pm$0.06 & y & [MHL2007] G282.2988-00.7769 1\\ 
G287.3716$^{\dagger}$ & 10:48:04.55 & -58:27:01.5  & 17,000 $\pm$ 5,800 & 15.0$^\star$ & 7.5 & 4.5$^{+0}_{-4}$   & y & 2MASS J10480455-5827015 \\ 
G301.8147$^{\dagger}$ & 12:41:53.86 & -62:04:14.6  & 22,000 $\pm$ 7,500  & 15.0$^\star$ & 6.8  & 4.4  & n & MSX6C G301.8147+00.7808\\
G034.8211$^{\dagger}$ & 18:53:37.88 & +01:50:30.5 & 24,000 $\pm$ 8,160 & 16.0$^\star$ & 6.6 & 3.5$^{+6.7}_{-0}$ & y & 2MASS J18533788+0150305\\
\hline
\end{tabular}
}
\tiny {\bf{Notes}}: For the four MYSOs that are highly embedded and therefore do not have available Gaia parallaxes, the bolometric luminosity, K-band magnitude and distance are taken from the RMS database. It is found that for massive objects the stellar luminosity dominates the total luminosity \citep[e.g., Herbig AeBes][$\sim$10\%]{Fairlamb2015}. Therefore, we assume that the bolometric luminosity can be represented by the stellar luminosity for these four sources. The mass is extrapolated following the methods adopted for main sequence objects \citep{Martins2005}. $^\star$ Note that masses derived with this approach come with a high uncertainty of 35 to 50\%. $^{\dagger}$ The properties of those sources are based on kinematic distances with typical uncertainty of 1.0 kpc, while the typical uncertainties on the reported luminosities is 34\% \citet{Mottram2011}. $^{a}$ Also known as PDS~37. $^{b}$ Also known as PDS~27. $^{c}$ These are the only two MYSOs in this sample with available and reliable measured parallaxes and stellar properties using Gaia \citep[EDR3, DR2;][]{Guzman2021,Wichittanakom2020}.   
\label{MYSOs_info}
\end{table*}

\subsection{GRAVITY observations}

Four of our targets, G282.2988, G287.3716, G301.8147, and G034.8211, were observed using the GRAVITY instrument \citep{Gravity_Coll2017,Eisenhauer2011} on the four 8.2-m UTs, which operates in the K-band. The observed spectral setup delivered interferometric observables of six-baselines in both low (fringe tracker channel) and medium (science channel) spectral resolution simultaneously. The observed projected baseline lengths, B, range between $\sim$40~m and 130~m, corresponding to angular resolutions $\lambda/2B$ between $\sim$ 5.7~mas and 1.7~mas at 2.2~$\mu$m, which at the average distance (3.4~kpc) of the present sample of MYSOs corresponds to a physical spatial scale of $\sim$6-20~au. The uv-plane for each MYSO is presented in Figure~\ref{fig:uvplaneG282}.

The interferometric observables were recorded on six baselines simultaneously on the fringe tracker (FT) and the science channel (SC). The SC records the interferometric observables at a medium spectral resolution of R$\sim$500 over the entire near-IR K-band window (1.99 $\mu$m--2.45 $\mu$m), which corresponds to a velocity resolution of 600~kms$^{-1}$. Typical individual integration times were between 5-30~s. The technical overview of the observations including the integration times and atmospheric conditions (i.e. coherence time, seeing) is given in Table~\ref{gravity_tech}.

For the reduction and calibration of the observations the GRAVITY standard pipeline recipes (as provided by ESO, version 1.1.2) were used with their default parameters.

\subsection{AMBER observations}

Two of the sources in our sample (G233.8306, G231.7986) were observed using the three beam combining instrument AMBER \citep{Petrov2007} at the VLTI. AMBER was the first generation beam-combiner of the VLTI that used to operate in the near-infrared {\it H} and {\it K} bands as spectro-interferometer, combining beams from three telescopes at three different spectral resolutions (35, 1500, 12000), and it was decommissioned in 2017. The selected configuration for G231 and G233 used the 8.2 meter aperture of three UTs (UT 2, 3, and 4), delivering projected baselines between $\sim$45~m and $\sim$88~m. The uv-plane is presented in Figure~\ref{fig:uvplaneAMBER}. The fringes were recorded spatially, i.e. each detector integration contains fringe pattern, also called an interferogram. The fringe patterns for each of the three baselines are detected co-spatially (incoherent addition of the light) and therefore encoded non-redundantly (i.e. with distinct fringe pattern frequencies) such that they can be recuperated by means of a Fourier transform of an interferogram \citep[see][]{RobbeDubois2007}. 

For this program, fringes could be obtained with AMBER set-up in its low spectral resolution mode which has R $=$ 35. No use could be made of the FINITO fringe-tracker as the deeply embedded sources are too faint for the H-band atmospheric window. Each observation of a program star was bracketed by a calibrator star. The meteorological conditions of the observing night are reflected in Table~\ref{AMBER_weather} by the coherence time ($\tau_{0}$) and seeing measurements from the Differential Image Motion Monitor (DIMM) on Paranal. The values in Table~\ref{AMBER_weather} indicate worse than average Paranal conditions and it should be clear that they were not favourable to obtain high quality spectro-interferometric measurements. For the further analysis of this dataset we focus on the visibilities extracted for the continuum. The raw AMBER dataset was reduced using the \texttt{amdlib} data reduction package \citep[version 3.0.9][]{Tatulli2007,Chelli2009}. During the data reduction we selected the best 20\% of the observed frames as has been previously suggested to provide robust visibilities \citep{Malbet2007}.

\subsection{Observational results}

For all six MYSOs of our sample (GRAVITY and AMBER) we could measure the visibilities of the 2.2~$\mu$m continuum. In addition, for the four sources observed with GRAVITY, we could further extract information on the spectra, the visibilities around the Br$\gamma$ emission, and the closure phases of multiple telescope triangles. We note that the GRAVITY dataset (4 sources) is superior to AMBER (2 sources) with respect to spectral resolution (500 versus 35) and uv-coverage (4 UTs versus 3 UTs), which allowed the extraction of the additional information. 

\label{obs_res}

\subsubsection{K-band spectra}

\begin{figure}
\begin{center}
\includegraphics[scale=0.38]{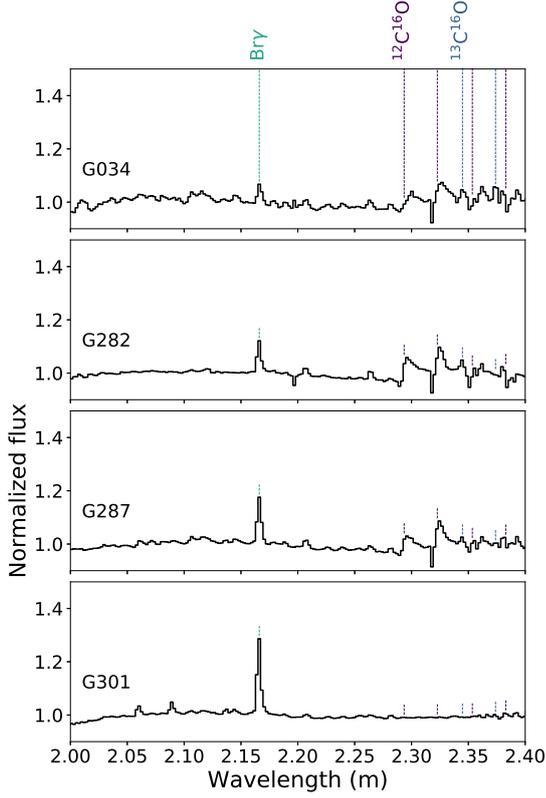} 
\end{center}
\caption{Normalised spectra of the MYSOs observed with GRAVITY around the 2.2~$\mu$m continuum level. The wavelength coverage contains the Br$\gamma$ which is seen in emission towards all sources, and the CO bandheads which are not detected towards G301. G301 shows also some weak emission within the 2.05-2.1$\mu$m range (Fe~II, He~I). The narrow absorption features are spectral artifacts caused during telluric correction.}
\label{fig:all_spectra}
\end{figure}

We present new NIR spectra of G282.2988, G287.3716, G301.8147, and G034.8211 as observed with GRAVITY (Figure~\ref{fig:all_spectra}). The observed wavelength coverage includes the Br$\gamma$ hydrogen recombination line emission at 2.167~$\mu$m, which is detected in emission in all sources. The GRAVITY spectra also cover the CO bandheads, which are found in emission towards all sources but G301 (Figure~\ref{fig:all_spectra}). Note that the CO bandheads were previously reported by \citet{Ilee2014} towards G282.2988 and G287.3716. In our study we do not focus on this molecular emission. 

For each source, we fit a Gaussian distribution to the line profile of the Br$\gamma$ emission and derive its line-to-continuum ratio at line peak (L/C), and the full width at half maximum (FWHM). G301.8147 shows the strongest Br$\gamma$ emission among the sources, with L/C at line peak of 1.29. G282.2988 is characterised by a Br$\gamma$ L/C at line peak of 1.17, followed by G287.3716 with a L/C at line peak of 1.12. G034.8211 is the source with the weakest Br$\gamma$ emission, and in particular it is characterised by a L/C at line peak of 1.06. The wavelength calibration uncertainties and the low spectral resolution, do now allow reliable measurements of peak velocities. 

We note that for all sources the FWHM ranged between $\sim$550-596~kms$^{-1}$, which is at the spectral resolution limits of the instrumental observing mode ($\sim$600~kms$^{-1}$), and therefore Br$\gamma$ is spectrally unresolved. Interestingly, the only source of the sample without signs of the CO bandheads (G301.8147) shows the strongest Br$\gamma$ emission. Increasing accretion rates \citep[e.g., stronger Br$\gamma$; see,][]{Mendigutia2011} or evolutionary stage \citep{Cooper2013a, Cooper2013b} are known to influence the detection rate of the CO bandheads towards MYSOs \citep{Ilee2018a}.

\subsection{Visibilities and phases}

We extracted the visibilities, differential and closure phases (CP) from the GRAVITY and AMBER observations around the 2.2~$\mu$m continuum and Br$\gamma$ emission. We note that AMBER combines three telescopes instead of four resulting in a single closure phase. In general, the visibility values are lower for angularly larger geometries. In practice, when comparing sources' geometries directly by means of visibilities, the baseline position angle comes into play. Relative brightness distributions of distinct emitting regions also affect the contrast levels per baseline length. Phases can be used to assess the degree of symmetry of the emitting region. Variations in differential phases between the continuum and a line emission indicate differences in the photocentre of the two emissions. On the other hand, a Non-axisymmetric brightness distribution of emission will result in non-zero closure phases. The differential phase of the Br$\gamma$ with respect to the continuum is typically 0$^\circ$ within the errorbars (see Figures~\ref{fig:vis_brg1}-\ref{fig:vis_brg4}), with possibly the only exception for the longest baselines towards G301. In this work, the differential phases are not taken into further consideration. We focus on the visibilities and closure phases of the continuum for all six MYSOs and of the Br$\gamma$ emission for four MYSOs (GRAVITY).   

\subsubsection{2.2~$\mu$m continuum}

All the sources in the sample (Table~\ref{MYSOs_info}) are fully or partially resolved, showing continuum visibilities V$_{cont}$ $<$1 even at the shortest baselines ($\sim$40~m). The measured visibilities generally decrease with increasing baseline length (see also Table~\ref{gravity_tech}). Before proceeding with a detailed geometric modeling of the emission we extract the measured visibilities of the continuum as averaged over a bandwidth of $\sim$ 0.03~$\mu$m bluewards and $\sim$ 0.03~$\mu$m redwards of the Br$\gamma$ emission\footnote{The sizes of the emission at 2~$\mu$m or 2.4~$\mu$m may deviate to some extent from what we report for the 2.2~$\mu$m.}. The measured visibilities, in addition to the angular size of an emitting region, are also linked to the position angle of the baselines and geometry (inclination) of the disc (e.g., G034.8211). 

In addition, we can assess the degree of symmetry or asymmetry of the emitting region by extracting the closure phases. For near axisymmetric brightness distributions the closure phases are always $\sim$0$^\circ$ (or 180$^\circ$), while any other measured value is indicative of a skewed intensity distribution. G034.8211 is the only MYSO in the sample which is characterised by asymmetric continuum emission at both the smallest ($\sim$1.7~mas) and intermediate scales ($\sim$3~mas) (Table~\ref{gravity_closure}).

\begin{table}[ht]
\caption{Closure phases (CP) of the 2.2~$\mu$m continuum emission observed with GRAVITY on the UTs. The associated triplets, together with the length and position angle (PA) of the longest baseline of the triplet are also reported.}
\small
\centering
\setlength\tabcolsep{2pt}
\begin{tabular}{c c c c c}
\hline\hline
Source & Triplet & Baseline & PA & CP$_{\rm cont}$ \\ & & (m) & ($^\circ$) & ($^\circ$)  \\
\hline\hline
{\bf{G034.8211}} & U3U2U1  & 102.4 & -143.3 & -8$\pm$4       \\
                 & U4U2U1  & 126.5 & -117 & 5$\pm$5     \\
                 & U4U3U1  & 126.5 & -117 & 12$\pm$3      \\
                 & U4U3U2  & 88.2 & -97.6 & 9$\pm$4         \\
\hline
{\bf{G282.2988}} & U3U2U1 &  93.6 & -147.7 & -0.7$\pm$0.3      \\
                 & U4U2U1 & 128.0 & -123.5 & 0.1$\pm$0.4      \\
                 & U4U3U1 & 128.0 & -123.5 & 0.6$\pm$0.4     \\
                 & U4U3U2 &  89.4 & -102.3 & -0.3$\pm$0.8     \\
\hline
{\bf{G287.3716}} & U3U2U1 & 92.9 & -148.8 & -1$\pm$2    \\
                 & U4U2U1 & 127.8 & -121.7 & 1$\pm$5     \\
                 & U4U3U1 & 127.8 & -121.7 & 3.5$\pm$2.5     \\
                 & U4U3U2 & 89.4 & -104.2 & 4$\pm$3     \\
\hline
{\bf{G301.8147}} & U3U2U1 & 90.7 & -148.0 & 6$\pm$10     \\
                 & U4U2U1 & 127.2 & -121.0 & 0$\pm$2      \\
                 & U4U3U1 & 127.2 & -121.0 & 1$\pm$3     \\
                 & U4U3U2 & 89.4 & -104.1 & 4$\pm$6      \\
\hline\hline
\end{tabular}
\label{gravity_closure}
\end{table}

\subsubsection{Br$\gamma$ emission}

The interferometric observables of GRAVITY as a function of wavelength of the Br$\gamma$ emission and the continuum around it for the four MYSOs are presented in Figures~\ref{fig:vis_brg1}-\ref{fig:vis_brg4}. A summary of the findings per source in given below.

\hspace{-0.5cm}{\bf{G282.2988:}} The Br$\gamma$ emission shows an increase in visibility of 1.3-6\% (highest difference at longest baseline; $\sim$130~m) with respect to the continuum, while the associated closure phases at longer baselines are $\sim$0.5$^\circ$-1$^\circ$ indicating a mostly symmetric emission of the ionised gas.\\ 
{\bf{G301.8147:}} The Br$\gamma$ emission is symmetric (CP$\sim$0$^\circ$) and shows an increase in visibility between 22-47\% (highest difference at short baselines; $\sim$50~m).\\ 
{\bf{G034.8211:}} The visibility of the Br$\gamma$ emission follows the visibility of the continuum at the long baselines (90~m, 130~m) and is higher than the continuum by $\sim$2-12\% for the rest of the baselines. In addition, the Br$\gamma$ emission shows closure phases up to $\sim$6$^\circ$-10$^\circ$ following those of the overall continuum emission.\\ 
{\bf{G287.3716:}} The Br$\gamma$ visibility increases by 7-10\% at two short baselines ($\sim$50~m) and shows no significant changes for the rest of baselines. The observed closure phases are $<$4$^\circ$ following the symmetric nature of the continuum (we note the large errors of 2$^\circ$-5$^\circ$).

We conclude that for all four MYSOs in the GRAVITY sample, Br$\gamma$ shows a similar or a higher visibility value compared to the continuum for different baselines, which is beyond the associated errors and therefore can be attributed to real geometrical effects. Hence, the observed changes in visibilities indicate that the ionised region traced via the Br$\gamma$ emission is of comparable size or more compact compared to the continuum emitting region. For G034.8211 and G301.8147 the Br$\gamma$ and continuum also show some variations in differential (i.e. differences in the photocentre of the continuum and that of the line emitting region) and closure phases (i.e., asymmetries, Figure~\ref{fig:vis_brg1}). The spatial distribution of the Br$\gamma$ emission appears to follow the morphology of the continuum for the symmetric sources (observed variations are within uncertainties and systematics), and it shows a more symmetric signatures compared to the continuum for G034.8211.

\section{Modelling the brightness distribution}

\label{geom_model}

To determine the size of the hot dust (2.2~$\mu$m continuum) and ionised gas (Br$\gamma$) emission, we adopt simple geometries to fit the observed visibilities of the sample of six MYSOs. We model the visibilities using the fitting software LITpro\footnote{LITpro is developed and maintained by the Jean-Marie Mariotti Center (JMMC) http://www.jmmc.fr/litpro} \citep{Tallon2008}.

\subsection{Size estimations - inner disc}

\label{size_lumin}

Observing and measuring the innermost radius of discs ($<$ 10 au) towards MYSOs is very important as this is the region where the interaction between the disc and the central star is most prominent, with material directly feeding the central star (e.g., via magnetospheric or boundary-layer accretion). The present dataset gives us access to scales of only few au at the typical distances of the current sample of MYSOs ($\sim$3.4~kpc). In addition, the K-band continuum flux can be mostly attributed to the thermal hot ($\sim$1500~K) dust emission, which is also the temperature at which silicate (Si) grains typically sublimate \citep{Kessler2007}, while this graphite (C) grains sublimate at higher temperatures \citep[2000~K;][]{Baskin2018}. The size of the 2.2 $\mu$m continuum emission has been traditionally used to directly probe the dust sublimation radius due to the radiation from the host star towards T Tauris and Herbigs \citep[e.g.,][]{Monnier2002}. Here, we present direct measurement of the K-band continuum size towards a sample of MYSOs, and we place MYSOs with K-band measurements in the context of a size-luminosity diagram for the first time \citep[Section~\ref{broad}; for M-band see][]{Grellmann2011}.

\subsubsection{Size of the 2.2~$\mu$m continuum emission}

\label{2micron}

To determine the characteristic size of the 2.2~$\mu$m continuum emission, we translate the observed visibilities of each MYSO to angular sizes. To do so, we apply three simple geometric models of a pre-defined brightness distribution, a Gaussian, a uniform disc, and a ring, and fit the observed to the predicted visibility curves. 

\begin{figure*}
\centering
\subfloat[]{\includegraphics[scale=0.26, angle = 90]{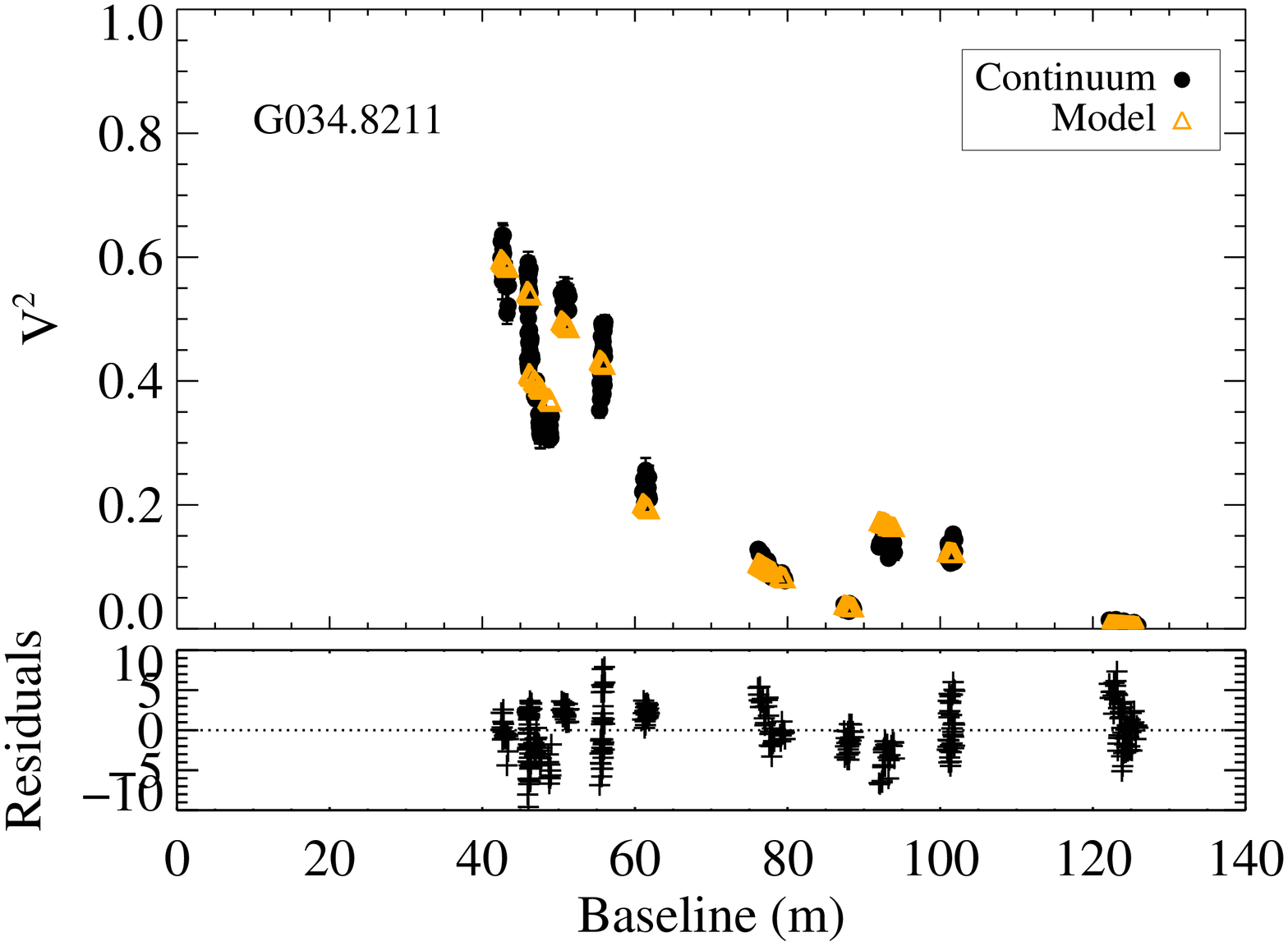}}
\subfloat[]{\includegraphics[scale=0.26, angle = 90]{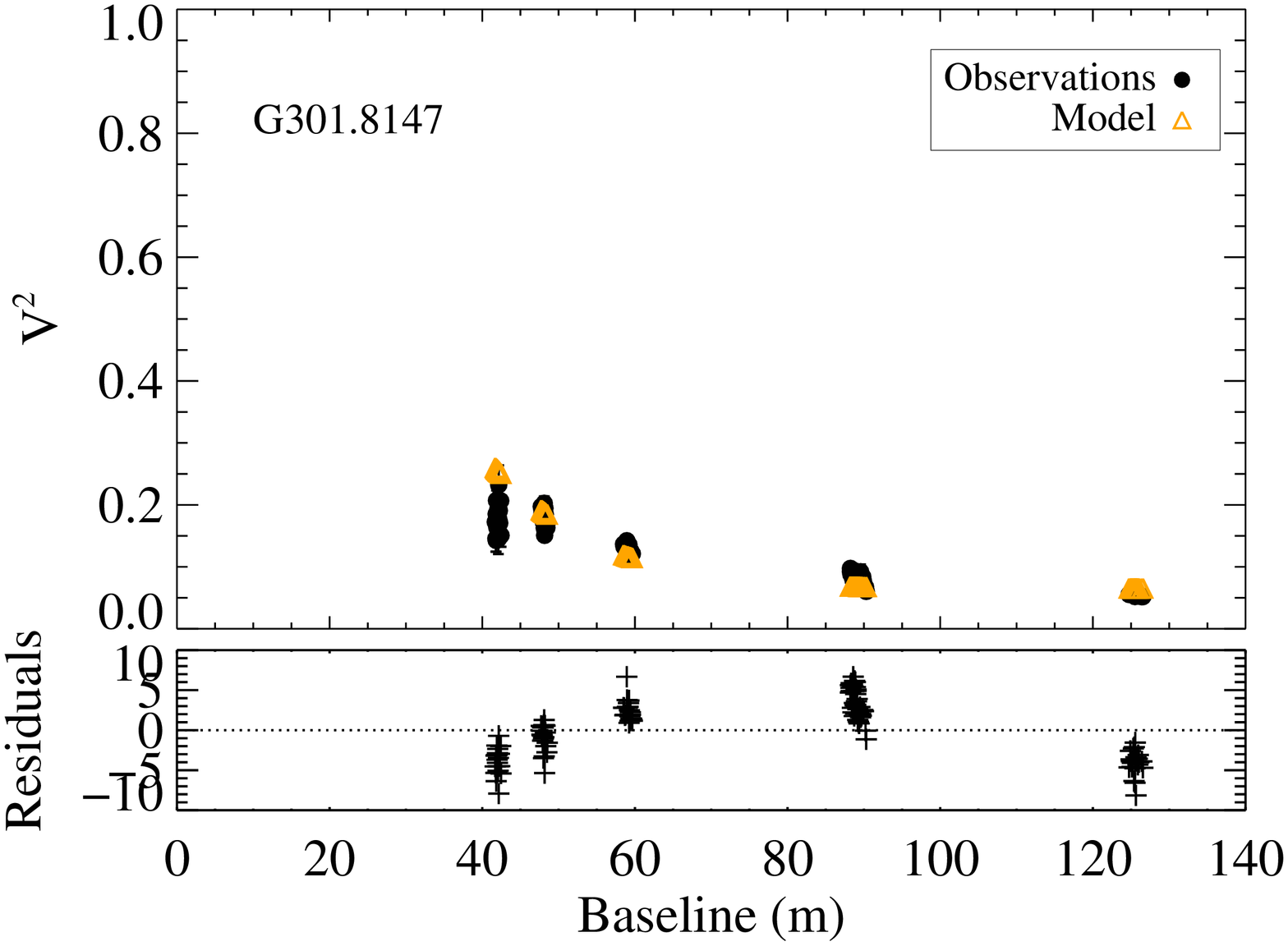}} 
\subfloat[]{\includegraphics[scale=0.26, angle = 90]{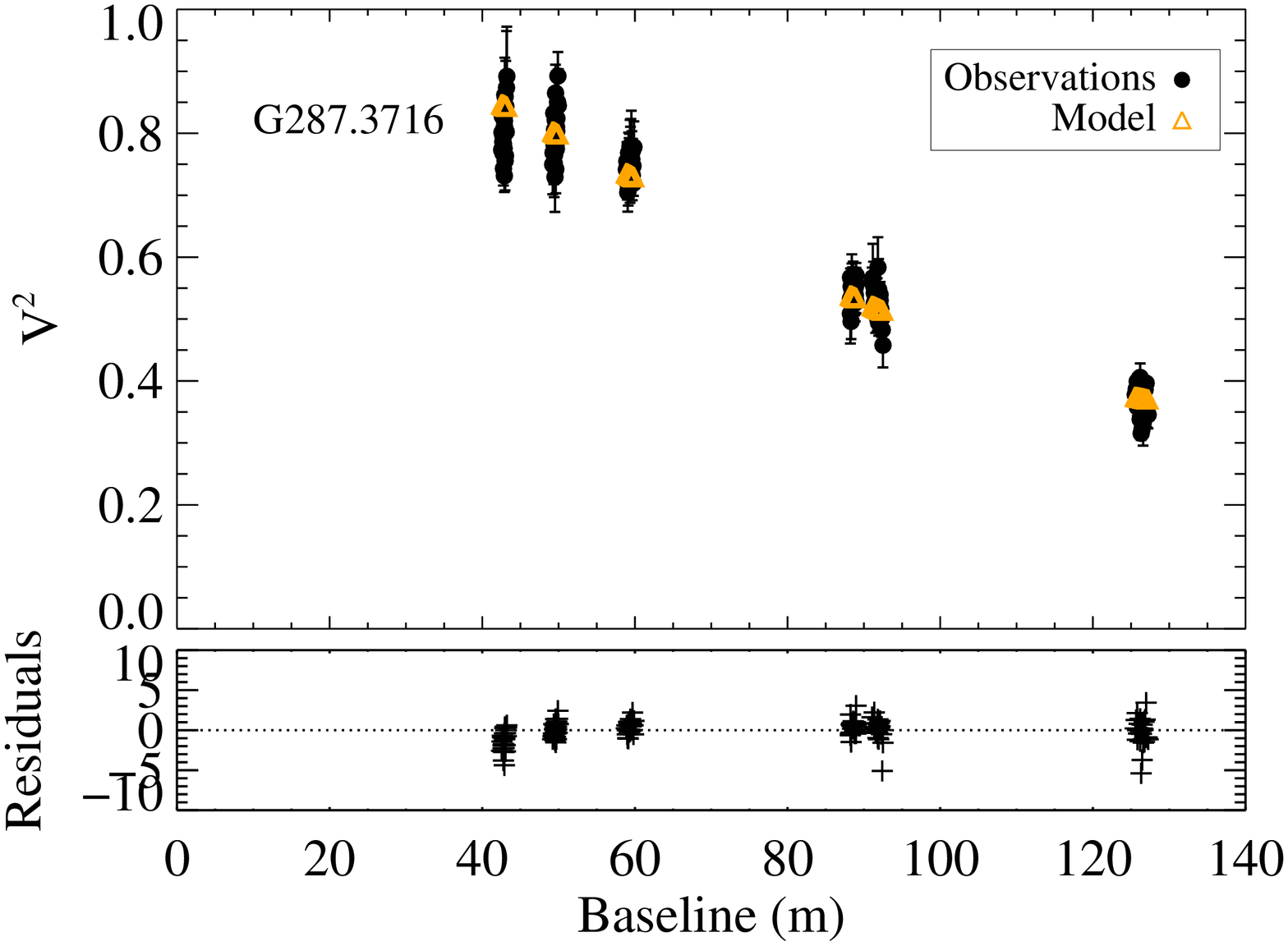}} \\
\subfloat[]{\includegraphics[scale=0.28]{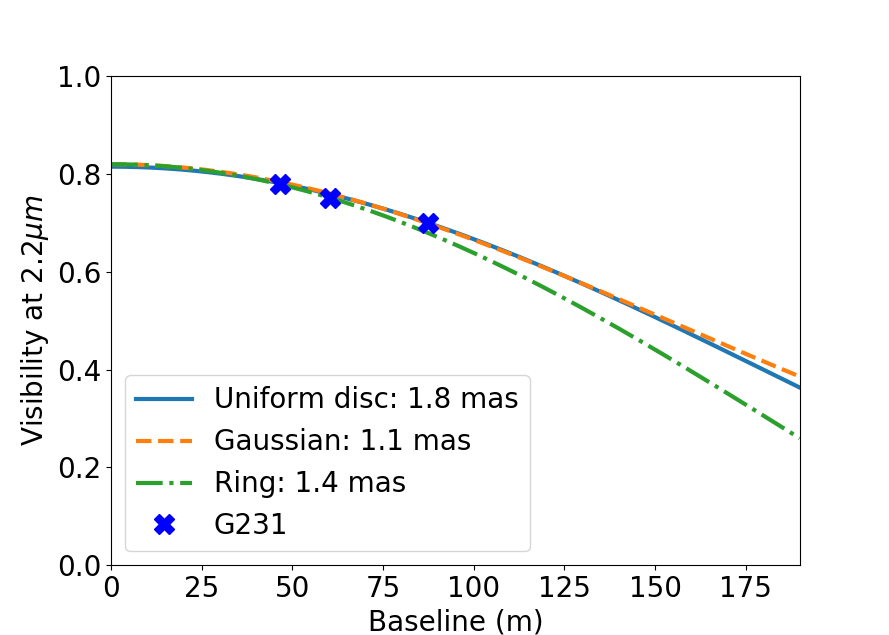}}
\subfloat[]{\includegraphics[scale=0.26]{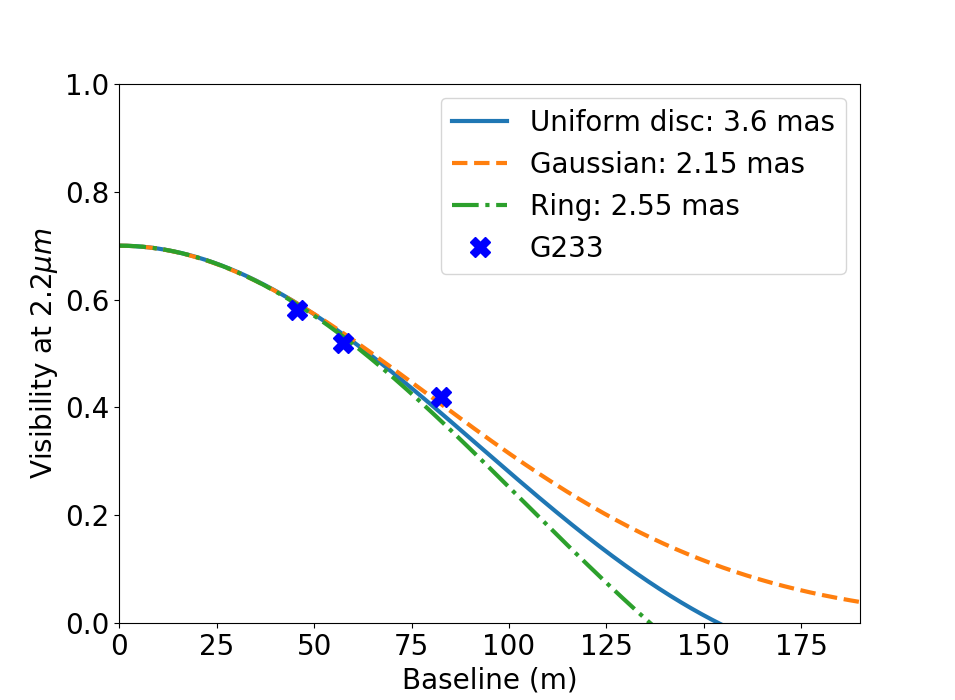}}

\caption{Modelled visibilities of the 2.2~$\mu$m continuum emission as function of baseline assuming a single disc brightness distribution overplotted with the observed ones as obtained with GRAVITY for (a) G034.8211, (b) G301.8147, and (c) G287.3816 as fitted using LITpro (Sect~\ref{2micron}) (d, e): Same as before but for observations obtained with AMBER towards G233 and G231, after fitting a Gaussian, a uniform disc, and a ring. The background flux contribution is found to be 16\% for G233 and $\sim$ 9\% for G231.}
\label{fig:LITpro_sizes}
\end{figure*}

We ran a grid of sizes between 0.2~mas and 10~mas. Figure~\ref{fig:LITpro_sizes} presents the observed visibilities obtained with GRAVITY and AMBER respectively, overplotted with the best fit modelled visibilities (lowest reduced $\chi^2$) for all MYSOs in our sample. The estimated sizes per source for the different brightness distribution are presented in Table~\ref{litpro_results}. For four out of six sources in our sample the addition of a background component (or halo) was required to explain the observed low visibility values ($<$ 0.8) at the shorter baselines. Lastly, for the sources in our sample for which pole-on geometries resulted a reduced $\chi^2$ $>$ 10, an additional centered point source was introduced in the fitting process. The flux weights of the additional components (1 for point source; 3 for background) are also presented in Table~\ref{litpro_results}.

We report the 2.2~$\mu$m emission sizes to be between 1.7~mas and 6.2~mas with typical errors between 1 and 6~\%. Depending on the adopted brightness distribution, the size estimations of a single source can vary up to a factor of $\sim$1.8 (e.g., G034.8211), while the reduced $\chi^2$s are mostly below 10. Overall, the lower the reduced $\chi^2$ is, the smaller the size variations among the models are. For one MYSO, G034.8211, the shape of the visibility curve does not show a gradual decrease in visibilities with increasing baselines (wiggly shape, Figure~\ref{fig:LITpro_sizes} a), and can only be fitted by assuming a flattened brightness distribution (reduced $\chi^2$ $<$ 30 versus $>$ 80). The best fit could be achieved for a flatten ratio of $\sim$1.5 (ratio between major and minor axes) regardless of the initial assumption of a disc, a Gaussian or a ring distribution. The position angle is between 20 and 34 degrees depending on the adopted brightness distribution. G282.2988 was the only MYSO for which a binary brightness distribution was necessary to improve the goodness of the fit (reduced $\chi^2$ $\sim$ 2 versus 12; see Sect.~\ref{bin_g282}). We additionally confirm the estimated sizes of the 2.2~$\mu$m continuum emission of the four MYSOs observed with GRAVITY, by performing an model independent image reconstruction where the closure phases are also taken into account (Appendix~\ref{image_rec}). The image reconstruction of G034.8211 revealed a similar elongated brightness distribution as the one retrieved from geometric modelling.

\begin{table*}[ht]
\caption{Sizes of the 2.2~$\mu$m continuum and Br$\gamma$ emission (listed as 2.2~$\mu$m / Br$\gamma$) towards G034.8211, G282.2988$^{b}$, G287.3716, G301.8147, G231.7986, and G233.8306 based on a single (i.e., Gaussian, disc) brightness distribution fit in LITpro. The flux weight 1, 2, and 3 correspond to the fractional flux contributions of the star (point source), dust (disc, gaussian, or ring) and that of a diffuse emission when a background was added, respectively.}
\centering
\setlength\tabcolsep{2pt}
\resizebox{\textwidth}{!}{
\begin{tabular}{l l l l l l l l l l}
\hline\hline
Source & Model & Flux w. 1 & Flux w. 2 & Flux w. 3 &  Diameter$\dagger$ & Flatten ratio & PA (minor axis) & Red.~$\chi^2$ & Measured size$^{a}$ \\ & & & & & [2.2~$\mu$m / Br$\gamma$] (mas) & [2.2~$\mu$m / Br$\gamma$] & [2.2~$\mu$m / Br$\gamma$] & [2.2~$\mu$m / Br$\gamma$] & (mas) \\
& & & &  &  & (degrees) &  &  \\
\hline
{\bf{G034.8211}} & Disc & n/a & 1.0 & n/a & 5.65$\pm$0.08 / 6.2$\pm$0.8 & 1.50$\pm$0.02 / 1.5$\pm$0.2 & 34$\pm$2 / 25$\pm$10 & 10 / 20 & \\
 & Gaussian & n/a & 1.0 & n/a & 3.46$\pm$0.02 / 3.42$\pm$0.06 & 1.52$\pm$0.01 / 1.51$\pm$0.04 & 20$\pm$1 / 17$\pm$4 & 20 / 17 &  3.47$\pm$0.04 \\
 & Ring & n/a & 1.0 & n/a & 6.2$\pm$0.4 / 3$\pm$1 & 1.50$\pm$0.1 / 1.5$\pm$0.3 & 23$\pm$14 / 24$\pm$17 & 30 / 35 & \\
{\bf{G301.8147}} & Disc & 0.26$\pm$0.04 & 0.22$\pm$0.03 & 0.52$\pm$0.08 & 5.22$\pm$0.1 / 2.27$\pm$0.07 & n/a & n/a & 3 / 6 & \\
 & Gaussian & 0.22$\pm$0.03 & 0.28$\pm$0.04 & 0.50$\pm$0.08 & 3.1$\pm$0.1 / 1.42$\pm$0.04 & n/a & n/a & 2.8 / 7 & 3.0$\pm$0.1 \\
 & Ring & 0.22$\pm$0.07 & 0.26$\pm$0.06 & 0.52$\pm$0.08 & 5.2$^{c}$ / 2.26$^{c}$ & n/a & n/a & 3 / 8 & \\
{\bf{G287.3716}} & Disc & 0.62$\pm$0.06 & 0.38$\pm$0.04 & n/a & 4.42$\pm$0.07 / 3.68$\pm$0.4 & n/a & n/a & 1.8 / 7 & \\
 & Gaussian & 0.52$\pm$0.05 & 0.48$\pm$0.05 & n/a & 2.34$\pm$0.05 / 1.28$\pm$0.05 & n/a & n/a & 1.76 / 2.3 & 2.15$\pm$0.03\\
 & Ring & 0.62$\pm$0.09 & 0.38$\pm$0.07 & n/a & 4.42$^{c}$ / 2.00$\pm$0.01 & n/a & n/a & 1.8 / 2.6 & \\
{\bf{G282.2988}} & Disc & 0.64$\pm$0.15 & 0.27$\pm$0.06 & 0.09$\pm$0.02 & 3.29$\pm$0.07/1.77$\pm$0.01  & n/a & n/a &  13/17  &  \\
 & Gaussian & 0.55$\pm$0.10 & 0.36$\pm$0.08 & 0.09$\pm$0.02 & 1.69$\pm$0.04/1.25$\pm$0.8  & n/a & n/a &  12/5 & 2.15$\pm$0.02 \\
 & Ring & 0.65$\pm$0.20 & 0.27$\pm$0.08 & 0.09$\pm$0.02 & 3.3$^{c}$ / 1.5$^{c}$  & n/a & n/a &  12/25  & \\
{\bf{G231.7986}} & Disc & n/a & 0.91$\pm$0.02 & 0.09$\pm$0.02  & 1.8$\pm$0.01 & n/a & n/a & 2 & n/a \\
 & Gaussian & n/a & 0.91$\pm$0.02 & 0.09$\pm$0.02  & 1.1$\pm$0.01 & n/a & n/a & 2 &  \\
 & Ring & n/a & 0.91$\pm$0.02 & 0.09$\pm$0.02  & 1.4$\pm$0.03 & n/a & n/a & 2.2 &  \\
{\bf{G233.8306}} & Disc & n/a & 0.84$\pm$0.05 & 0.16$\pm$0.04 & 3.6$\pm$0.03 & n/a & n/a & 2.3 & n/a \\
 & Gaussian & n/a & 0.84$\pm$0.05 & 0.16$\pm$0.04 & 2.15$\pm$0.03 & n/a & n/a & 2 &  \\
 & Ring & n/a & 0.84$\pm$0.05 & 0.16$\pm$0.04 & 2.55$\pm$0.05 & n/a & n/a & 2.6 &  \\
\hline
\end{tabular}
}
\tiny {\bf{Notes}}: $^{a}$ The measured size of the reconstructed 2.2~$\mu$m images (WISARD) as a result of a 2D Gaussian fit (Sect.~\ref{image_rec}) \\ $^{b}$ A binary component resulted a better fit on the visibilities (see Sect.~\ref{binary_stat}) \\ $^{c}$ When the fitting procedure resulted in a ring of 0 mas diameter, to estimate the size of the emission the ring width was used instead \\ $\dagger$ For ring brightness distributions this corresponds to the internal diameter of a normalized uniform ring

\label{litpro_results}
\end{table*}

\subsection{Br$\gamma$ emission}

\label{size_e}

The origin of the Br$\gamma$ emission observed in the close vicinity of (M)YSOs \citep{Mendigutia2011,Murakawa2013,Davies2010,Kraus2008,Pomohaci2017,Lumsden2012} can be explained by theory \citep[e.g., jets, magnetospheric accretion, disc wind;][]{Ferreira1997,Tambovtseva2014,Tambovtseva2016} and can be further tested by observationally constraining the spatial origin of the emission compared to the size of the hot dust (2.2~$\mu$m emission). The GRAVITY dataset allows us to estimate the size of the ionised gas by measuring visibilities at the central spectral channel of the Br$\gamma$ emission towards 4 MYSOs and directly compare it to that of the continuum emission. We note that the low spectral resolution of this dataset ($\sim$600~kms$^{-1}$) does not allow us to study a possible stellar wind, disc-wind or a jet origin of the Br$\gamma$ emission kinematically \citep[typical FHWM of 100~kms$^{-1}$-200~kms$^{-1}$;][]{Bunn1995}.

\subsubsection{Size of the Br$\gamma$ line emitting region}

\label{size_brg_d}

To estimate the size of the ionised emission, we used the measured calibrated visibilities at the spectral channel which corresponds to the peak of the Br$\gamma$ line profile. Before modelling the visibilities of any line emission, we need to correct for the continuum visibility and flux contributions \citep[see also,][]{Malbet2007}. In particular, we applied Equation~\ref{contin}

\begin{equation}
V_{line+cont} = \frac{V_{cont} \times F_{cont} + V_{line} \times F_{line}}{F_{cont} + F_{line}},
\label{contin}
\end{equation}

\hspace{-0.5cm}where the total visibility ($V_{line+cont}$) is a function of continuum and line visibilities ($V$) and fluxes ($F$), and solved for $V_{line}$.

\begin{figure}
\includegraphics[scale=0.4, angle = 90]{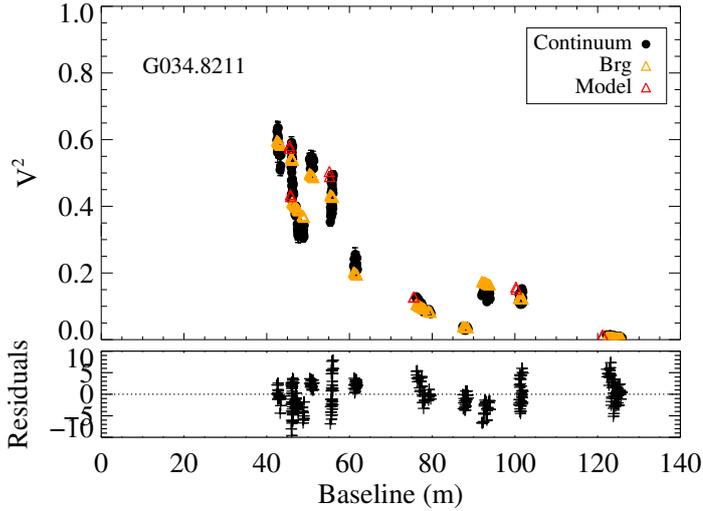}
\caption{Modelled visibilities and closure phases of the 2.2~$\mu$m continuum after assuming a flattened disc, overplotted with the observed ones towards G034.8211. The observed visibilities of the Br$\gamma$ emission are also overplotted in red, demonstrating the co-planar origin of the continuum and the ionised gas emitting regions.}
\label{fig:LITpro_G034_Brg}
\end{figure}

We derive the size of the Br$\gamma$ emission by fitting the interferometric observables with a simple geometrical model. In particular, we fit three different brightness distributions, a Gaussian, a uniform disc and a ring, for a range of sizes (0.2~mas--10~mas), similarly to how we treated the 2.2~$\mu$m continuum emission. Figure~\ref{fig:LITpro_G034_Brg} shows an example (G034) of the observed visibilities of the continuum and the Br$\gamma$ overplotted with the corresponding best fit of the continuum. The estimated sizes per source for a given brightness distribution are presented in Table~\ref{litpro_results}. The best fit of each adopted distribution demonstrates that the ionised gas systematically originates from a similar or smaller region (up to $\sim$10\%) compared to the continuum for three out of four MYSOs observed with GRAVITY. 

Arguably, underlying photospheric Br$\gamma$ absorption is known to affect the measured continuum level and Br$\gamma$, and therefore the size estimates of the Br$\gamma$ emission could be overestimated (Equation~\ref{contin}). Such contributions may become more significant for G301 and G282, for which the resolved out emission accounts for up to $\sim$50\% and 10\% respectively (Table~\ref{litpro_results}), and less for G034 and G287, for which the stellar contribution is unresolved \footnote{In \citet{Pomohaci2017}, it is argued than when dealing with MYSOs and sources which are characterised by large continuum excess, the contribution of the photospheric absorption is negligible.}. We note, that although such contributions may affect the absolute measurements of the Br$\gamma$ sizes (we therefore provide upper limits), the qualitative findings of the smaller Br$\gamma$ sizes with respect to the continuum emission will remain. A detailed study of those effects is beyond the scope of this study.

\section{MYSOs in a broader context}

\label{broad}

\subsection{Size-luminosity diagram}

\begin{figure}
\includegraphics[scale=0.4]{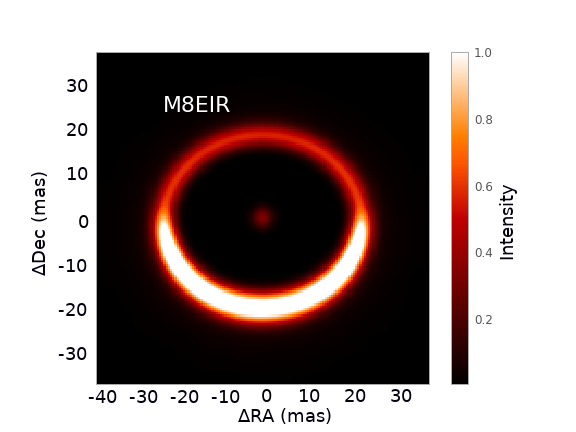}
\caption{Model image of the 2.2~$\mu$m continuum emission towards M8EIR \citep[based on][]{Frost2021} convolved at 1.7~mas resolution (GRAVITY/VLTI). The distance to M8EIR is 1.3 kpc.}
\label{fig:abi_model}
\end{figure}

To proceed with the size-luminosity relation we adopt the sizes of the ring brightness distribution determined in Sect.~\ref{2micron}. Even though a ring does not always reflect the best fit of the interferometric observables, it is preferred for the purpose of this section for consistency with other studies of this nature. In particular, a ring is generally assumed to reflect the size of the inner disc where dust sublimates \citep[see also,][]{Monnier2002}. In addition, the adopted brightness distribution can be supported by theory, since the hot dust emission is expected to stem from a narrow disc annuli at distances very close to the star \citep[see also Figure~\ref{fig:abi_model} and][]{Kraus2010,Stecklum2021}. The morphology and properties of the inner rim around low and intermediate mass YSOs has been the topic of multiple studies \citep{Isella2005,Tannirkulam2007,Kama2009,McClure2013}. Here, we directly compare the measured ring sizes of the five MYSOs in our sample (excluding G282 which can be best fitted as a binary), with their associated luminosities \citep[Table~\ref{MYSOs_info}; taken from][]{Mottram2011,Guzman2021,Wichittanakom2020}. We investigate the location of MYSOs in the size-luminosity diagram with respect to the dust sublimation radius as predicted from theory. In particular, we investigate the dust sublimation radius as predicted by a) oblique heating of an optically thick flat disc \citep[classical disc;][]{Hillenbrand1992,Millan2001a,Monnier2005}, b) direct disc heating from the star in the presence of an optically thin cavity ignoring backwarming effects \citep{Tuthill2001}, and c) direct disc heating from the star in the presence of optically thin cavity, but this time taking into account backwarming effects from the hot dust \citep[self-irradiation;][]{Dullemond2001}. Lastly, we compare our results with studies of the MYSOs' less massive counterparts (Herbig AeBes, T~Tauris).

\subsubsection{MYSOs}

\label{mysos}

\begin{figure*}
\centering
\subfloat[]{\includegraphics[scale=0.3,trim={0 0 0 3.0cm},clip]{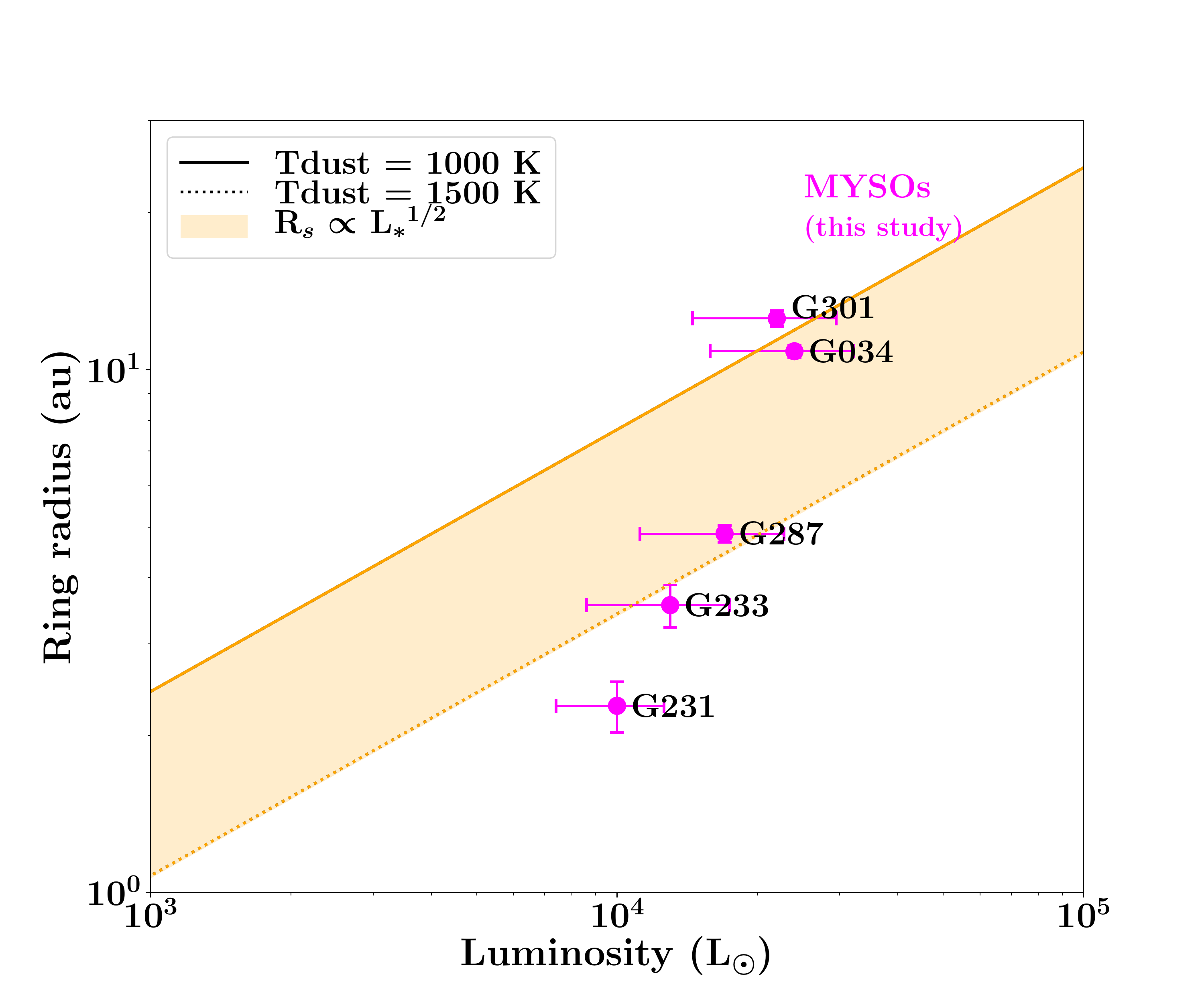}}
\subfloat[]{\includegraphics[scale=0.3,trim={0 0 0 3.0cm},clip]{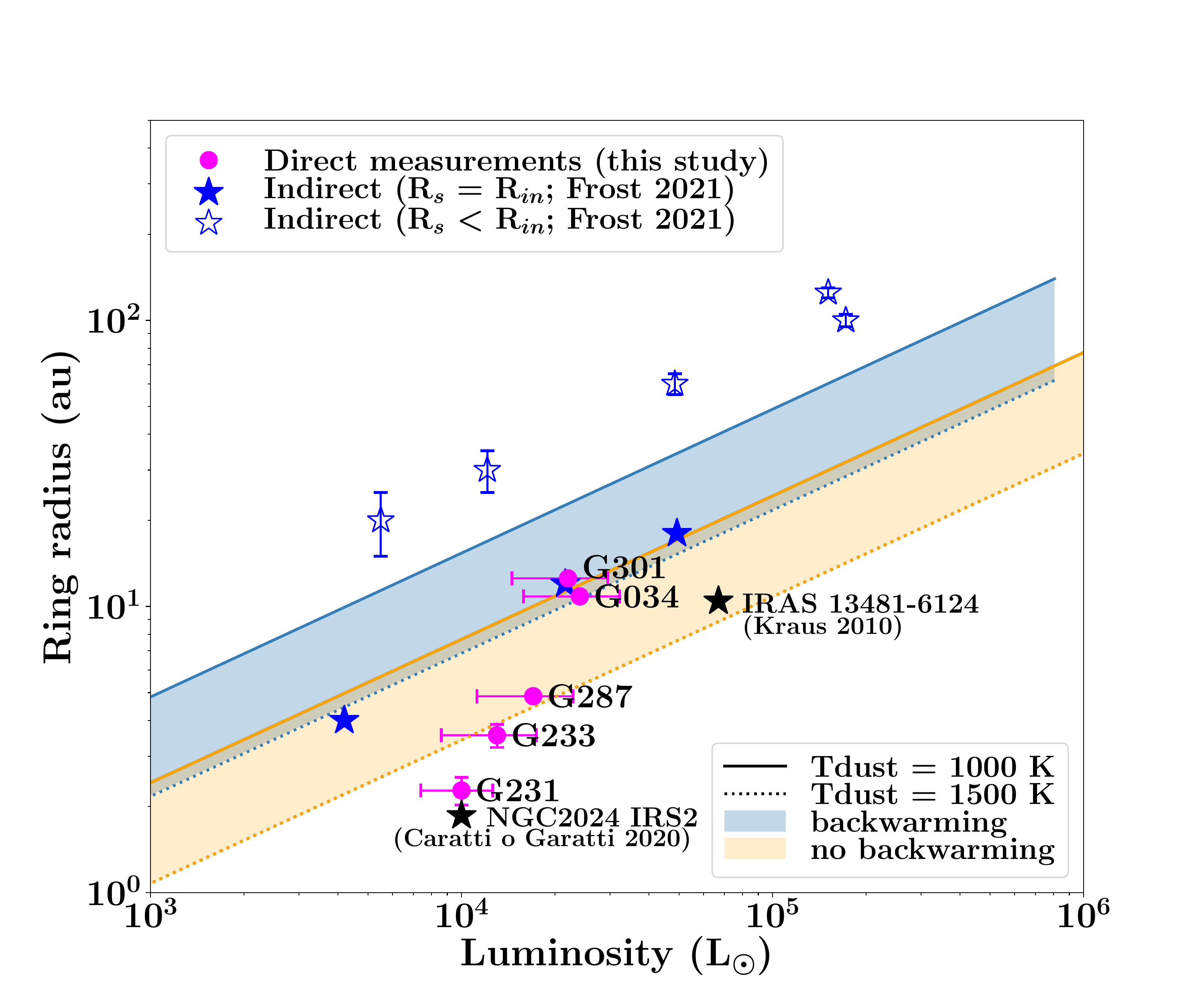}} \\
\subfloat[]{\includegraphics[scale=0.3,trim={0 0 0 3.0cm},clip]{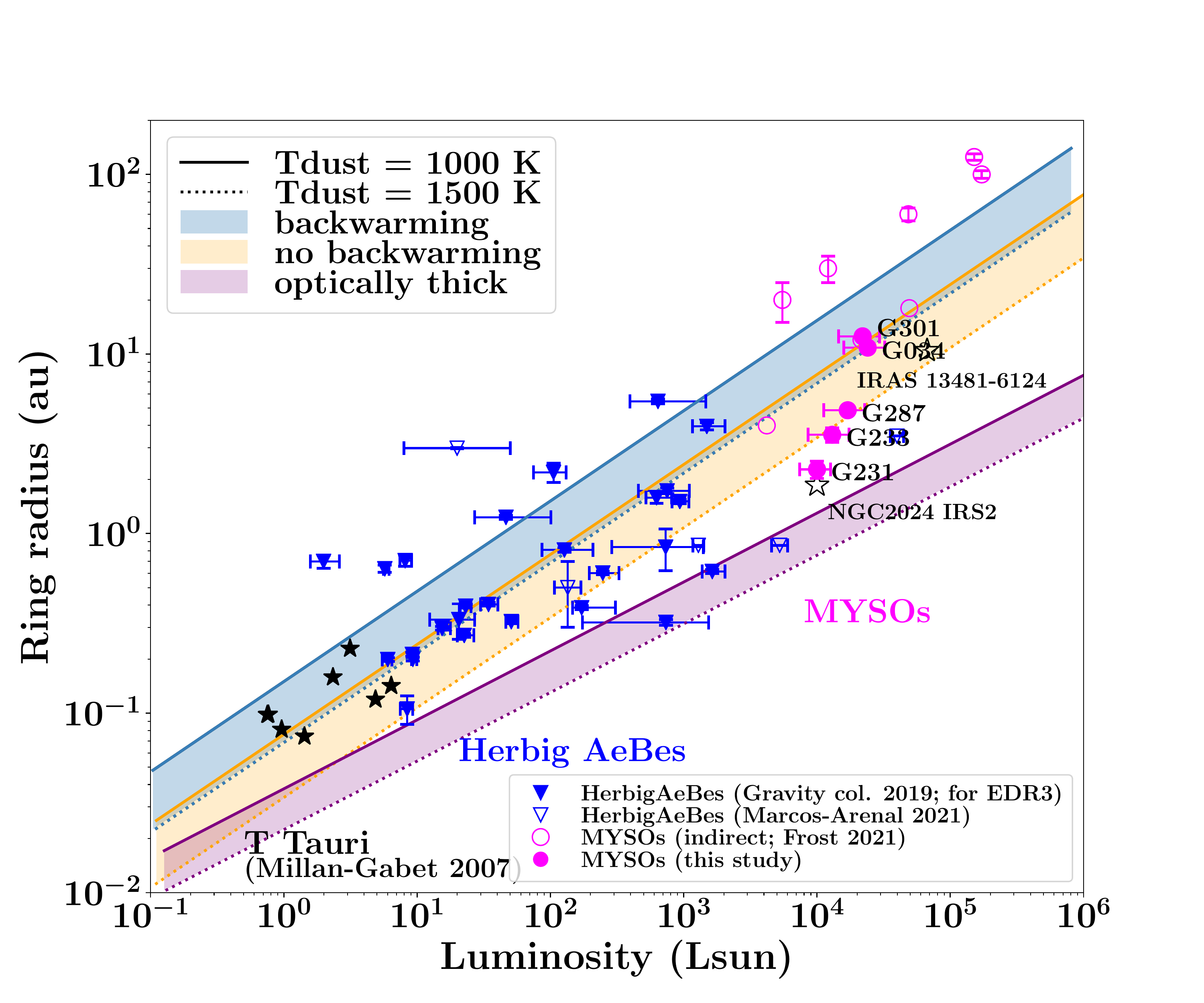}}

\caption{a) NIR size-luminosity diagram of our observed sample of MYSOs. The measured inner ring size R$_{in}$ of the observed MYSOs (GRAVITY, AMBER; based on Table~\ref{litpro_results}) is plotted with respect to the stellar luminosity. The shaded area represents the dust sublimation radius R$_{s}$ as predicted by the presence of a disc with an optically thin cavity for a range of temperatures. The adopted dust sublimation temperatures are 1000~K (solid line) and 1500~K (dotted line). b) Same as before but overplotted with two MYSOs with K-band measurements from literature \citep[filled black stars; IRAS 13481-6124, NGC2024 IRS2,][]{Kraus2010,Caratti2020} and with indirect measurements \citep[star symbols; modelled sizes presented in][]{Frost2021}. The open and filled stars represent the sources for which the modelled inner radius was found to be larger or equal to the R$_{\rm s}$ respectively. The dust sublimation radius as predicted by models of different dust temperatures with (blue area) and without (yellow area) backwarming effects is also overplotted. c) Same as before but including the available measurements of Herbig~AeBes and T~Tauris from literature. The dust sublimation radius as predicted by the classical scenario of a flat optically thick disc is also overplotted.}
\label{fig:my_MYSOs_size_lum}
\end{figure*}

By plotting the measured 2.2~$\mu$m sizes of the present sample of MYSOs (excluding G282 because of its binarity) as a function of their stellar luminosities (Figure~\ref{fig:my_MYSOs_size_lum}; a), it becomes apparent that there is a large scatter on the inner disc radii at the luminosity range (1.1$\times$10$^{4}$~L$_{\sun}$-2.4$\times$10$^{4}$~L$_{\sun}$) of this sample of MYSOs. To better explain the observed sizes with respect to stellar luminosities we overplot the predicted dust sublimation radius of a disc with an optically thin inner cavity \citep[R$_{\rm s}$ $\propto$ L$_{*}$$^{1/2}$;][]{Tuthill2001} for a range of dust sublimation temperatures \citep[1000~K-1500~K][]{Kessler2007,Boley2013}. With the exception of G231, the rest of MYSOs appear to follow what is predicted by the models. G231 is the only source in our sample, which appears significantly smaller and its size cannot be explained by the optically thin scenario (Figure~\ref{fig:my_MYSOs_size_lum}; a). 

To investigate the size-luminosity relation towards a larger sample of MYSOs, in addition to our direct K-band inner radii measurements, we overplot the other 2 MYSOs from literature with K-band interferometric observations and available 2.2 $\mu$m continuum size measurements \citep[IRAS 13481-6124, NGC2024 IRS2,][]{Kraus2010,Caratti2020}. Moreover, we overplot indirect measurements of the inner disc radii of eight MYSOs (Figure~\ref{fig:my_MYSOs_size_lum}; b) derived by \citet{Frost2021}. \citet{Frost2021} performed advanced radiative transfer modelling to simultaneously fit high angular resolution interferometric observations at mid-infrared (MIDI/VLTI), images (VISIR/VLT, COMICS/Subaru), and spectral energy distributions towards their MYSO sample. A good fit of the MIDI visibilities of five of those additional MYSOs at the shorter wavelengths could only be achieved when their inner radii was set to be significantly larger than the sublimation radius predicted from the optically thin models. To verify this finding we extracted the modelled images at the wavelength of interest (K-band; 2.2~$\mu$m) and convolved them to match the spatial resolution of GRAVITY/VLTI (1.7~mas). The measured size of the modelled 2.2~$\mu$m continuum emission for all eight sources is in agreement \citep[within 0.5-2~au, therefore with the errors reported in][]{Frost2021} with the inner radius predicted by models based on longer wavelength observations (e.g., M8EIR; Figure~\ref{fig:abi_model}). When these MYSOs are placed in a size-luminosity diagram, as we show in Figure~\ref{fig:my_MYSOs_size_lum}, self-irradiation (blue shade) cannot explain their inner radii sizes, but direct K-band interferometric measurements are necessary to independently confirm this finding.

The combined sample of MYSOs starts revealing a trend of increasing inner disc radii with increasing luminosity, and in particular the sizes scale with the square-root of the stellar luminosity. We further discuss the location of MYSOs with respect to models in Sect.~\ref{on_the_size}.

\subsubsection{Herbig AeBes and T~Tauris}

To directly compare the size-luminosity relation of the MYSOs to YSOs of lower mass, we collect and overplot the measured sizes of the 2.2~$\mu$m emission of known T Tauris and Herbig AeBes in Figure~\ref{fig:my_MYSOs_size_lum} (c). The initial luminosities and inner sizes are obtained from \citet{Millan2007}, \citet{Monnier2002}, \citet{Pinte2008} (and references therein), and the most recent GRAVITY YSO survey \citep{Perraut2019} and the five new Herbig measurements presented in Marcos-Arenal et al. 2021. The luminosities and sizes of all objects are scaled taking into account the new distances obtained using Gaia EDR3 parallaxes \citep{Guzman2021}, while no binarity at the traced scales was reported for these sources.  

In Figure~\ref{fig:my_MYSOs_size_lum} (c) we present the size-luminosity diagram with respect to the dust sublimation radius as predicted by three different disc models for dust sublimation temperatures of 1000~K and 1500~K. Figure~\ref{fig:my_MYSOs_size_lum} shows that with the exception of few Herbig AeBes which appear smaller than the rest (classical disc regime), the observed inner radius of most YSOs follow the distribution of the dust sublimation radius of an optically thin disc, increasing with the square-root of the luminosity. This trend can be seen within a wide range of luminosities. A more detailed discussion on the sizes of different classes of objects with respect to different models is presented in Sect.~\ref{on_the_size}.

\subsection{On the binarity}

\label{binary_stat}

To inform and differentiate among the theories of high-mass binary formation, studies need to provide observational information on the frequency, separation, and mass ratios of binary MYSOs. This work targets separations of few au to few 100s au scales, and provides a bridge between Adaptive Optics assisted imaging targeting 600~au to 10,000~au separations \citep{Pomohaci2019} and a high resolution spectroscopic survey of MYSOs which probes indirectly the closest separations at sub-au scales \citep[Shenton et al. in prep; for massive young stars see also,][]{Apai2007}.

\subsubsection{Our sample with respect to binarity}

In this study, we undertake the first companion search at milli-arcsecond scales towards a sample of MYSOs. We investigate whether our sample of six MYSOs at 2.2~$\mu$m with VLTI (GRAVITY or AMBER) contains close binaries at mas separations, covered by the field of view and angular resolution of the present interferometric K-band observations. The VLTI-UT configuration at 2.2~$\mu$m is sensitive to binary separations of 0.5 milli-arcsecond up to the single telescope diffraction limit ($\sim$70~mas). For a typical distance of 3.4 kpc this corresponds to 1.7 au and 240 au, respectively. Any wider pairs should already be traced by the NaCo imaging survey. We note that GRAVITY is sensitive to companion detections of a maximum magnitude difference $\Delta$K $\sim$ 5 mag within a 3-50~mas range of separations \citep[see also,][]{Sanchez_Bermudez2017}, similar to the one achieved in \citet{Pomohaci2019} at 1-3$''$ separation range. For the given sensitivity limits we would not be able to detect subsolar mass companions.  

With the exception of G034.8211, all objects were also part of a survey of 32 MYSOs using Adaptive Optics assisted high resolution K-band imaging with NaCo on the 8.2~m VLT \citep{Pomohaci2019}. Three objects in our sample (G301.8147, G287.3716, and G282.2988) were found to be parts of multiples with wide companion separations between 1.8$''$ and 2.9$''$, while the other two objects showed no multiplicity. We note, that 30\% of the full sample of 32 objects was observed to be binary with physical separations between 400 and 46000 au (0.6$''$-3.1$''$).

\subsubsection{Finding binaries - method} 

\label{bin_g282} 

To investigate binarity among the MYSOs in our sample, we proceeded with evaluating the addition of an off-centre source component in the geometrical models presented in Section~\ref{geom_model} with respect to the resulted $\chi^2$. Our models revealed that single brightness distributions are sufficient ($\chi^2$ $<3$ for four objects with pole-on geometries) to reproduce the interferometric observables for most MYSOs. Modelling G233 and G231 as single sources resulted in reduced $\chi^2$ of $\sim$ 2, which in combination with the limited amount of visibility measurements, make the inclusion of an off-centre point source unjustified. For G301 and G287, which come with a higher number of visibility / closure phase measurements, a reduced $\chi^2$ as low as 1.8-3 could be achieved without the need of adding an off-centre source. Introducing an off-centre component for G034 did not result in an improvement of the fit ($\chi^2$ $>$ 10). 

\begin{figure}
\includegraphics[scale=0.3, angle = 90]{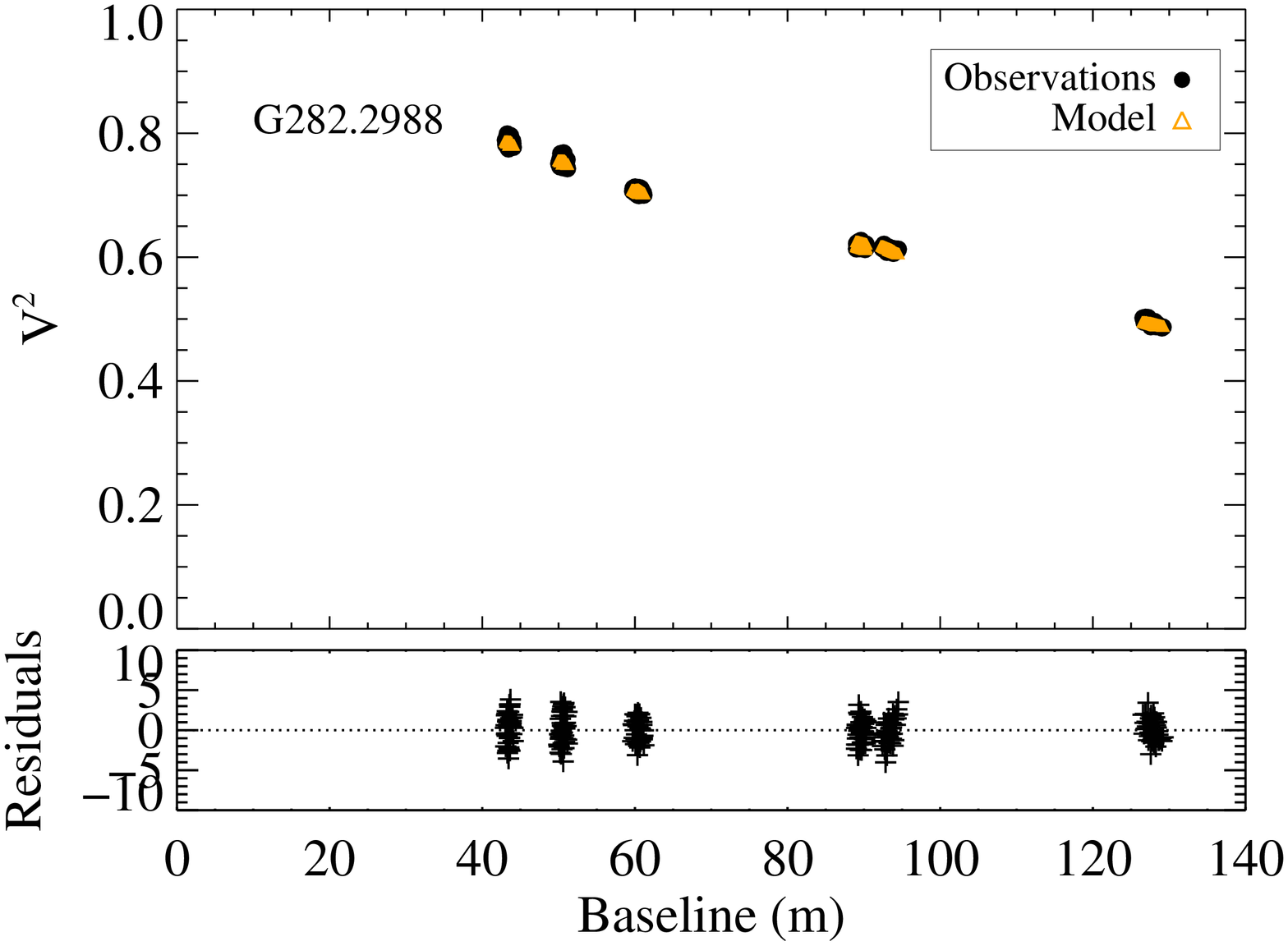}\par
\includegraphics[scale=0.3, angle = 90]{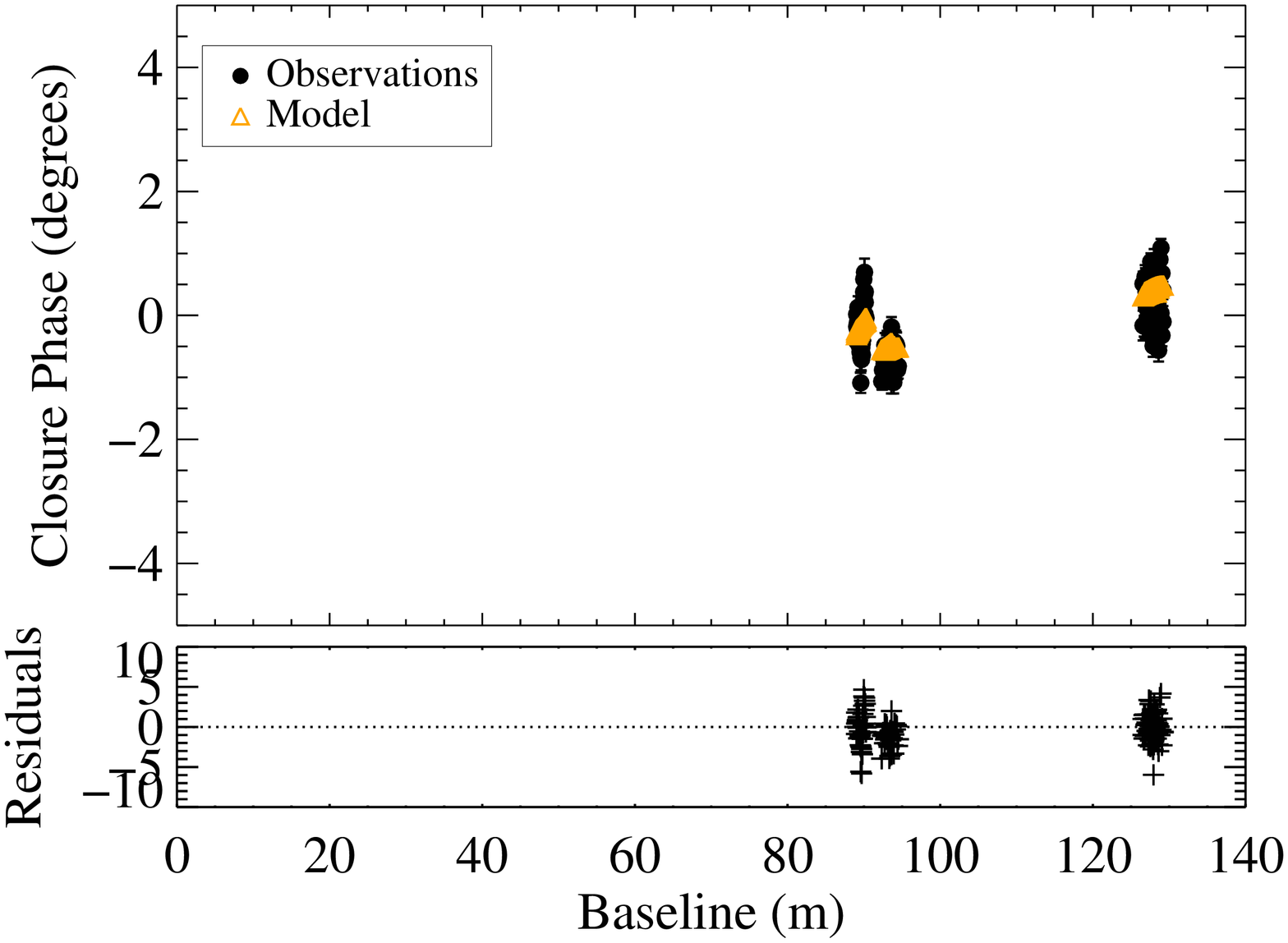}\par
\caption{Modelled visibilities and closure phases of G282.2988 as fitted using a binary model in LITpro. The best fit reveals a component separation of 23.7 mas \citep[consistent with][]{Koumpia2019} and primary to secondary flux ratio of $\sim$ 10.}
\label{fig:LITpro_G282_1}
\end{figure}    

G282.2988 (known also as PDS~37), is the only object for which both an elongated disc or the addition of an off-centred source improved the fit significantly (Figure~\ref{fig:LITpro_G282_1}), by reducing the $\chi^2$ from 12 to 2-3. The source was recently identified as binary at shorter H-band wavelengths using PIONIER/VLTI \citep[PDS37;][]{Koumpia2019}. The K-band modelling reveals a companion at 23.7$\pm$0.1~mas with a primary to secondary flux ratio of 10.4$\pm$1.2 and a position angle PA of 260$\pm$2$^{\circ}$. The fainter secondary companion is found to be $\sim$5 times more extended than the primary resulting in a $\chi^2$ of 3 compared to $\sim$70 when a point source is considered. Although the brightness is a good proxy for mass at long wavelengths, the emission at 2.2~$\mu$m at such embedded environments is still affected by significant extinction. The presence of a large disc surrounding the fainter object may hint to a more massive but embedded nature of the companion. Both the reported separation and the PA are consistent with \citet{Koumpia2019}, making the binary nature of G282 a favorable geometry over that of an elongated disc.

\subsubsection{Binary fraction}

In our sample of six MYSOs (AMBER and GRAVITY), we find an MYSO binary fraction of 17$\pm$15\% in K-band. Two of the sources in our sample are observed only with AMBER, and therefore are limited to only three measured visibilities for baselines between $\sim$40~m and 80~m. We note that the interferometric binary signal for G282.2988 becomes apparent for baselines $>$ 80~m ($<$ 2.8~mas; $\sim$ 10~au), and would have been flagged as a non-binary source in our AMBER observations as the angular resolution provided by AMBER on UTs would not have been sufficient to resolve it. Therefore, this limitation should be considered for the two sources of our sample observed with AMBER. G231.7986 (PDS~27) in particular, which is found to be a single brightness distribution based on AMBER observations, was found to show strong binary interferometric signatures on PIONIER H-band observations \citep{Koumpia2019}. PDS~27 is the only source in our sample presented in \citet{Perraut2019}. In that study, PDS~27 shows closure phase variations up to few degrees, but the authors argued that there was no need to model a binary companion. We note that the size of PDS~27 as obtained with AMBER (1.8~mas) is consistent to the size obtained with GRAVITY \citep[1.66~mas][]{Perraut2019}. The fact that a binary was not detected in the case of PDS~27 with AMBER or GRAVITY could be a result of the different fluxes (i.e., different masses, evolutionary stages) of the candidate companions in K- or H-band. If we take into account a more homogeneous observational dataset, by focusing on the GRAVITY sample alone, the MYSO binary fraction is 25$\pm$21\%. A more thorough discussion on the statistics and comparison with literature is presented in Sect.~\ref{on_bin}.

\section{Discussion}

\label{discussion}

We discuss the relation between the luminosity of the central star and its associated NIR size towards MYSOs. In addition, we discuss our findings on MYSO binarity at milli-arcsecond scales (2-300~au) for this sample of 6 MYSOs. 

\subsection{On the size-luminosity relation}

\label{on_the_size}

We observe a large scatter of about an order of magnitude in the NIR size at similar luminosities for the classified Herbig Be stars and MYSOs (10$^{4}$~L$_{\sun}$ $<$ L$_{*}$ $<$ 10$^{5}$~L$_{\sun}$). Numerous factors have been explored to explain the different sizes of the K-band continuum emission (i.e. radius of the inner rim) at a certain luminosity, both in terms of predictions and observations. The predicted dust sublimation sizes were found to be affected by the grain sizes, opacities and composition \citep[e.g.,][]{Monnier2002}, the nature of the accretion \citep[e.g., turbulence][]{Kuchner2002}, or photoevaporation \citep{Danchi2001}. Observations on the other hand can introduce uncertainties on the interpretation of the luminosity \citep[e.g., multiplicity;][]{Hartmann1993} and NIR size measurements (i.e., geometry of the brightness distribution). More recently, Marcos-Arenal et al. 2021 performed an investigation of the size-luminosity correlation and the observed size scatter towards Herbig AeBes. Here, we investigate the influence of backwarming effects and accretion on the location of MYSOs on the size-luminosity diagram.

\subsubsection{Backwarming effects}

It is known that for geometries where the vertical height of the inner rim is not significantly smaller than the radius of the inner rim, the backwarming by the circumstellar hot dust is not negligible, resulting in an increase in the dust sublimation radius \citep[self-irradiation;][]{Dullemond2001}. Here, we explore this mechanism with respect to the location of the different classes of objects in the size-luminosity diagram. The addition of backwarming results in a larger sublimation radius compared to the dust sublimation radius predicted by the classical flat disc with an optically thick inner region of gas, and the disc with an optically thin cavity of gas without backwarming effects (Figure~\ref{fig:my_MYSOs_size_lum}; c).

The sizes of the T Tauri stars in Figure~\ref{fig:my_MYSOs_size_lum} (c) can be better explained with the dust sublimation radius predicted by a disc with an optically thin cavity, where the inner rim is directly heated by the stellar radiation, and taking into account backwarming. The Herbig Aes and the more luminous Herbig Bes follow the size distribution as predicted by an optically thin disc, but this time neglecting the backwarming effects. This finding indicates that the small grains are more prominent backwarming heating mechanism for low-mass stars but they become less important with increasing mass. The situation is less clear for the regime of the more luminous Herbig Be stars and MYSOs. Although most of the sizes of the inner rim can be explained by the optically thin disc with or without the backwarming effect, five of the MYSOs with indirect measurements are larger and cannot be explained by none of the models, while one MYSO and four of the Herbig Be stars appear undersized. The Herbig exceptions rather follow the size distribution expected by the classical scenario of a flat optically thick disc with oblique heating \citep{Hillenbrand1992}. 

The two most luminous MYSOs in our sample (G034 and G301) have a larger K-band size (i.e. inner radius) than the rest, and are located in the lower temperature regime (1000~K) of an optically thin disc, when backwarming effects are neglected, or the higher temperature regime (1500~K) of an optically thin disc if backwarming effects are included (Figure~\ref{fig:my_MYSOs_size_lum} b). Given that a temperature of 1000~K is very low for dust sublimation to occur, we consider the scenario of backwarming effects becoming more important for these two MYSOs more likely. In contrast, the inner disc sizes of G287 and G233 follow the distribution of the dust sublimation predicted by an optically thin disc assuming a dust sublimation temperature of 1500~K, and neglecting backwarming effects. G231 appears to have an undersized inner radius compared to the rest of the MYSOs which cannot be explained by either of the two optically thin scenarios, while it appears oversized compared to the predictions of an optically thick disc. 

 We conclude that introducing backwarming effects in the discs surrounding young stellar objects appear to have a prominent role in explaining their location on the size-luminosity diagram, but there is no clear trend/explanation on why this mechanism seems to be more prominent for some objects with respect to others. Differences in dust grain compositions and more complex inner disc geometries (e.g., a flat, exposed inner rim) may contribute to the differences in backwarming contributions. However a detailed exploration of those effects would require a dedicated modelling study. 

\subsubsection{Evolution and accretion} 

To investigate the possible influence of the evolutionary status on the different locations of the MYSOs on the size-luminosity diagram, we retrieved the available evolutionary class of G034 and G231 from \citet{Cooper2013a} and \citet{Cooper2013b}. The classification scheme presented in \citet{Cooper2013a}, is based on the morphology of the NIR spectra and defines three Types (with their own subtypes) of (M)YSOs: Type I, II and III. Sources of Type I are the youngest (redder among subtypes), and show strong H$_{2}$ emission and no ionised lines. Type III sources are the oldest (most blue subtype), and they show strong H~I lines, prominent fluorescent \ion{Fe}{ii} emission at 1.6878 $\mu$m emission, and, if any, weak H$_{2}$ emission. Based on that scheme, G034 was classified as a class~II object, while G231 was classified to a more evolved, IIIb type. Further near infrared spectra (see Figure~\ref{fig:nir_spectra}) of G287, G231, G233, G282 and G301 were obtained using IRIS-2 on the AAT in a wider project related to RMS between 2006 and 2008. The spectral resolution in the H and K bands was $\sim$2400, allowing a clearer identification of weak lines than the data presented in \citet{Cooper2013a} and \citet{Cooper2013b}. Four of the objects, G231, G233, G282 and G301 show strong Br series, indicating a class III designation in Cooper's evolutionary scheme. Of these only G282 shows evidence of H$_{2}$ emission, which makes it class IIIa, while the other three MYSOs are class IIIb. The class III's also show evidence for a stronger ionising continuum through the presence of fluorescent FeII - in particular the IIIb's show both the 1.688 and 2.089 micron lines. G233, G282 and G301 also show [FeII] emission, showing that shocked gas in an outflow type process is still present. G287, like G034 can be classed as a type II, given the stronger H$_{2}$ emission and relatively much weaker HI emission. 

Following the same reasoning as for Herbigs, we would expect the less evolved with higher accretion rate objects, G034 and G287, to follow a regime of smaller inner disc radii in the size-luminosity diagram, and the rest, more evolved MYSOs, to follow a regime of larger sizes. This is in contrast to what we observe for four out of six sources, with G287 and G301 being the only two MYSOs following that expectation. Based on these observations, it appears that similar to Herbigs, the different evolutionary stages alone cannot explain the observed discrepancy in the measured 2.2~$\mu$m sizes. 

In addition, three MYSOs with indirect measurements \citep{Frost2021} follow the same relation as our more luminous MYSOs (optically thin disc with backwarming effects), while five of those objects are systematically larger and cannot be explained by the models. The larger sizes of the inner radius compared to the dust sublimation radius for those MYSOs, are attributed to increasing inner holes with age due to photoevaporation or the presence of binary companions \citep{Frost2019,Frost2021}. Figure~\ref{fig:my_MYSOs_size_lum} (c) shows that the distribution of the locations of the oversized Herbigs and MYSOs appears to follow a trend, although it cannot be explained from the current models. We note that even if one assumes a dust sublimation temperature of 2000~K as an upper limit, the predicted sizes of the inner radius would get smaller increasing the discrepancy even more. To verify that the observed trend is real and investigate its origin in more detail, direct measurements of the inner sizes (traced in the K-band) of the sample presented in \citet{Frost2021} are necessary.

To explain the observed scatter in sizes at given luminosities (Figure~\ref{fig:my_MYSOs_size_lum}), we also investigate the impact of accretion rates. High accretion rates may result in optically thick gaseous environments, shielding the dust in the inner radius. In that context, objects with higher accretion rates would be characterised by a smaller inner radius compared to the rest at a given luminosity, therefore, deviating from the general optically thin trend. This argument could be used to explain the one undersized MYSO and the undersized Herbigs \citep{Muzerolle2004}. Both in theory and observations the low-mass star formation is characterised by accretion rates of 10$^{-9}$-10$^{-7}$~M$_{\sun}$~yr$^{-1}$ \citep[T-Tauris regime,][]{Ingleby2013,Hartmann1998}, while high-mass star formation generally requires accretion rates that are 3-4 orders of magnitude higher \citep{Hosokawa2009}. We note that star formation is characterised by a variable rather than a steady accretion \citep{Vorobyov2009}. The observed discrepancy in the size-luminosity diagram for massive objects of similar luminosities could be indicative of less evolved massive young stars which are naturally characterised by higher accretion rates. If we extend this argument to Herbigs, less evolved objects could then explain their location at the regime where sublimation occurs for the optically thick disc scenario. Marcos-Arenal et al. 2021 investigated the size-luminosity distribution of HAeBes with respect to the nUV Balmer excesses, the H$\alpha$ and accretion luminosities, and the mass accretion rates, and they did not report a clear trend. In addition, the presence of the CO bandheads emission which were previously found to originate from an inner gaseous disc \citep{Ilee2014,Caratti2020} could act as a shielding mechanism for the dust allowing it to survive at distances closer to the central star, and therefore resulting in a smaller inner radius of the dusty disc (i.e., smaller 2.2 $\mu$m size). Based on the NIR spectra of this sample, the presence of CO could explain the small sizes of G287 and G233, but not the large size of G304, while similarly the absence of this molecular emission could explain the large size of G301 but not the small size of G231.  

In conclusion, we find that the dust inner rim radius of MYSOs, when directly measured via the 2.2~$\mu$m emission does not show a clear trend with respect to the stellar luminosity. When MYSOs are treated as a class and compared to low luminosity T Tauri, Herbig Ae and most of the Herbig Be stars, then we observe a general trend of increasing inner rim radius with the square-root of the stellar luminosity. This finding is suggestive of a universal trend in the observed size-luminosity diagram, indicating that the sizes of inner regions of discs around young stars scale with luminosity independently of the stellar mass, and are consistent with the dust sublimation radius predicted by models.

\subsection{Origin of the Br$\gamma$ emission}

\begin{figure}
\begin{center}
\includegraphics[scale=0.3]{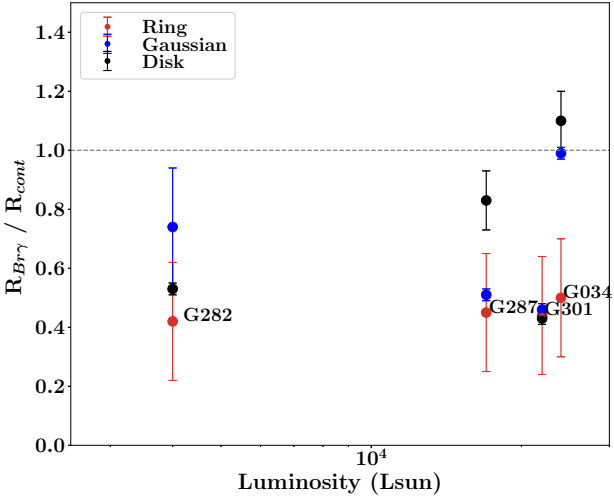} 
\end{center}
\caption{Measured ratio of the Br$\gamma$ emitting size (R$_{\rm Br\gamma}$) over the size of the 2.2~$\mu$m continuum emission (R$_{\rm cont}$) plotted as function of luminosity. The Br$\gamma$ emission originates from a smaller region compared to the dust continuum (Table~\ref{litpro_results}). The different colors correspond to the different adopted brightness distribution: Gaussian (blue), ring (red) or disc (black).}
\label{fig:Brgamma_continuum}
\end{figure}

The VLTI/GRAVITY observations indicate that the Br$\gamma$ emitting region is
similar or smaller in size than the region where the hot dust
resides. Figure~\ref{fig:Brgamma_continuum} demonstrates this by means of the ratio of the
Br$\gamma$ to the 2.2 $\mu$m continuum sizes as function of source
luminosity. This result we would like to put into context. 

Heating and ionisation by shocks and radiation occurs in the gaseous
structures that make up the accretion environment in young stars. The
Br$\gamma$ transition is therefore a prime diagnostic in young, embedded
stars and often resolved in spectro-interferometric observation at
100~m baselines. In accreting low-mass stars, the Br$\gamma$ transition
is compact and traces the magnetospheric accretion columns \citep{Garcia2020,Bouvier2020}. 
In this case, the emission is located well within the dust sublimation radius of the disc. Stepping
up in mass however, the magnetosphere becomes less important as field
strengths decrease while at the same time the star is hotter. As a
result, the Br$\gamma$ emission in Herbig Ae stars is no longer
restricted to the magnetosphere but observed to be more extended
albeit still smaller than the dust sublimation radius \citep{Garcia2015, Caratti2015, Mendigutia2015, Ellerbroek2015, Kurosawa2016}. Whether the emission is restricted to the protoplanetary disc of the HAe star or
if it subtends a larger angle remains unanswered so far. In some
late B-type PMS stars, where the line emission region is smaller than
the dust continuum, a disc-wind originating in the gaseous parts of
the inner (0.2 au) disc is favoured \citep{Kreplin2018}. In mid-B type stars Br$\gamma$
and the dust occupy a similar emission region \citep{Hone2019}. 

A break with the ``Br$\gamma$ smaller than dust continuum'' trend is found
in the early B-type Herbig Be star MWC\,297 \citep[17\,M$_{\odot}$][]{Vioque2018}, the ionised emission was found to be 40\% larger 
than the hot dust continuum \citep{Malbet2007}, while the kinematics are consistent with that of a disc wind at scales
of a few au \citep{Hone2017,Weigelt2011}. Having settled on the ZAMS, the generation and ionisation of a disc wind, extending beyond the disc's hot dust, 
in MWC\,297, a B\,1.5IV star, is perhaps not surprising. On the other hand, what could be considered surprising is the observed situation in MYSOs, 
where the ionised emission is sytematically more compact than dust emission.

MYSO examples for which Br$\gamma$ could be spectro-astrometically
mapped by means of Integral Field Unit observations and closure phases
indicate bipolar geometry at high (500~km/s) velocities. Both W33A VLA1 \citep{Davies2010} and IRAS13481-6124 \citep{Caratti2016}
demonstrate the origin of Br$\gamma$ in fast, collimated jets and/or in collimated winds at the base of the jets. This
picture is extended to deeper embedded sources, where fast ionised
jets can be mapped in the radio, as done in the MYSO sources Ceph\,HW2,
GGD\,27, and G345.4938 \citep{Curiel2006,Masque2015,Guzman2016}. High shock
velocities would likely destroy any molecule within, constrasting this
outflow component from any molecular disc emission. Notably, at high accretion rates (M$_{acc}$ $>=$10$^{-3}$ M$_{\odot}$/yr), 
the accreting MYSO is expected to be bloated \citep[][]{Hosokawa2010}, and therefore cool, which can arguably prevent the direct ionisation of the disc. However, \citet{Simon1983} showed that winds of MYSOs can be so dense that hydrogen is collisionally excited to its n$=$2 state, which makes its ionisation from cooler stars possible \citep[see also,][]{Koumpia2020,Drew1998,Drew1998a}. Indeed, a bloated star could also explain the narrow single line profiles of Br$\gamma$ observed in MYSOs in various studies \citep[e.g.,][]{Pomohaci2017}, which is in contrast to the relatively broader and double-peaked lines predicted by disc models around hot main sequence stars \citep{Sim2005}. Indeed, a slightly bloated object would have a larger disc (inner) radius and thus lower rotational velocities (i.e., narrow line profiles).

A compact Br$\gamma$ emission as the one we find, can also be a result of shocks from a jet or a disc-wind. Exceptionally disc instabilities that allow accretion to proceed will allow the formation of a very compact gaseous disc which can be shock-ionised close to the star \citep[G345.4938,][]{Guzman2020}. Jets were recently traced via radio thermal emission and found to be abundant in high-mass star formation \citep[up to 84\%,][]{Purser2021}. In our sample of MYSOs we find that Br$\gamma$ emission originates from a smaller area, but co-planar to that of the continuum, therefore if a jet is the underlying mechanism, we most likely trace the base of the jet.

We surmize that an ionised disc emission does not require the star to become hot enough, i.e. to settle on the main sequence once the accretion rate goes down. Br$\gamma$ being significantly smaller than the hot dust continuum (the sublimation radius basically) could also trace jet emission in MYSOs, or rather the base of the jet in a magneto-centrifugal disc-wind. The relatively small size ($\sim$3-10~au) of the Br$\gamma$ emitting region measured here roughly matches the MYSO jet collimation region, which is usually located at the Alfven radius, from few to several tens of au from the source \citep[see e.g. Fig. 13 in][]{Kolligan2018,Staff2019}, depending on the stellar mass and age. Although the full details are not known yet, with our GRAVITY findings we can constrain the geometry so that axi-symmetric models are to be favoured.

\subsection{On the binarity}

\label{on_bin}

In the current study we investigate binarity in a sample of six MYSOs (8.6~M$_{\sun}$-15.4~M$_{\sun}$) targeting separations between $\sim$2~au and 300~au (assuming a distance of 3.4 kpc) and found that one object out of the six can be better modelled with a binary geometric model (binary fraction: 17$^{+21}_{-17}$\% with a 70\% confidence interval). Here, we compare our results to those observed for a wide range of masses, evolutionary status and targeted separations and to those predicted by theory \citep[for a thorough review on multiplicity see,][]{Duchene2013}.

\subsubsection{Observed binary fractions}

Our aim here is to compare our findings on MYSOs with those of different classes of objects. To do so it is important to look at the statistics tracing a similar range of separations.

\citet{Kraus2011} targeted a sample of low-mass YSOs (0.25~M$_{\sun}$-2.5~M$_{\sun}$) at separations 3-5000~au and reported a binary fraction between 63-76\% (decreasing with mass). This fraction drops to 55\% at comparable to our study separations (up to $\sim$300~au). \citet{Connelley2008} investigated binarity of embedded Class~I solar type objects targeting separations between 50-4500~au and reported a binary fraction as high as 55\%, dropping down to 15\% for more evolved Class~I objects \citep[for classification of low-mass protostars see,][]{Lada1987}. Once again, if we focus on separations up to $\sim$300~au, the Class~I binary fraction drops to $\sim$13\%. When one moves to the main sequence solar-type stars, \citet{Raghavan2010} reported a binary fraction of 44\%, which drops down to 28\% for separations between $\sim$ 2~au and 300~au. For further statistics on low mass Class 0 protostars see \citet{Chen2013} and \citet{Tobin2016}.

Moving on more massive objects, \citet{Baines2006}, studied the intermediate-mass Herbig Ae/Be stars reporting a binary fraction of $\sim$70\% at 50-750 au separations, and an increasing binary fraction with increasing mass. The non-coverage of the 2~au-50~au separations do not allow a direct comparison with our sample, but at first instance the binary fraction is at least 2 times higher than what we report. \citet{Pomohaci2019} targeted a sample of MYSOs at wide separations (from 600~au and up to 10,000s~au) and report a multiplicity fraction of 31\%. Although the traced separations are 2 orders of magnitude larger than our study, the reported statistics are similar and within the errors. \citet{Karl2018} studied 16 massive young stars (Trapezium), and report a decrease of companions ($<$ 30\%) at separations between 1-100~au, which is in alignment with our findings \citep[for spectroscopic close massive binaries this fraction is as low as 12\%,][]{Apai2007}. We note that the multiplicity of the Trapezium which is probably very dynamically evolved \citep{Allison2009}, which may explain the low fraction of binaries at 1-100~au separations. 

Lastly, massive main sequence stars (O- type) are reported to have a fraction of 53\% at 2-200 au separations \citep{Sana2014}, which is more than a factor of two higher than what we find. More recently, Frost et al. (in prep.) investigate binarity towards a sample of 37 B-type MS stars using PIONIER on the VLTI (H-band observations) and find a very high binary fraction at $\sim$2-180 mas separations (sub-au to few hundreds au). 

When examining the binary fractions at face value for similar ranges of separation (Figure~\ref{fig:Binarstatistics}, Table~\ref{bin_sample}), the MYSO binary fraction is at least a factor of 2-3 lower than the low mass T~Tauris and high-mass main sequence O stars, but similar to what is found towards less evolved Class~I objects, and wider MYSO binaries. We note that the large difference in stellar masses between Class I low-mass objects and MYSOs introduces limitations regarding the natural separations of the binary components, therefore exploring the same ranges of separation may point to different stage of dynamical processes in the evolution of those objects. Massive stars are more commonly found in binaries with separations of up to few hundreds au when they are in a more evolved main sequence phase ($\sim$ 53\%), compared to their forming and young stages ($<$ 30\%). So at first glance, the observed statistics suggest an increase of the massive binary fraction with evolution, which contradicts both observational findings on low mass objects and theoretical predictions {\citep[e.g.,][]{Reipurth2014}. 

We investigate this finding further, and we take a closer look at the specific observations and techniques used in each study. The direct comparison of various samples and methods is limited mainly because the statistics are based on non-uniform observations and techniques. As also demonstrated in Table~\ref{bin_sample}, the targeted studies are characterised by different sample sizes, mass ranges, and sensitivity limits. The present study in particular is based on a sample of only 6 objects, and as a consequence the associated statistics suffer from a large uncertainty. In addition, our observations do not provide a uniform threshold for companion detection within the interferometric field of view. The limited uv-coverage combined to the specific SNR for each source produces a rather separation-dependent contrast threshold for detection \citep[for quantitative studies on those effects see,][]{Absil2011,Davies2018}. We note that the observed uv-coverage alone could result missing out $\sim$ 50\% of companions at the smallest scales ($\sim$ 0.5-4 mas).

The studies based on K-band observations are similar in detection sensitivity ($\Delta$K $\sim$ 4-5) with the exception of that of solar-type MS stars ($\Delta$K $\sim$ 2.5), which can partially explain the observed drop in their binary fraction as they move from the pre-main sequence phase. In Table~\ref{bin_sample} we can see that although the sample presented in \citep{Sana2014} targets very similar separations as our study, we trace a mass range which goes up to the lowest limit of the mass traced in that study. Taking this into consideration is more sensible to compare our fraction of that of MS B-type stars which cover similar mass ranges as our study. We note that the separations traced by Frost et al. (submitted) cover our separation ranges. In addition, both studies on MS OB stars are characterised by similar sensitivity in H-band, while our study uses K-band observations. The different ages and filters between the current study and those studies, probably suggests that the instrumental sensitivity corresponds to a different range of physical masses of the candidate companions. Even after one takes all these limitations into consideration it is still difficult to attribute the striking difference between the very high binary fraction of MS B-type stars and the low binary fraction of MYSOs (17\%) on observational biases alone. Lastly, \citet{Oudmaijer2010} studied massive stars in MS, and in particular a sample of Be stars and ``normal'' B stars in K-band targeting $\sim$30-2400~au separations. The binary fraction of the combined sample is $\sim$30\% (29$\pm$5). 

To summarise, to be able to provide a confidence level on the reported fractions, and confirm or discredit the observed increase in binary fraction with age, it is necessary to survey homogeneous samples, using similar observational techniques and targeting stars at similar distances. Future studies should adopt an approach similar to the one presented in Figure~\ref{fig:Binarstatistics}, but making use of large sample observations that strictly trace i) the same primary mass ranges in each bin, ii) the same companion mass ranges sampled in each bin, and iii) a comparable dynamical evironment of the samples (ie. no, or the same, external influence on multiplicity).

When we turn our focus on the binary fraction of MYSOs in the K-band at different scales but similar sensitivity $\Delta$K (this study and \citet{Pomohaci2019}), we observe that the binary fraction is about the same (within the errors, 17$^{+21}_{-17}$\% to 31$\pm$8\%) and can be probably considered flat within the entire separation range of 2-10,000~au. We note that this comparison is among the most robust ones we have.

\begin{table*}
\caption{Stellar samples targeting binary companions at $\sim$ 2~au and 300~au separations. The binary fractions at those separations are also reported.}
\small
\centering
\setlength\tabcolsep{2pt}
\begin{tabular}{c c c c c c c c}
\hline\hline
& \multicolumn{2}{c}{Embedded} &
\multicolumn{2}{c}{PMS} &
\multicolumn{2}{c}{MS} \\
\hline
 & MYSOs & Class~I & Young OBs  & YSOs & O-type & Solar type \\ 
\hline\hline
Sample  & 6 & 267 & 16 & 152 & 279 & 454 \\
Mass (M$_\odot$) & 8-16 & 0.5 - 100 (L$_\odot$) & 6.7 - 39 & 0.25 - 2.5 & $>$ 15  & $\sim$1 \\
Method  & Interferometry (I)  & AO imaging  & Interferometry  & Aperture Masking (AM) & (I+AM)  & All combined  \\
Sensitivity  & $\Delta$K $\sim$ 5  & $\Delta$L $\sim$ 4  &  $\Delta$K $\sim$ 5 &  $\Delta$K $\sim$ 4 & $\Delta$H $\sim$ 5 &  $\Delta$V$<$3, $\Delta$K$<$ 2.5 \\
Binary fraction & 17$^{+21}_{-17}$\% & 15$\pm$2\% & 30$\pm$11\% & 55$\pm$6\% & 53$\pm$5\% & 28$\pm$2\% \\
\hline\hline
\label{bin_sample}
\end{tabular} 

\tiny {\bf{Notes}}: The references of each class of objects are: This study (MYSOs), \citet[][; Class~I]{Connelley2008}, \citet[][; Young OBs]{Karl2018}, \citet[][; YSOs]{Kraus2011}, \citet[][; MS OBs]{Sana2014}, \citet[][; MS Solar type]{Raghavan2010}. 

\end{table*}

\begin{figure*}[h]
\begin{center}
  \includegraphics[scale=0.5]{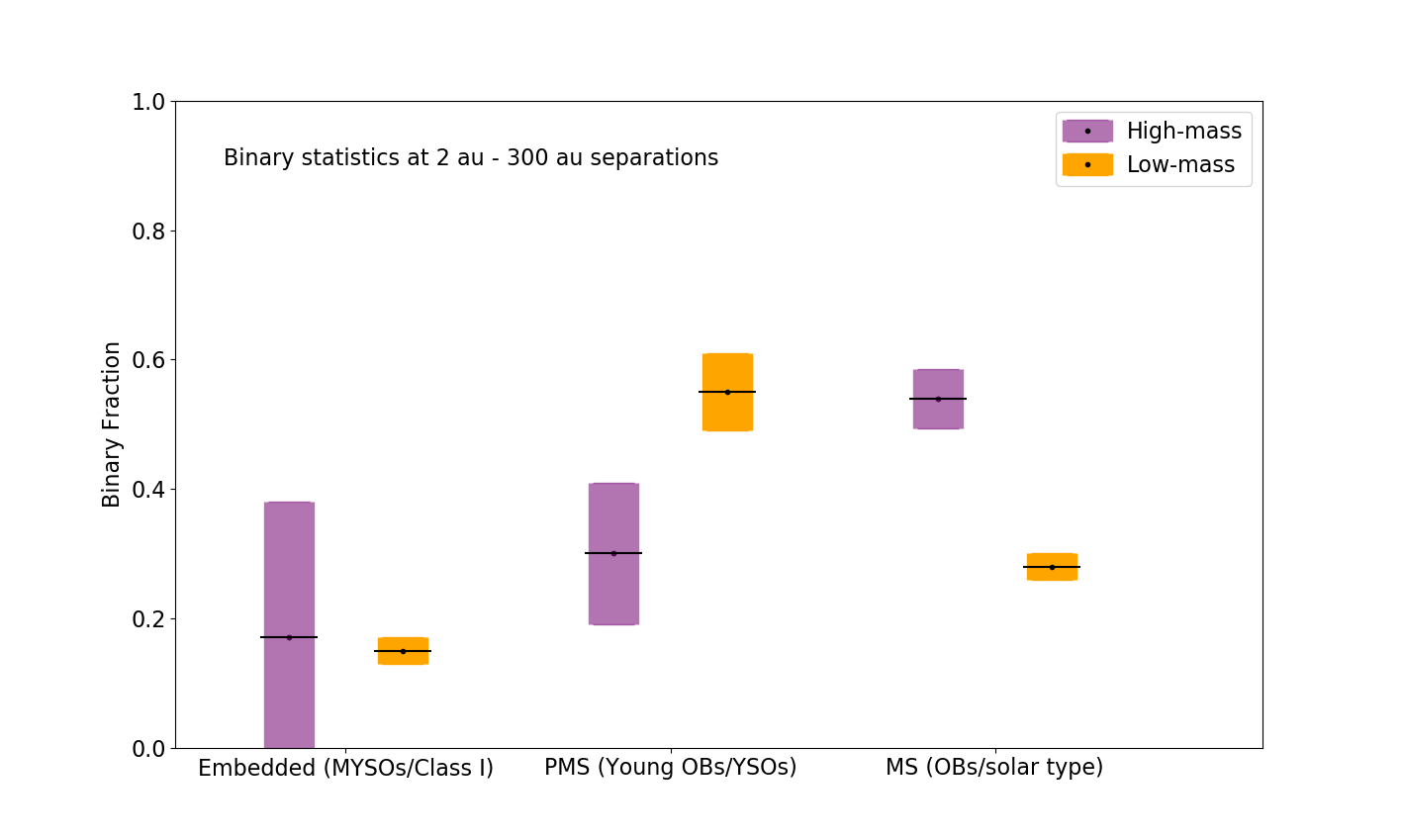}
\end{center}
          \caption{Binary statistics (see Table~\ref{bin_sample}) for companions located between 2~au and 300~au for low and high mass stars of various evolutionary stages: i) embedded: low-mass Class~I vs MYSOs (8~M$_{\odot}$ - 16~M$_{\odot}$), ii) PMS: YSOs (0.25~M$_{\odot}$ - 2.5~M$_{\odot}$) vs Young OBs, iii) MS: OBs vs solar-type.}
   \label{fig:Binarstatistics}
       \end{figure*}

\subsubsection{Comparison with theory}

Some of the most prominent theories on binary formation are those of the core accretion and fragmentation, disc fragmentation, and stellar migration or capture. Numerical simulations of core collapse fragmentation can predict multiples with separations of several hundreds of au \citep[][]{Myers2013}. \citet{Meyer2018} predicts tighter binaries with unequal components and early-stage separations of a couple of hundreds au and down to less than tens of au as the system evolves, through accretion disc fragmentation. Such systems are tighter than what disc fragmentation theories previously predicted \citep[100~au-1000~au;][]{Kratter2006}. Tighter binaries can also be a result of a capture \citep{Bonnell2005} or magnetic braking during accretion \citep{Lund2018}, or external stellar interactions \citep{Bate2002}; We note that most of these theories start with separations of several hundreds of au before they evolve to tighter systems. Recently, \citet{Ramirez2021} present evidence for an inward migration of stars as a function of a cluster age, indicating that massive binaries start their lives in wide pairs ($\sim$100~au) before they evolve into tighter massive binaries after about 1.5 Myr. We note that this timescale is more than an order of magnitude larger than the age of PDS~37 \citep[$\sim$0.06~Myr;][]{Vioque2018}. \citet{Sana2017} favoured such a hardening scenario to explain the lack of close companions at birth ($\sim$12\%), starting off with pairs at $>$ 0.5~au separations. Observational studies of massive binaries in the PMS phase are crucial to verify, distinguish and inform the several formation theories in place \citep[e.g.,][]{Moe2017}.  

Taking all this into consideration, our findings in terms of separations ($<$100~au) appear to be more consistent with the theoretical predictions of disc fragmentation, which, as it evolves, favours a system of a main high-mass component and accreting low-mass companions at tight separations \citep{Meyer2018}. We note that the inclination of the binary and the individual components could not be extracted from the geometric modelling alone and projection effects cannot be excluded (i.e., the physical separation of the system may be larger). The knowledge of the orbital inclination of such systems and that of their circumstellar/circumbinary discs is very valuable information in constraining and distinguishing among the proposed theories. In particular, disc fragmentation predicts mostly coplanarity among the orbital plane and the discs of the stellar systems \citep{Kratter2006}, which is also what was observed towards Herbig stars \citep{Wheelwright2011}. On the other hand, binary formation via capture predicts random orientations between discs and orbital plane.

\section{Summary}

\label{conclusions}

We present the first interferometric survey in K-band of six massive YSOs. Our study increases the MYSOs with K-band interferometric measurements by a factor of 4. Below, we summarise our findings on the characteristic sizes of the hot dust at 2.2~$\mu$m and ionised gas (Br$\gamma$), and on the high-mass binarity at milli-arcsecond scales using spatial information. 

\begin{itemize}
\item We spatially resolve the crucial star/disc interface in a sample of MYSOs in K-band, and finally confirm observationally the prominence of au-scale discs in high-mass star formation.
\item The K-band continuum emission is spatially resolved for all MYSOs in our sample. The 2.2~$\mu$m measured characteristic size of MYSOs shows a large scatter for the given range of luminosities, but it is overall consistent with the location of the inner rim (i.e., dust sublimation radius) of a disc with an optically thin cavity. 
\item When the inner sizes of MYSOs are compared to those of lower mass Herbig AeBe and T Tauri stars, they seem to follow a universal trend at which the sizes scale with the square-root of the stellar luminosity. Such a trend indicates that similar radiative processes take place at inner regions of young stars independently of their mass.
\item The measured continuum and Br$\gamma$ visibilities of G034.8211 could be only fitted with a flattened/elongated disc geometry, indicative of a close to edge-on geometry. For the rest of MYSOs in our sample we find no signatures of a flattened brightness distribution in their visibility curves. 
\item We find that the Br$\gamma$ emission is comparable or more compact in size with respect to the thermal emitting dusty region, and the two emissions are spatially aligned. This new finding gives credence to disc winds/disc accretion models to describe the geometry of the inner parts of MYSOs, which appear to be prominent in massive star formation.
\item G282.2988 is the only MYSO in our sample of six, which required a binary geometry to fit the interferometric observables. Therefore, we report an MYSO binary fraction of 17$\pm$15\% in the K-band at the traced scales (few au to few hundreds au). This fraction is comparable to what was previously reported at 600-10,000~au scales, indicating a flat fraction for a wide range of separations.
\item Based on the present statistics, MYSO binaries at 2~au-300~au separations are less common than massive main sequence stars on similar scales. This finding contradicts the observational findings towards low-mass stars and the theoretical predictions of a decrease in multiplicity with evolutionary stage, but is not deprived from observational biases and further investigation is needed.
\end{itemize}

In this paper, we adopt a simple approach to understanding the NIR size distribution of MYSOs with respect to other classes of lower mass YSOs for given dust destruction predicted by three basic disc models. We find that MYSOs follow similar behaviour to that of low luminosity T Tauri, Herbig Ae and most of the Herbig Be stars. In particular, the 2.2~$\mu$m size of MYSOs can directly be related to the dust destruction radius predicted by an optically thin scenario where the inner rim is directly heated by the central star. A more detailed physical modelling, where the interferometric disc sizes of this sample of MYSOs will be fitted simultaneously with their SEDs, and NIR spectra (covering the Br$\gamma$, NaI and CO emission), is necessary to form a more detailed picture on the innermost environment of those enigmatic objects, and will be the aim of future studies. 

The present study is the first attempt to address the multiplicity of MYSOs in milli-arcsecond scales, and in particular via direct spatial measurements. Multi-wavelength observations (e.g., PIONIER; H-band, MATISSE; M-, N- bands on VLTI) of large sample of MYSOs will be a great asset in providing binary statistics and in fully identifying and characterising companions.

\begin{acknowledgements}
We would like to thank the anonymous referee for provid-ing helpful comments and suggestions that improved the paper. E.K. is funded by the STFC (ST/P00041X/1). J.D.I. acknowledges support from the Science and Technology Facilities Council of the United Kingdom (STFC) under ST/T000287/1. A.C.G. has received funding from the European Research Council (ERC) under the European Unions Horizon 2020 research and innovation programme (grant agreement No. 743029). We would like to thank Jacques Kluska for stimulating discussions when preparing the paper. This publication is based on observations collected at the European Southern Observatory under ESO programme(s) 0102.C-0838, 0103.C-0459 (GRAVITY), and 092.C-0064 (AMBER). This research has made use of the SIMBAD data base, operated at CDS, Strasbourg, France. The GRAVITY data reduction was undertaken using ARC3, part of the High Performance Computing facilities at the University of Leeds, UK. This research has made use of the \texttt{AMBER data reduction package} of the Jean-Marie Mariotti Center\footnote{Available at http://www.jmmc.fr/amberdrs}.
\end{acknowledgements}

\bibliographystyle{aa} 
\bibliography{ref}


\begin{appendix}

\section{Observations}

\begin{figure*}
\begin{multicols}{2}
\includegraphics[width=\linewidth]{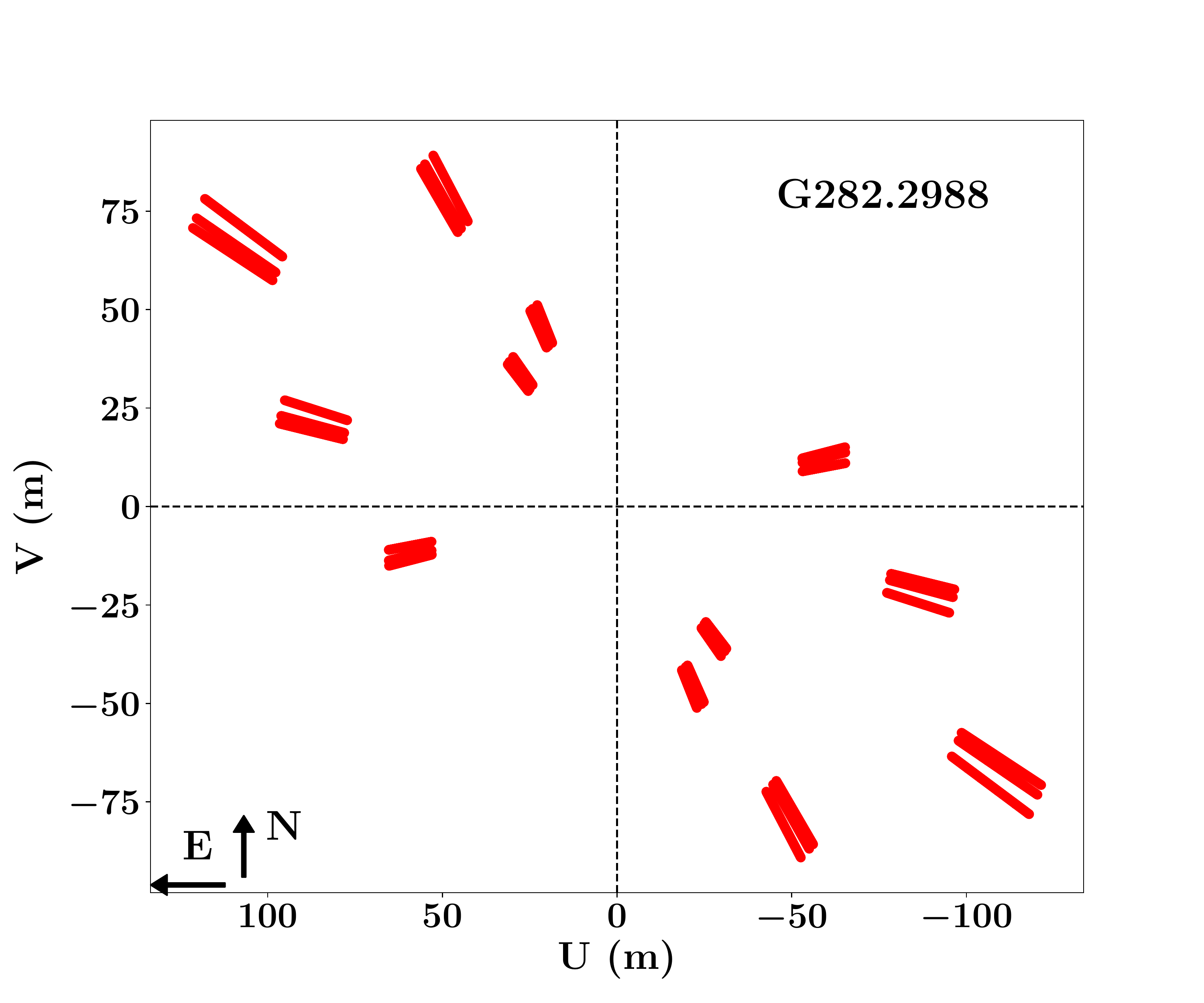}\par
\includegraphics[width=\linewidth]{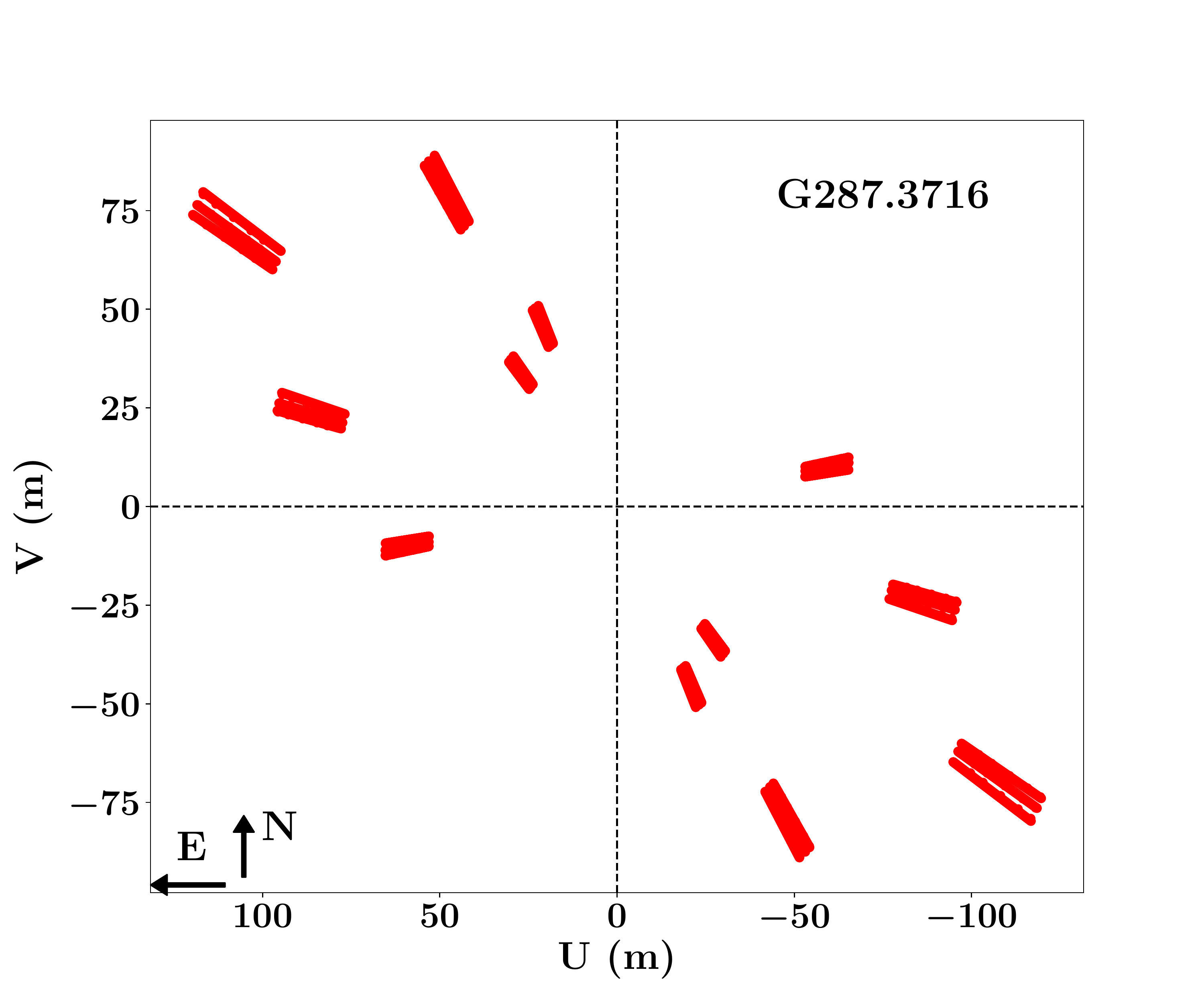}\par
\includegraphics[width=\linewidth]{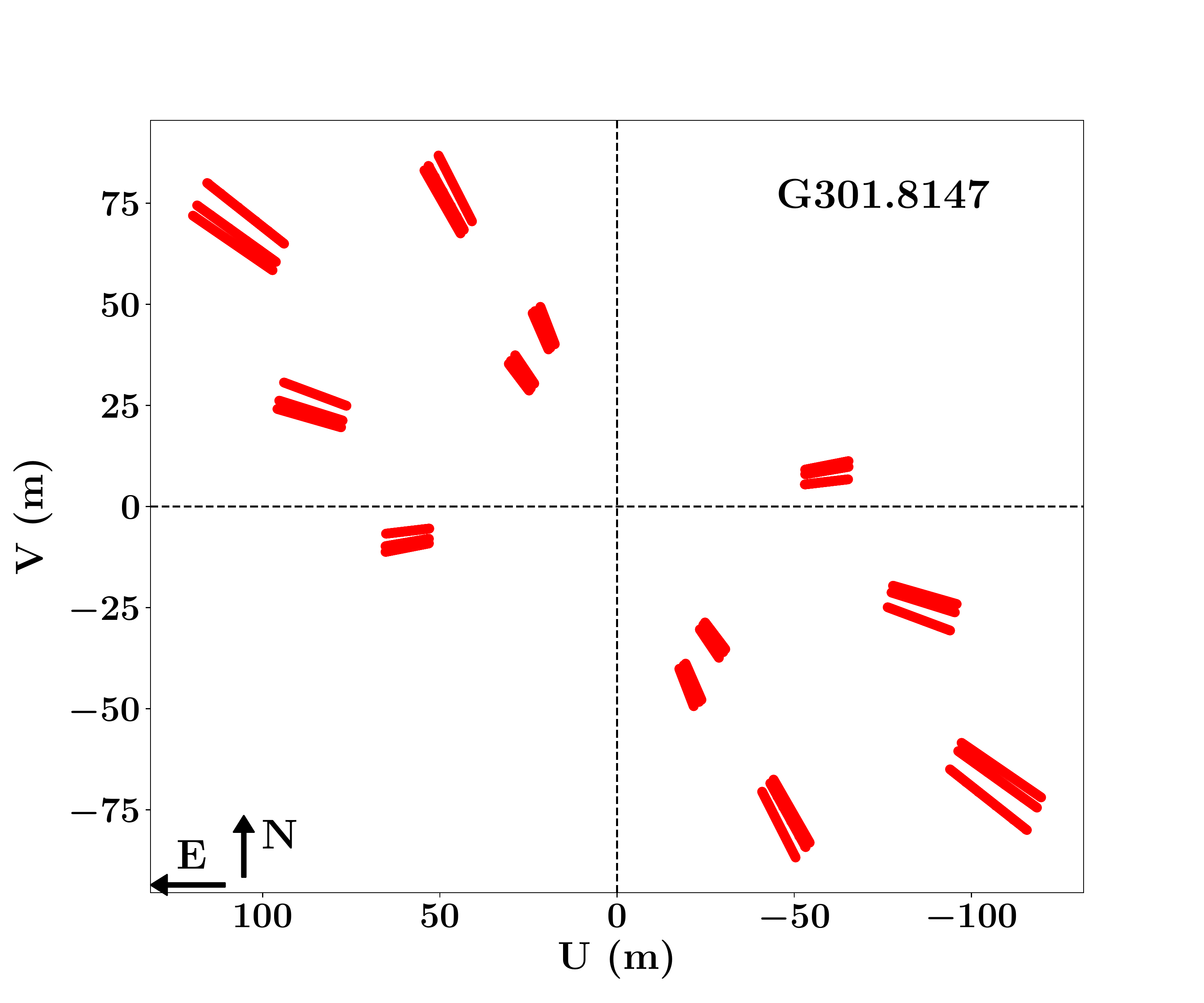}\par
\includegraphics[width=\linewidth]{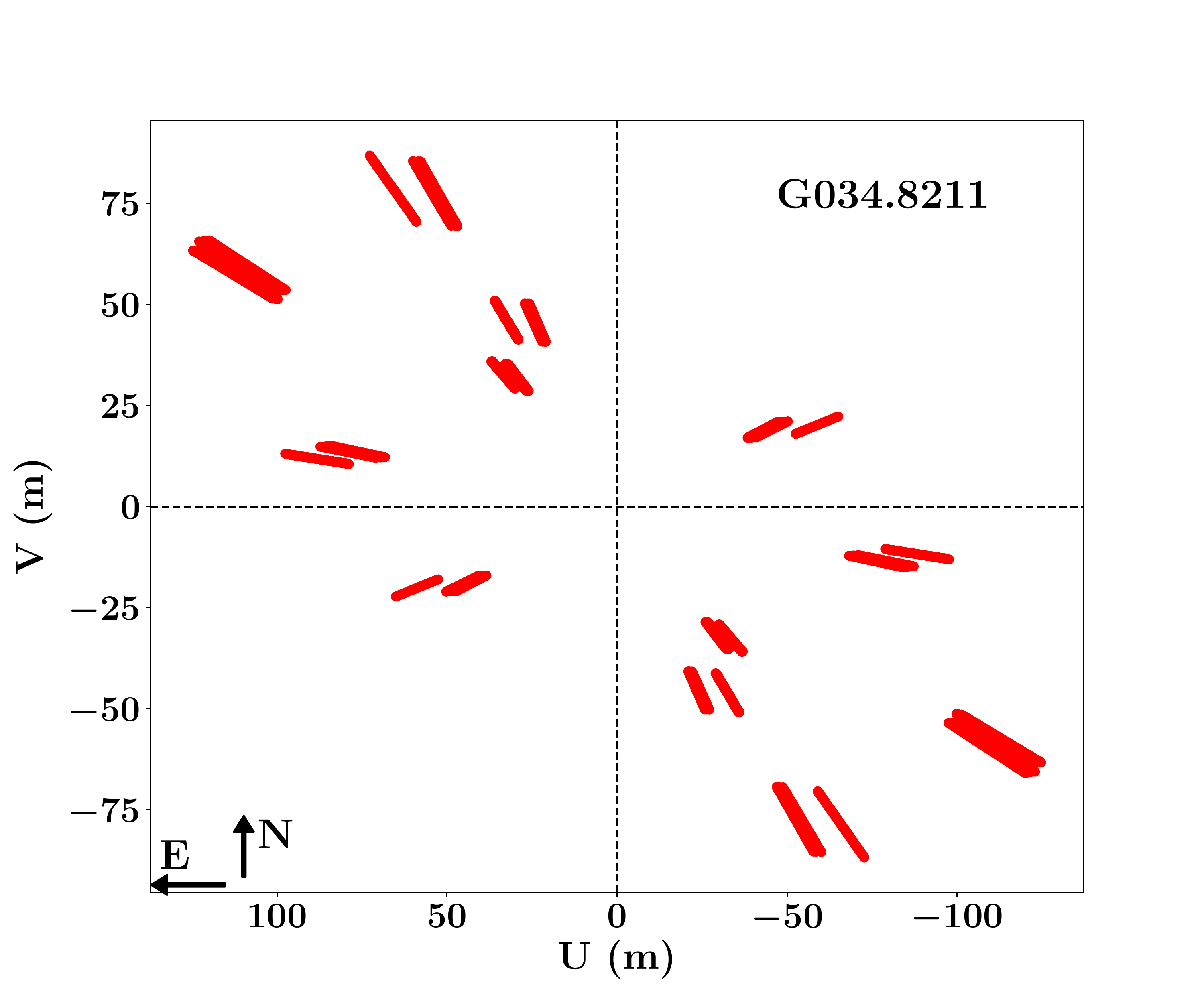}\par
\end{multicols}
\caption{uv-plane coverage of VLTI/GRAVITY of G282.2988, G287.3716, G301.8147, and G034.8211. North (N) is taken as position angle PA=0$^\circ$ and East (E) as PA=90$^\circ$.}
\label{fig:uvplaneG282}
\end{figure*}

\begin{figure*}
\begin{multicols}{2}
\includegraphics[width=\linewidth]{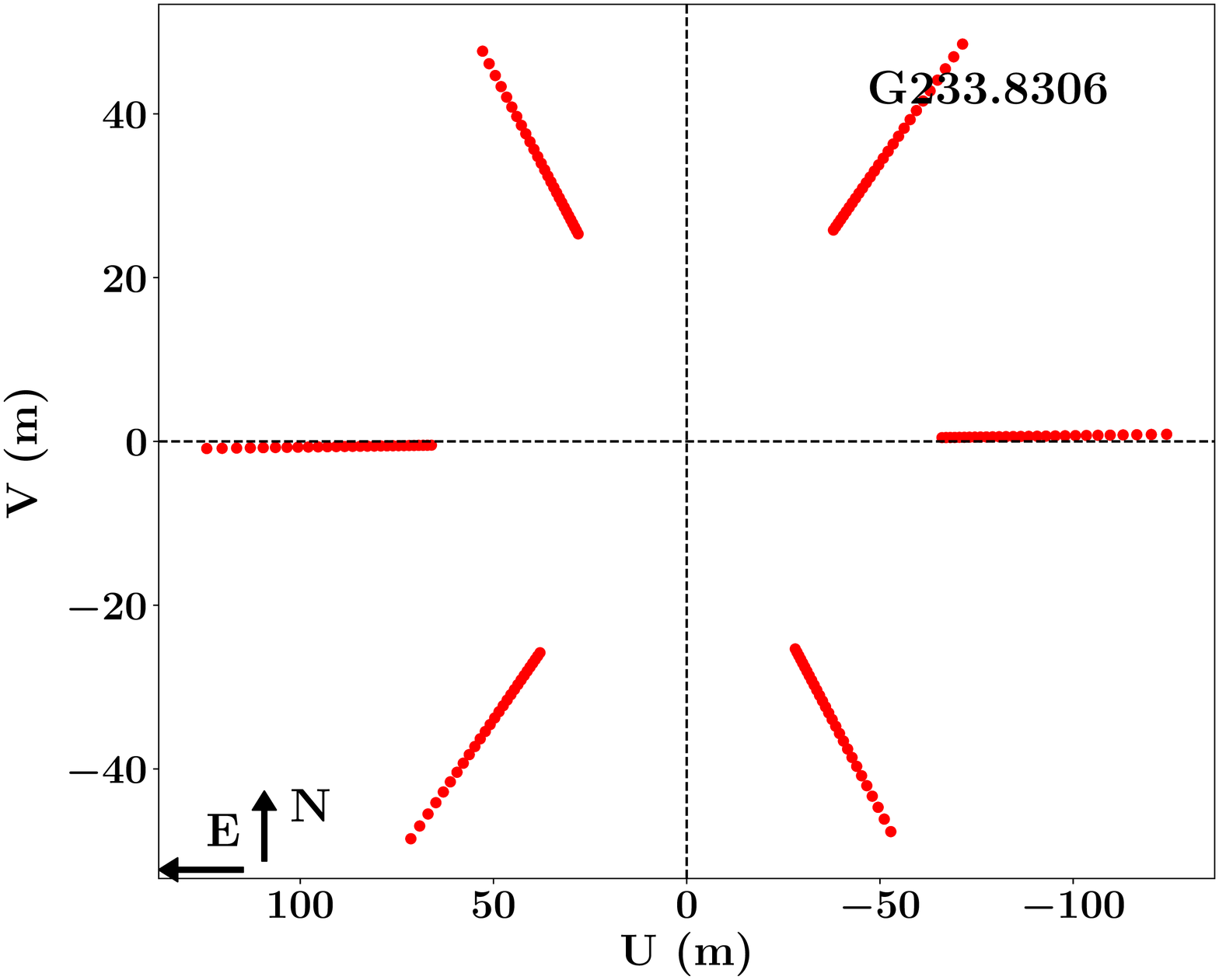}\par
\includegraphics[width=\linewidth]{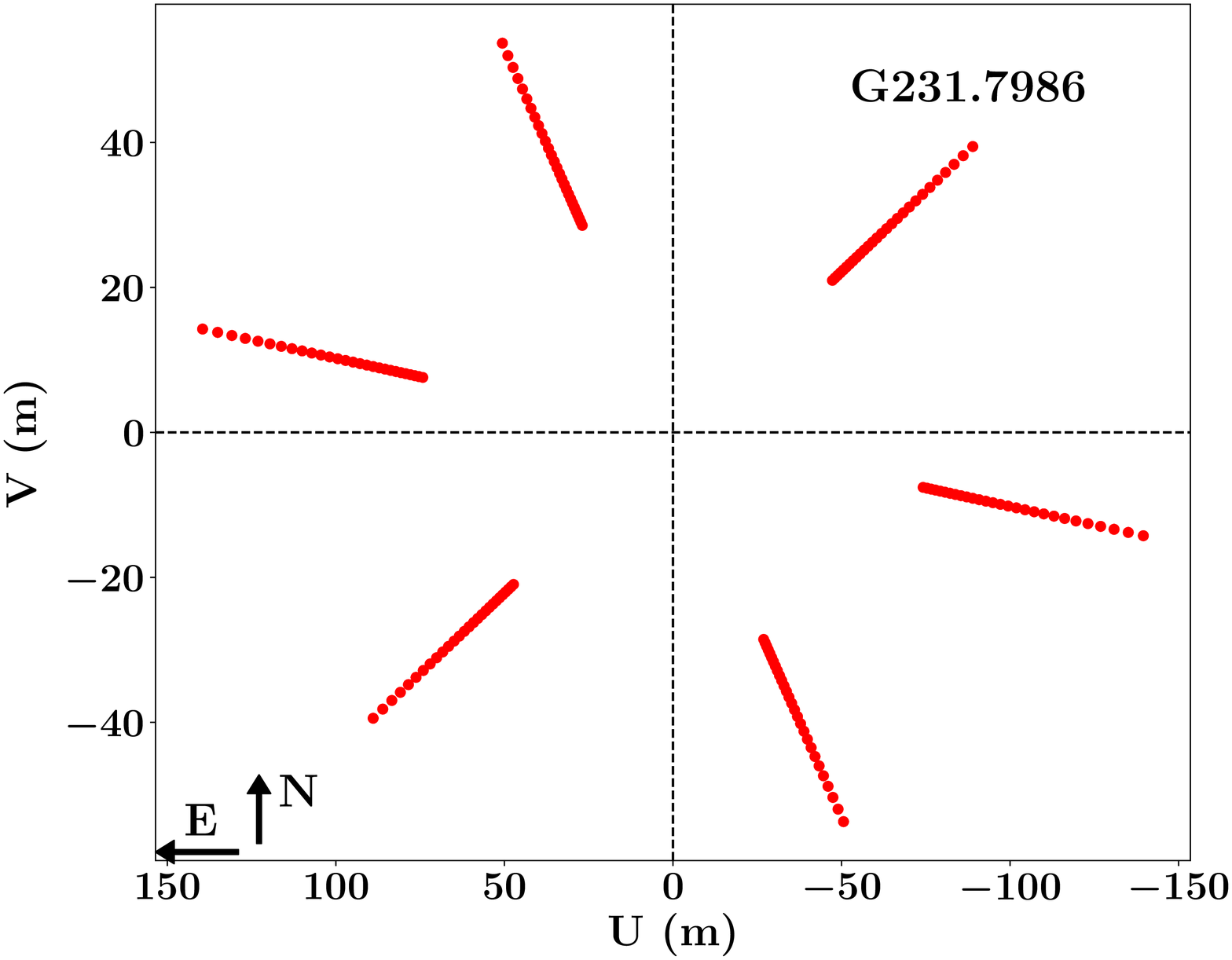}\par
\end{multicols}
\caption{uv-plane coverage of VLTI/AMBER of G233.8306, and G231.7986. North (N) is taken as position angle PA=0$^\circ$ and East (E) as PA=90$^\circ$.}
\label{fig:uvplaneAMBER}
\end{figure*}

\begin{figure*}[h]
\includegraphics[scale=0.5]{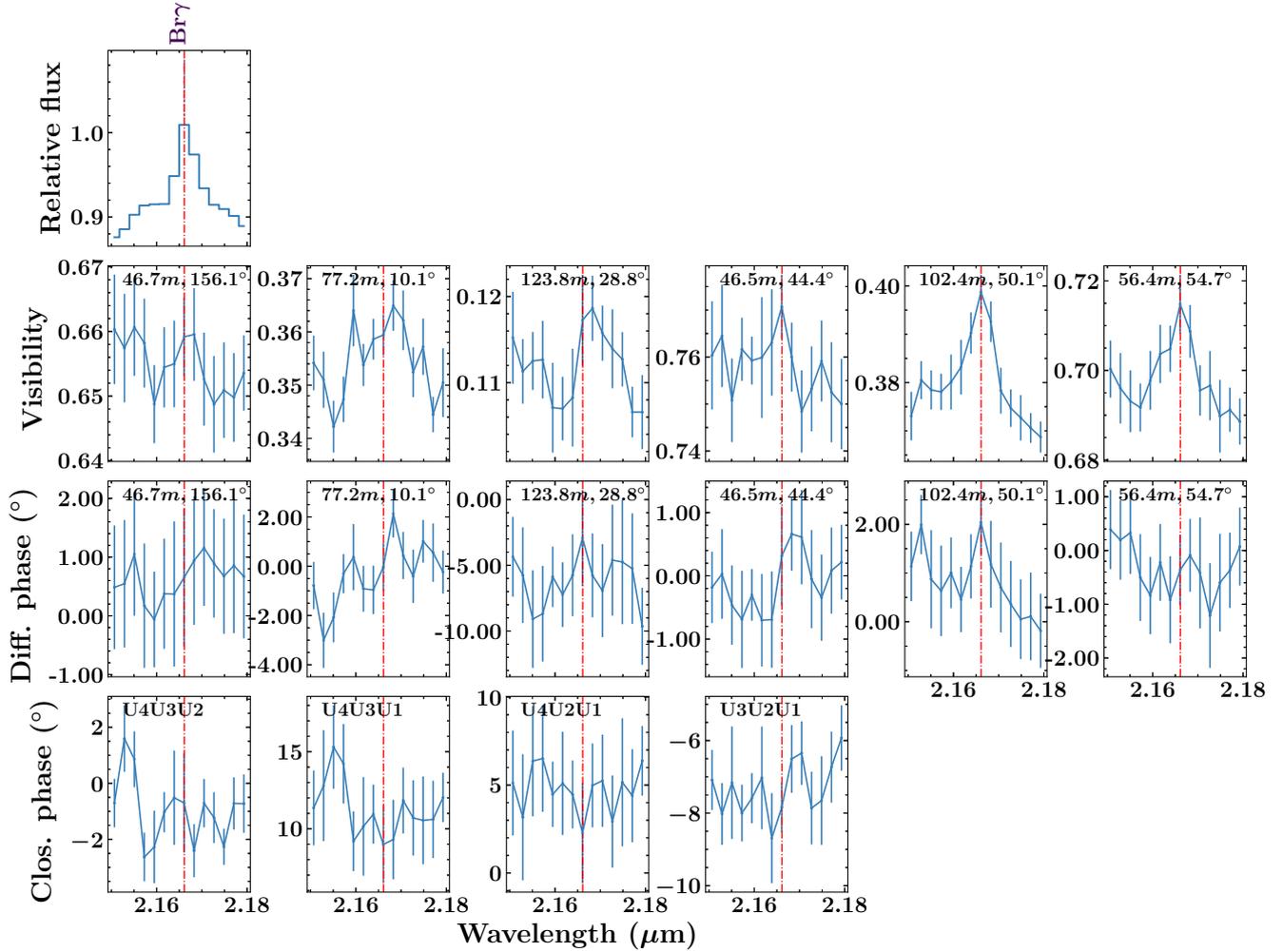} 
          \caption{Relative flux, visibility, differential and closure phase as a function of wavelength around the Br$\gamma$ emission (and continuum around it) towards G034 using GRAVITY on the UTs. The baseline length, and the position angle for all observed 6 baselines and 4 triplets of the UT configuration are also given.}
   \label{fig:vis_brg1}
       \end{figure*}

\begin{figure*}[h]
\includegraphics[scale=0.5]{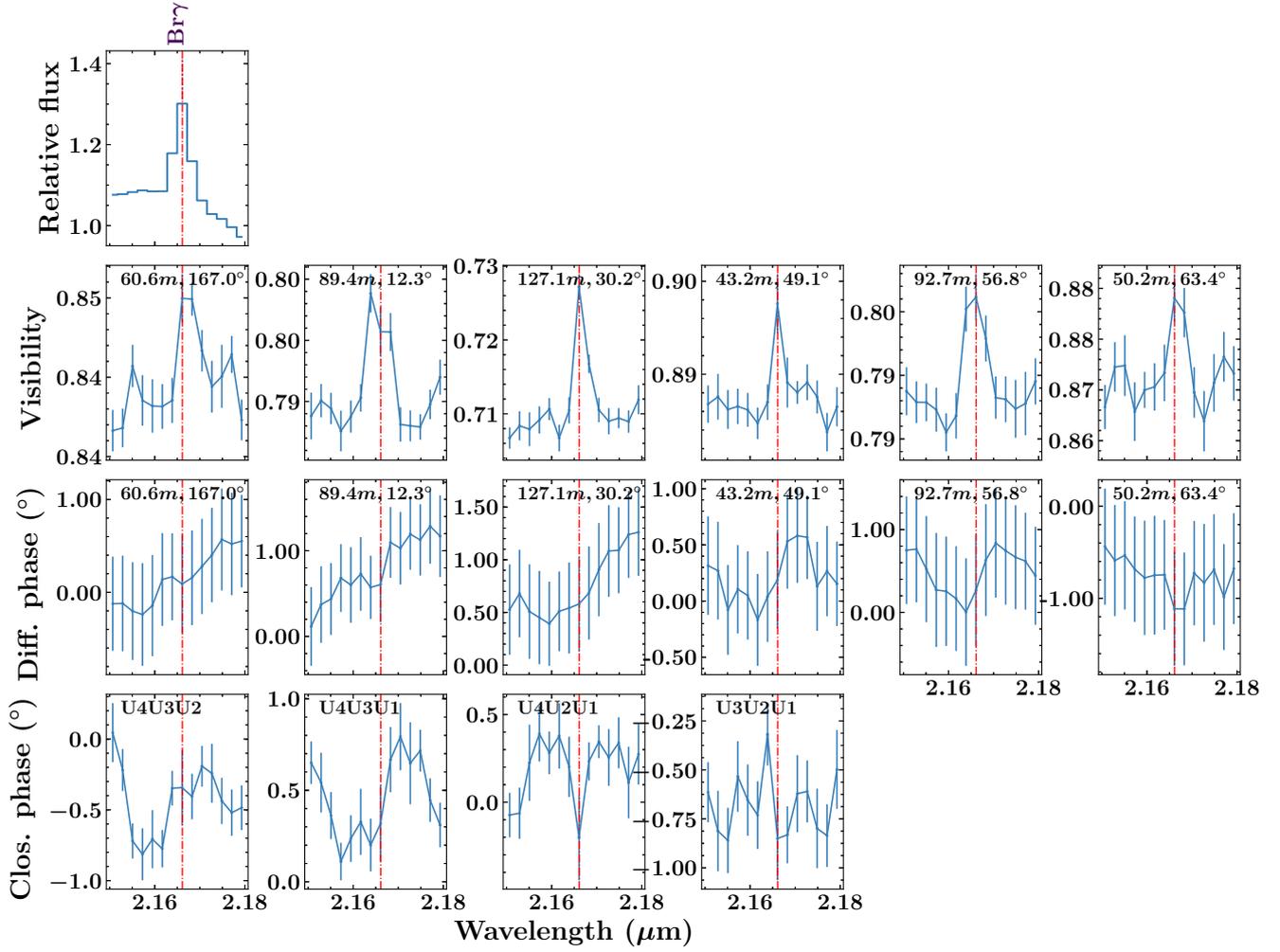} 
          \caption{Same as Fig.~\ref{fig:vis_brg1} but for G282.}
   \label{fig:vis_brg2}
       \end{figure*}

\begin{figure*}[h]
\includegraphics[scale=0.5]{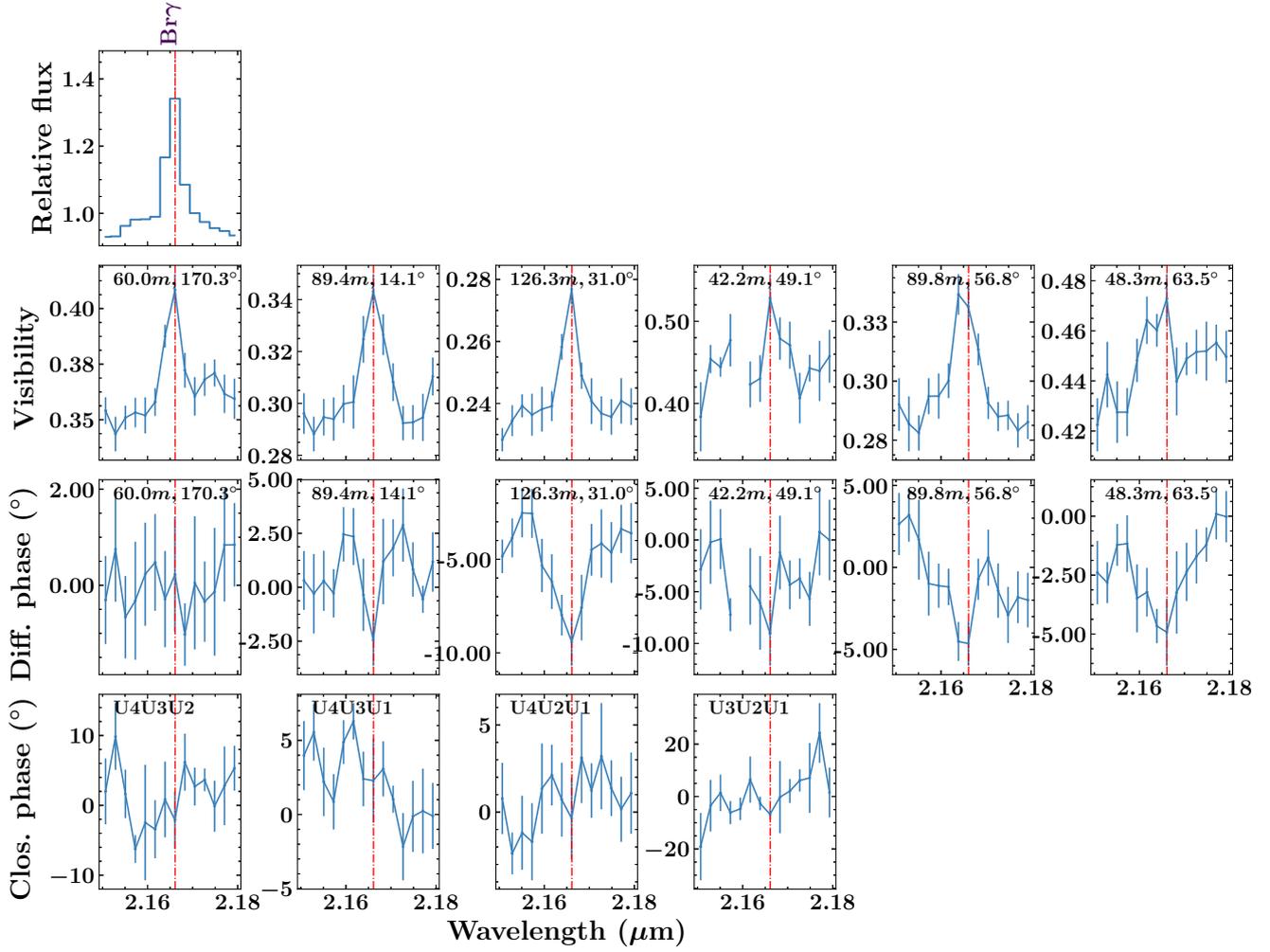} 
          \caption{Same as Fig.~\ref{fig:vis_brg1} but for G301.}
   \label{fig:vis_brg3}
       \end{figure*}

\begin{figure*}[h]
\includegraphics[scale=0.5]{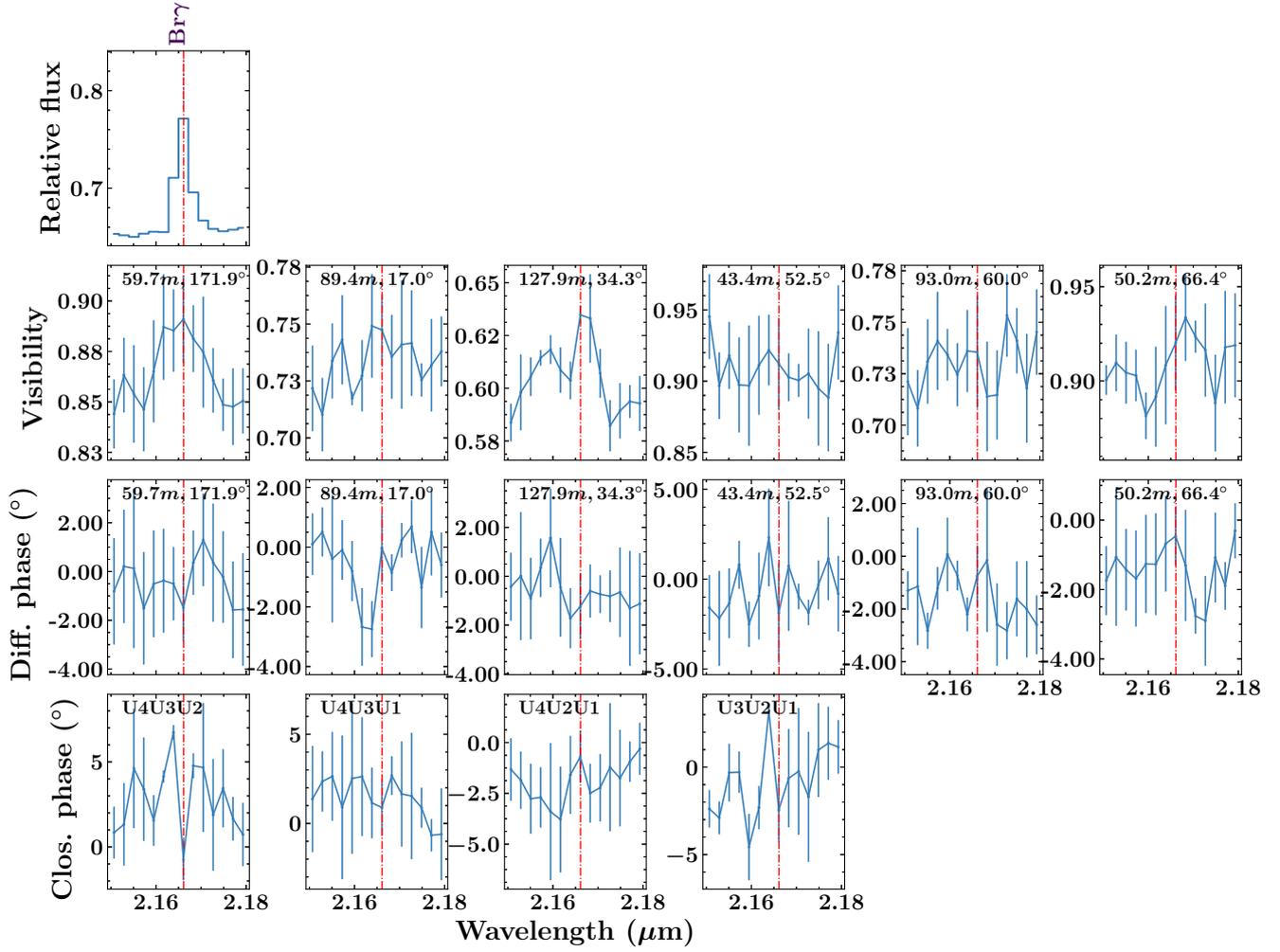} 
          \caption{Same as Fig.~\ref{fig:vis_brg1} but for G287.}
   \label{fig:vis_brg4}
       \end{figure*}

\begin{table*}
\caption{Observed MYSOs with their associated calibrators. The spectral types, K-magnitudes and K-band sizes (uniform disc) of the calibrators are also reported.}
\small
\centering
\setlength\tabcolsep{2pt}
\begin{tabular}{c c c c c}
\hline\hline
Science target & Calibrator & K & Spectral  & Size  \\ & & (mag) & type & (mas)  \\
\hline\hline
G282.2988  & HIP 49790 & 5.6 & G5 E & 0.35 \\
G287.3716  & HIP 55010  & 6.9 & F2IV/V & 0.16 \\
G301.8147  & HIP 61444  & 5.4 & F2/3 (III) & 0.32  \\
G034.8211  & HD 175312  & 5.05 & K2III &  0.49 \\
G231.7986  & HIP 35922  & 6.4 & A7IV & 0.2  \\
G233.8306  & HIP 36392  & 5.6 & G0V & 0.32  \\
\hline\hline
\end{tabular} 

\tiny {\bf{Notes}}: The spectral types, K-magnitudes and uniform disc K-band diameters of the observed calibrators are from the JMMC SearchCal \citep{Bonneau2011}. The calibrators were also used as a telluric standard and during the normalisation of the spectra.

\label{calibrators}
\end{table*}

\begin{table*}[ht]
\caption{Technical overview of the GRAVITY observations on UTs of the four observed MYSOs at the beginning of their observing run. The DIT is the individual exposure time, and $\tau_{coh}$ is the coherence time in visible. The visibilities of the continuum are also reported.}
\small
\centering
\setlength\tabcolsep{2pt}
\begin{tabular}{c c c c c c c c c c c c c }
\hline\hline
Source/ & Date & Station & Baseline & PA & DIT & $\tau_{\rm coh}$ & Seeing & V$_{\rm cont}$  \\ Config.& & & (m) & ($^\circ$) & (s) & (ms) & (arcsec) &  \\
\hline\hline
{\bf{G034.8211}}/ &  &  &  &  &  &  &  &   \\
U1-U2-U3-U4 & 2019-04-23 & U4U3 & 62.2 & 161.1 & 5 & 3 & 1.1 & 0.456$\pm$0.002   \\
& & U4U2 & 88.6 &  7.5 &  &  &  & 0.175$\pm$0.002  \\
& & U4U1 &124.9 & 27.1 &  &  &  & 0.062$\pm$0.007  \\
& & U3U2 & 43.0 & 47.7 &  &  &  & 0.776$\pm$0.003  \\
& & U3U1 & 93.2 & 55.9 &  &  &  & 0.374$\pm$0.002  \\
& & U2U1 & 51.0 & 62.8 &  &  &  & 0.721$\pm$0.003  \\
\hline
{\bf{G282.2988}}/ &  &  &  &  &  &  &  &   \\
U1-U2-U3-U4 & 2018-12-20 & U4U3 & 59.9 & 170.4 & 10  & 12 & 0.45 & 0.8448$\pm$0.0003 \\
 & & U4U2 &  89.4 & 15.8 &  &  &  & 0.7900$\pm$0.0004  \\
 & & U4U1 & 128.0 & 33.5 &  &  &  & 0.7040$\pm$0.0005   \\
 & & U3U2 &  43.6 & 51.9 &  &  &  & 0.8898$\pm$0.0004   \\
 & & U3U1 &  93.6 & 59.4 &  &  &  & 0.7833$\pm$0.0005   \\ 
 & & U2U1 &  50.7 & 65.9 &  &  &  & 0.8708$\pm$0.0005   \\
\hline
{\bf{G287.3716}}/ &  &  &  &  &  &  &  &   \\
U1-U2-U3-U4 & 2019-04-23 & U4U3 & 59.7 & 171.9 & 30  & 5 & 0.49 & 0.850$\pm$0.005   \\
 & & U4U2& 89.4 & 16.9 &  &  &  & 0.728$\pm$0.005   \\
 & & U4U1&127.8 & 34.3 &  &  &  & 0.598$\pm$0.004   \\
 & & U3U2& 43.4 & 52.4 &  &  &  & 0.914$\pm$0.009   \\
 & & U3U1& 92.9 & 59.9 &  &  &  & 0.717$\pm$0.006   \\
 & & U2U1& 50.2 & 66.4 &  &  &  & 0.904$\pm$0.005   \\
\hline
{\bf{G301.8147}}/ &  &  &  &  &  &  &  &   \\
U1-U2-U3-U4 & 2019-01-22 & U4U3 & 59.3 & 174.1 & 10  & 19 & 0.54 & 0.363$\pm$0.002   \\
 & & U4U2 &  89.4 & 18.0 &  &  &  & 0.307$\pm$0.002   \\
 & & U4U1 & 127.1 & 34.6 &  &  &  & 0.230$\pm$0.002   \\
 & & U3U2 &  42.6 & 52.4 &  &  &  & 0.436$\pm$0.010   \\
 & & U3U1 &  90.7 & 59.8 &  &  &  & 0.273$\pm$0.004   \\
 & & U2U1 &  48.7 & 66.3 &  &  &  & 0.414$\pm$0.003   \\
\hline\hline
\end{tabular}
\label{gravity_tech}
\end{table*}

\begin{table*}
\caption{Technical overview of the AMBER observations on UTs (U324 configuration) of the observed MYSOs at the beginning of their observing run on the night of 13 March 2014. The DIT is the individual exposure time, and $\tau_{coh}$ is the coherence time.}
\small
\centering
\setlength\tabcolsep{2pt}
\begin{tabular}{c c c c}
\hline\hline
Source  & DIT & $\tau_{coh}$  & seeing  \\  & (ms) & (ms) & (arcsec) \\
\hline\hline
G231.7986  & 26 & 2.5 & 1.3 \\
G233.8306  & 26 & 3.0 & 1.0  \\
\hline\hline
\end{tabular} 
\label{AMBER_weather}
\end{table*}

\section{Image reconstruction}
\label{image_rec}

In addition to the simple geometric modelling of the visibilities, we investigate asymmetries (making use of the closure phases) of the brightness distribution of the 2.2~$\mu$m continuum emission. To this end we perform model-independent image reconstruction towards a sample of MYSOs for the first time, and in particular towards the sources observed with GRAVITY. The quality of the image reconstruction increases with: i) the ratio between the emitting region and the angular resolution, ii) the brightness of the source, and iii) the coverage of the uv-plane. In our case, we are limited by all these 3 factors (e.g., Figure~\ref{fig:uvplaneG282}), and therefore the resulting images are highly influenced by the beam shape of the observations and cannot trace detailed structures of the emission. Nevertheless the image reconstruction algorithms make it possible to independently measure the size of the 2.2~$\mu$m continuum emission and to make use of the closure phases to trace possible asymmetries, and to investigate structures that simple geometrical model fitting may have missed (See Sect.~\ref{2micron}, Table~\ref{litpro_results}). 

Figure~\ref{fig:image_reconstruction_G034} presents the resulting image reconstruction using WISARD \citep{Meimon2004,Meimon2008}, which is implemented in JMMC OImaging tool \footnote{Available at http://www.jmmc.fr/oimaging; part of the European Commission's FP7 Capacities programme (Grant Agreement Number 312430)} towards G282.2988, G287.3716, G301.8147, and G034.8211. For direct comparison with the geometric modelling, we present the measured sizes of a 2D Gaussian distribution of the reconstructed images in Table~\ref{litpro_results}. The measured reconstructed sizes of G034, G301, and G287, differ by 0.2\%, 3\%, and 8\% respectively, compared to the sizes obtained by geometric modelling after applying a gaussian brightness distribution. This difference is up to 27\% for G282. We attribute the observed asymmetries towards G287.3716, G301.8147 to image reconstruction artifacts, since they cannot be justified by the small observed closure phases of the sources.

\begin{figure*}
\begin{multicols}{4}
\includegraphics[scale=0.2,width=\linewidth,trim={2.0cm 2.0cm 2.0cm 4.3cm},clip]{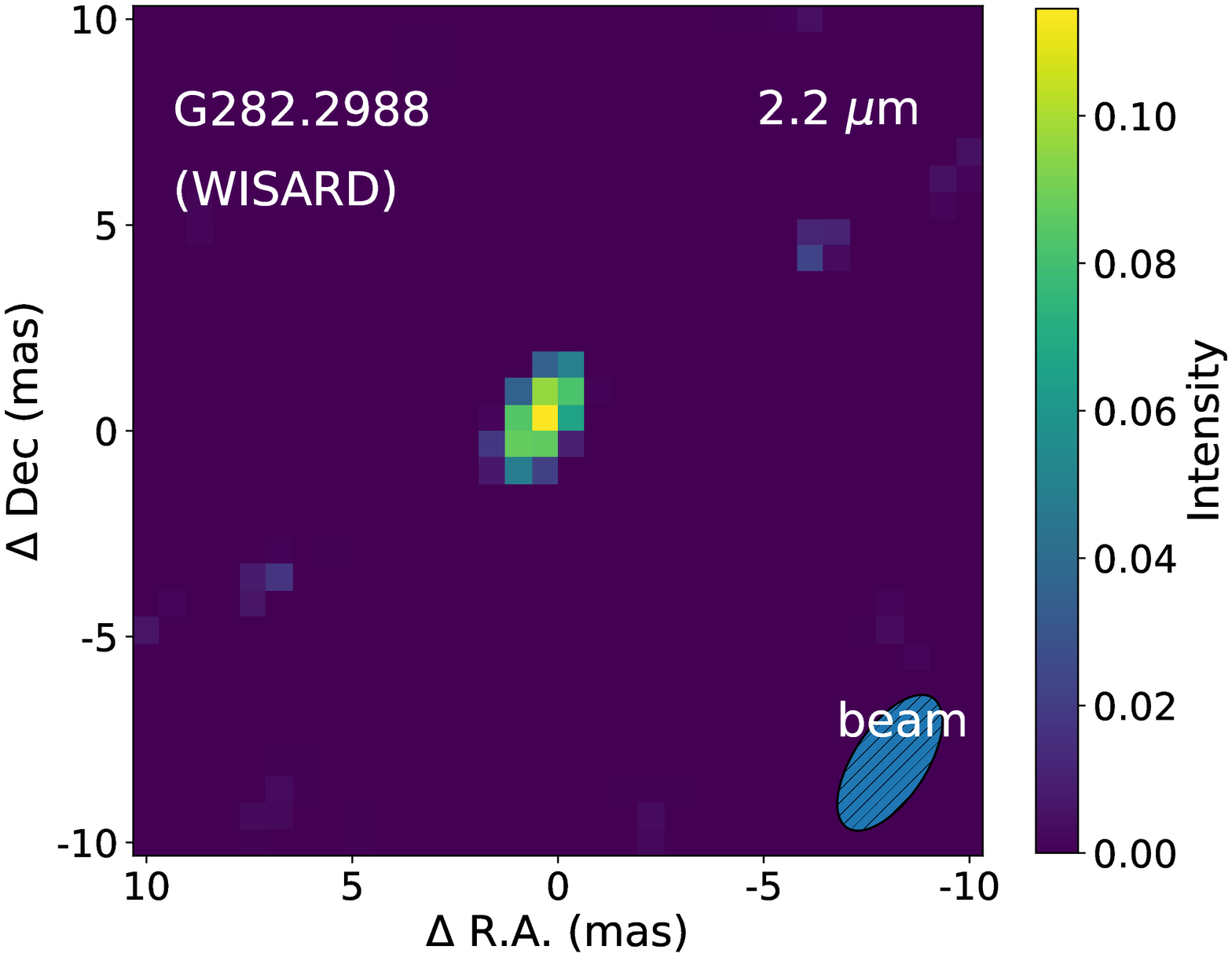}\par 
\includegraphics[scale=0.2,width=\linewidth,trim={2.0cm 2.0cm 2.0cm 4.3cm},clip]{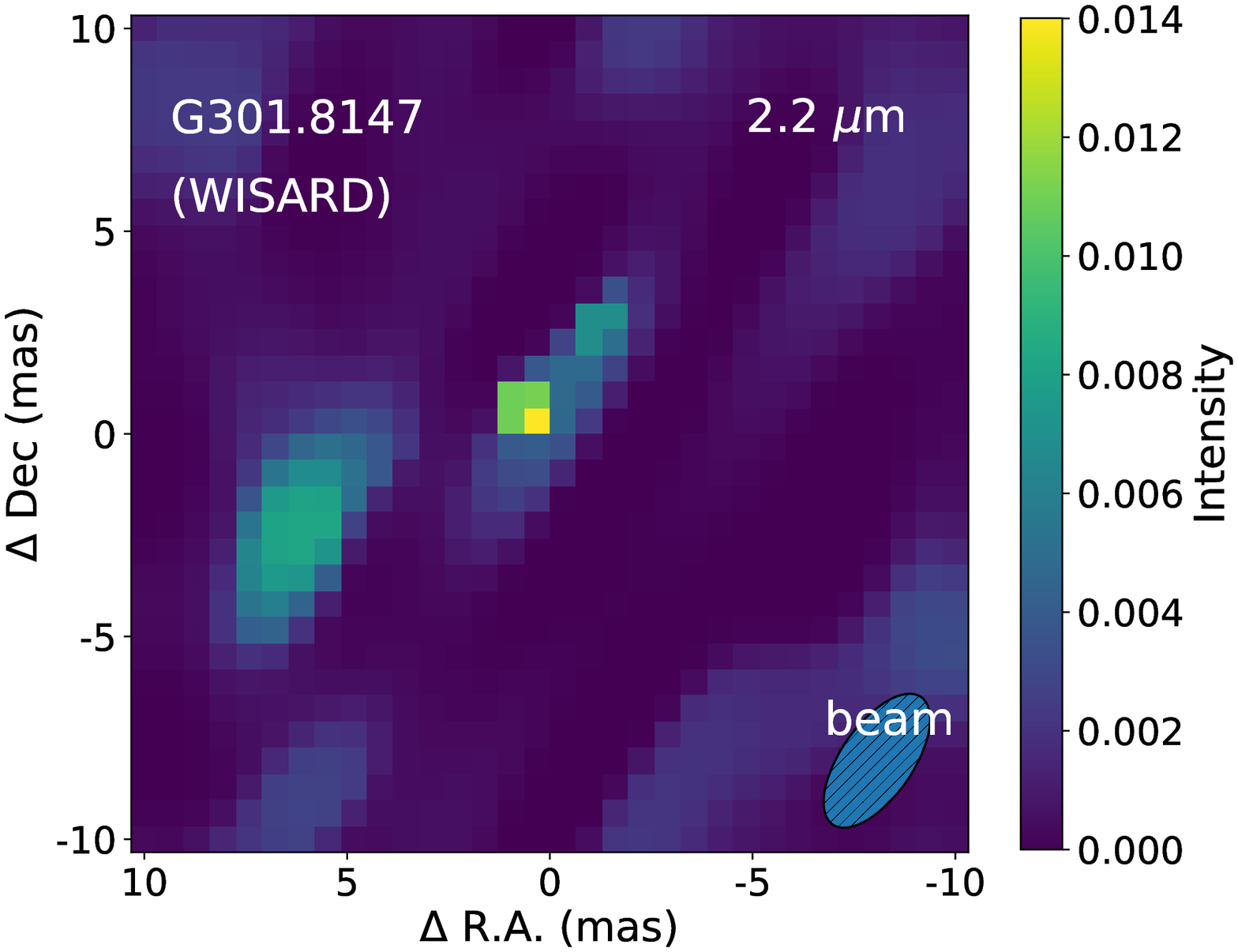}\par
\includegraphics[scale=0.2,width=\linewidth,trim={2.0cm 2.0cm 2.0cm 4.3cm},clip]{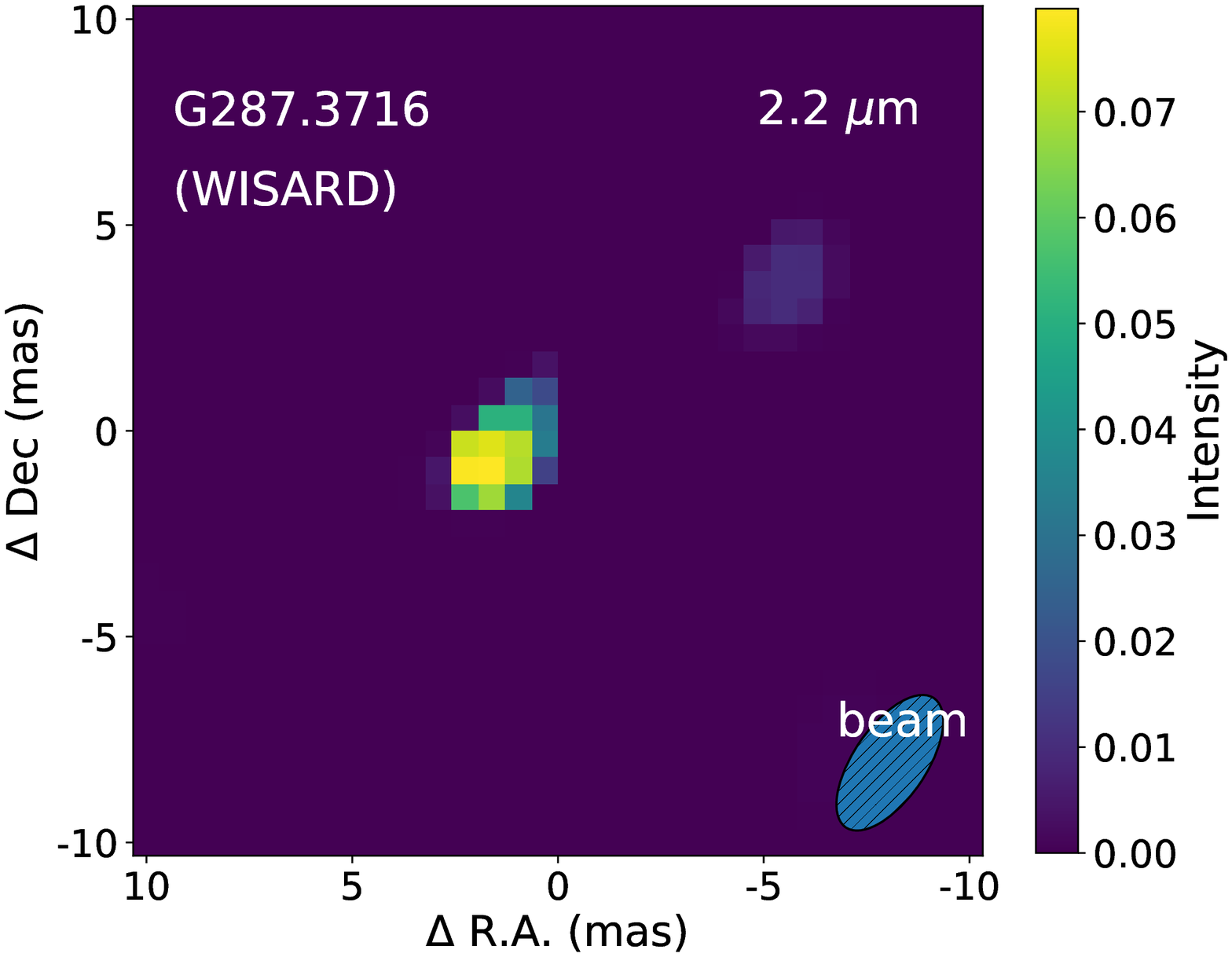}\par
\includegraphics[scale=0.2,width=\linewidth,trim={2.0cm 2.0cm 2.0cm 4.3cm},clip]{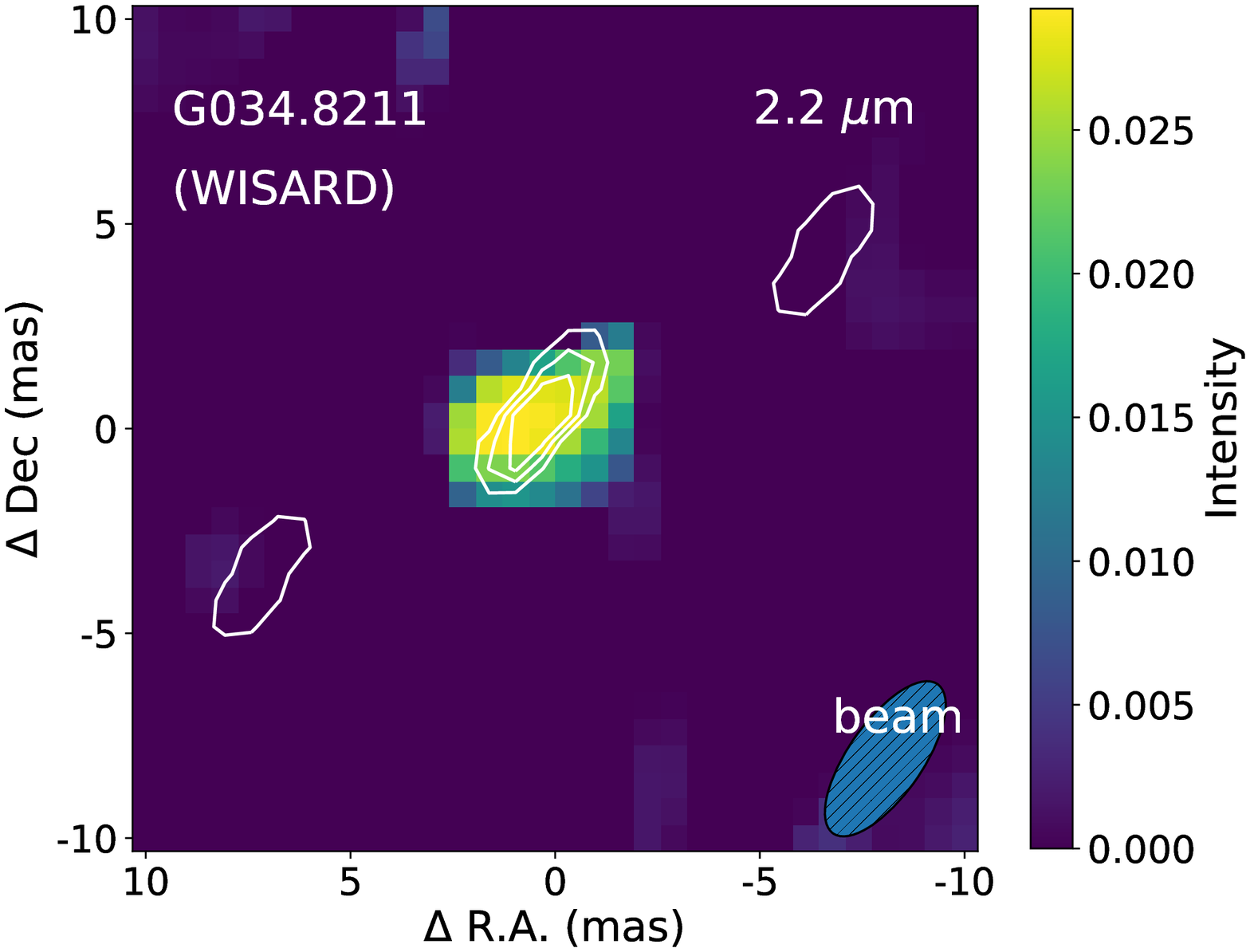}\par
\end{multicols}
\begin{multicols}{4}
\includegraphics[scale=0.2,width=\linewidth,trim={2.0cm 2.0cm 2.0cm 4.3cm},clip]{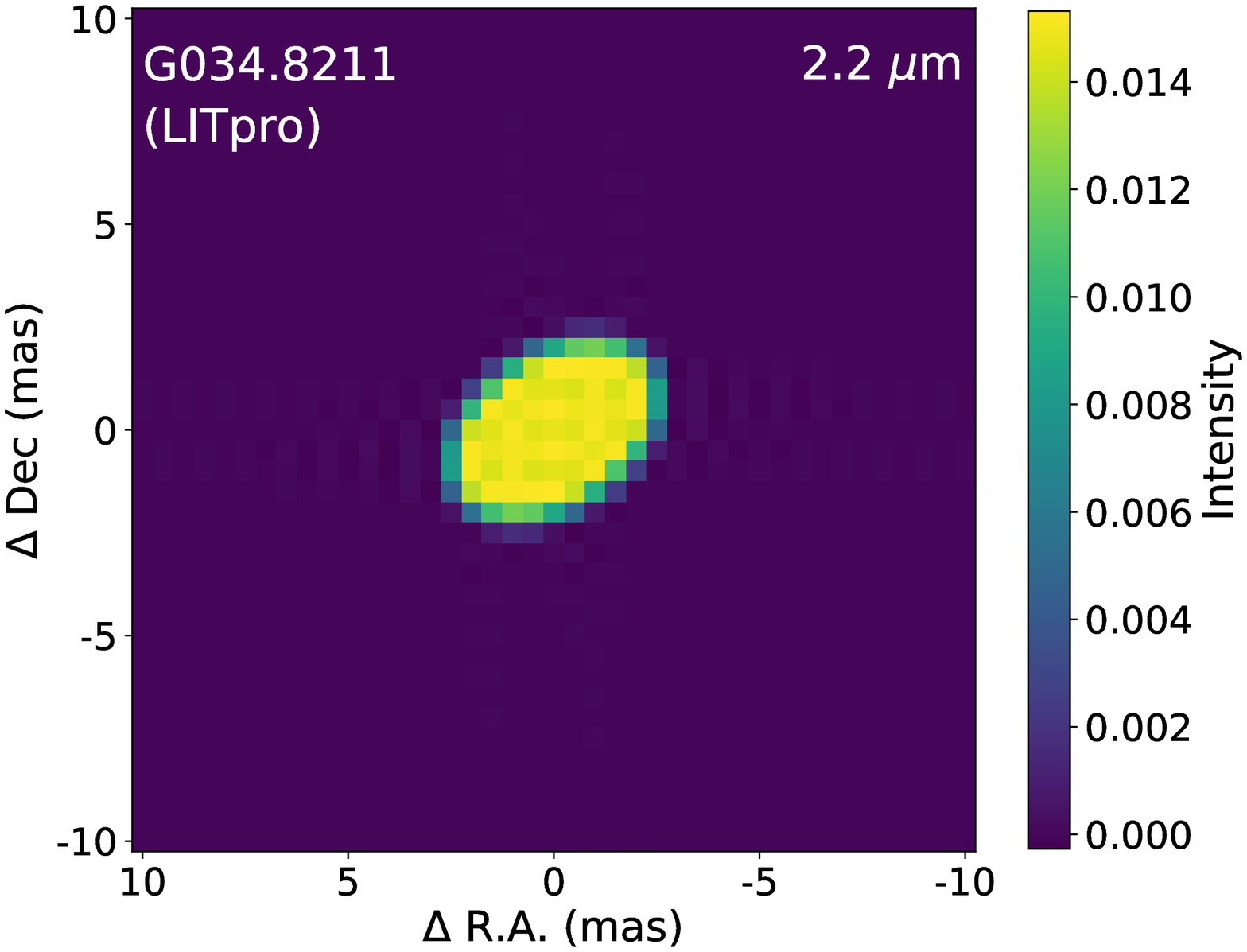}\par
\includegraphics[scale=0.2,width=\linewidth,trim={2.0cm 2.0cm 2.0cm 4.3cm},clip]{G034_continuum_WISARD_2_test_beam_reconstructed.eps}\par
\includegraphics[scale=0.2,width=\linewidth,trim={2.0cm 2.0cm 2.0cm 4.3cm},clip]{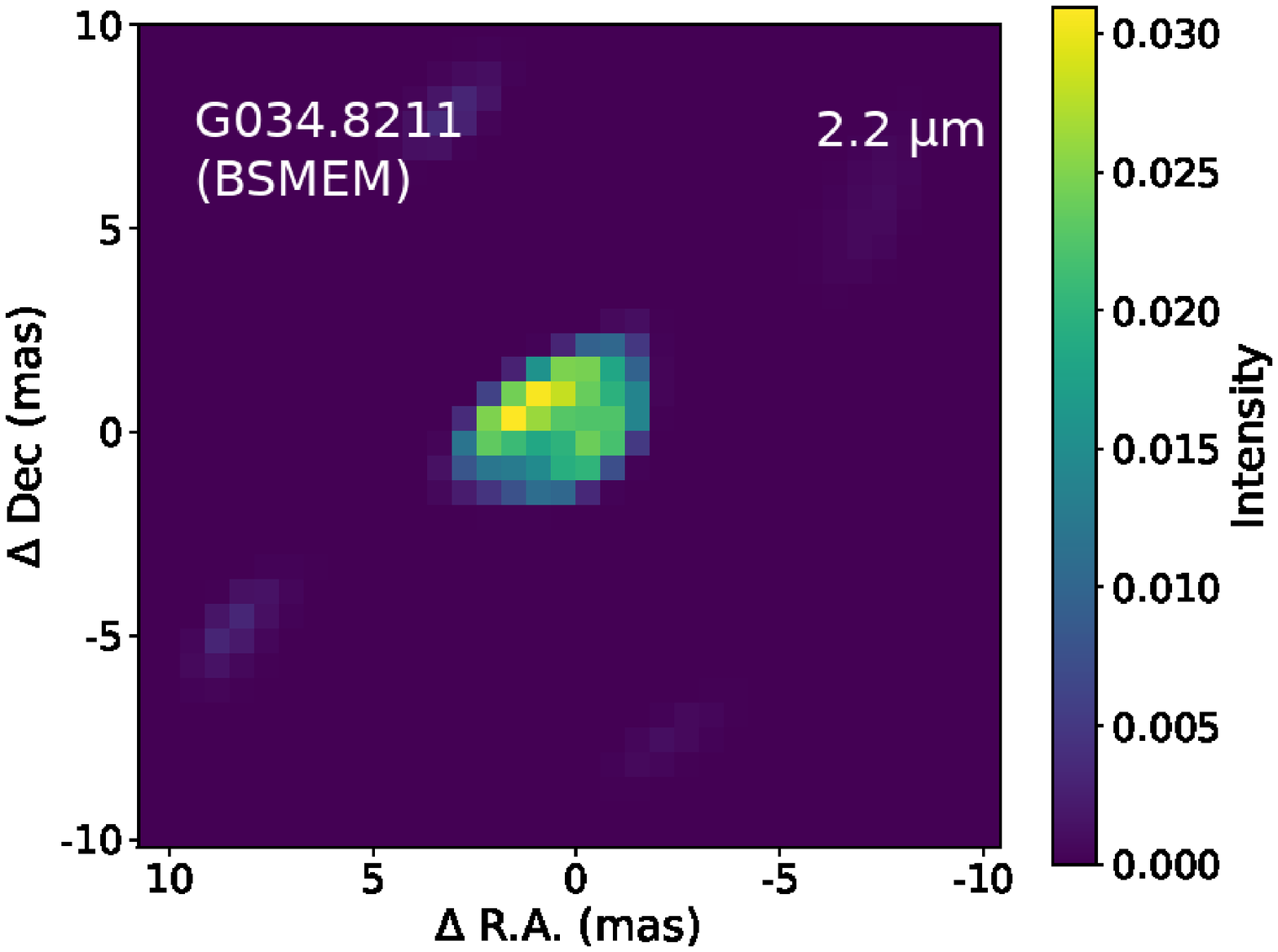}\par
\includegraphics[scale=0.2,width=\linewidth,trim={2.0cm 2.0cm 2.0cm 4.3cm},clip]{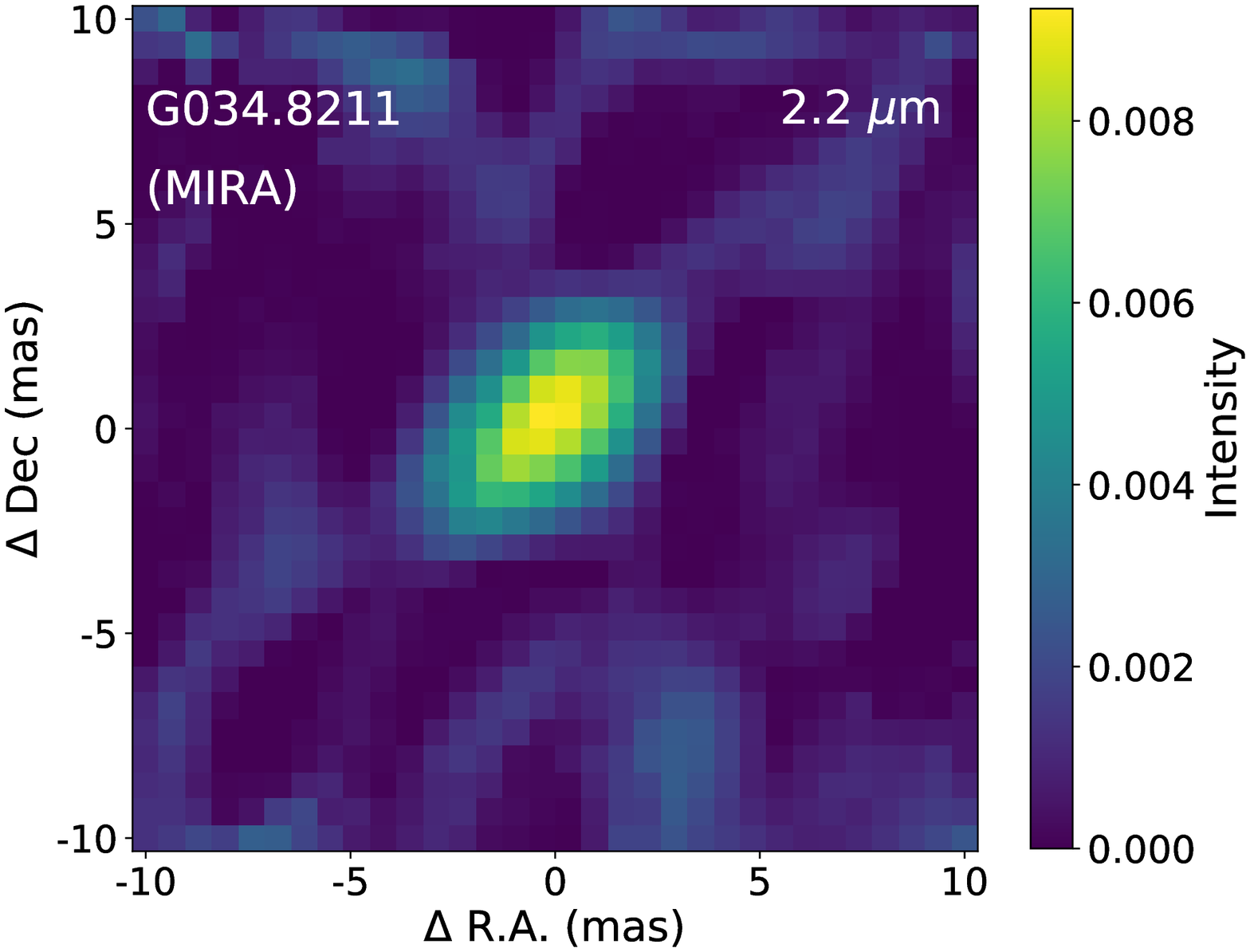}\par
\end{multicols}
\caption{Top: Image reconstruction of the 2.2~$\mu$m continuum towards G282.2988, G287.3716, G301.8147, and G034.8211 using WISARD. Bottom: Image reconstruction of the 2.2~$\mu$m emission towards G034.8211 using three different algorithms (WISARD, BSMEM, and MIRA). In addition, we present the image resulted independently by applying a simple geometric model (bottom left). The white contours represent the reconstructed image of the calibrator star for evaluation of the achieved spatial resolution.}
\label{fig:image_reconstruction_G034}
\end{figure*}

One source of our sample, G034.8211, appears to be the only source with a ratio between its size and the angular resolution higher than 1. The measured ratio of G034.8211 is 1.1, while for the other three sources is between 0.67 and 0.93. Therefore G034.8211 makes a strong case to investigate the image reconstruction further. Wanting to limit the algorithm dependency on the reconstructed images, we proceed to a thorough comparison of WISARD with the Multi-aperture ImageReconstruction Algorithm \citep[MIRA;][]{Thibaut2008} and BSMEM \citep{Baron2008}. A detailed description of the initialisation of the image reconstruction algorithms is presented in \citet{Koumpia2020}.

As we see in Figure~\ref{fig:image_reconstruction_G034} all three algorithms produce very similar elongated disc structures and show a striking similarity with what the geometric modelling revealed for this MYSO. We note that the reconstructed image of G034 using the BSMEM algorithm reveals some asymmetric emission within the inner $\sim$3mas, in the form of a ``midplane shadow''. The observed asymmetry may depict the observed changes in closure phases of the continuum emission towards the source (up to 20 degrees at the smallest scales; Table~\ref{gravity_closure}). We note that similar midplane shadows have been observed at larger scales (hundreds of au) towards a number of low mass T~Tauris sources in the NIR \citep[e.g., IM~Lup, RXJ~1615, MY~Lup, DoAr 25][]{Avenhaus2018,Garufi2020}, and have been attributed to opacity effects and scattered stellar light. Lastly, such asymmetries can be also attributed to an unresolved binary component at $<$10~au separations, or ongoing fragmentation being the cause of an observed asymmetry in the brightness distribution.     

In conclusion, the image reconstruction results in measured sizes which are consistent within 10\% (for three out of four sources) with the sizes resulting from the geometric modelling presented in Sec.~\ref{geom_model} after applying a Gaussian brightness distribution.

\clearpage

\section{NIR spectra}

Figure~\ref{fig:nir_spectra} presents the near infrared spectra of G287, G231, G233, G282 and G301. These spectra were obtained using IRIS-2 on the
AAT between 2006 and 2008. The spectral resolution in the H and K bands was $\sim$2400. 

\begin{figure*}
\begin{multicols}{2}
\includegraphics[scale=0.23]{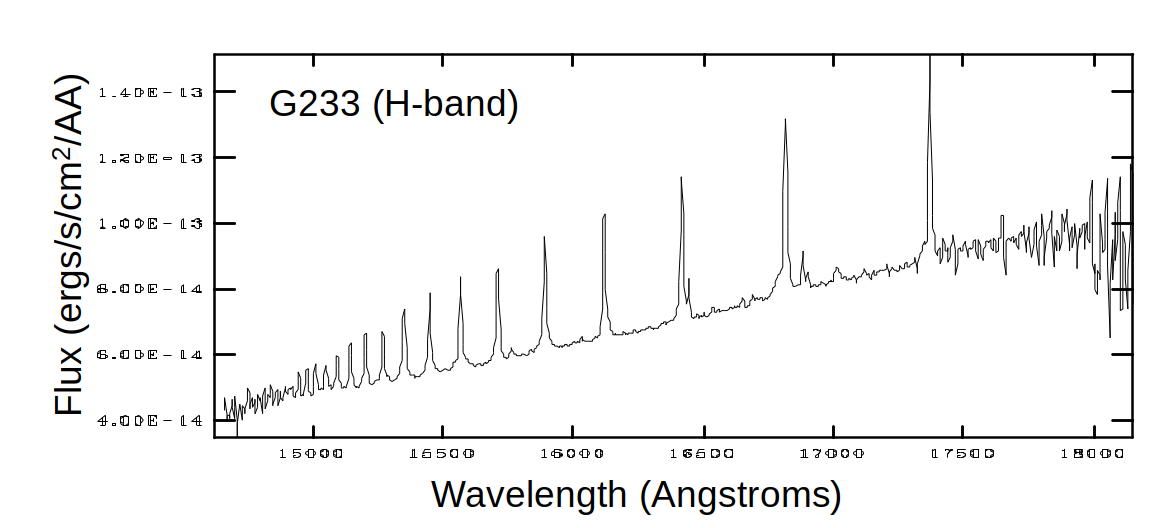}\par
\includegraphics[scale=0.23]{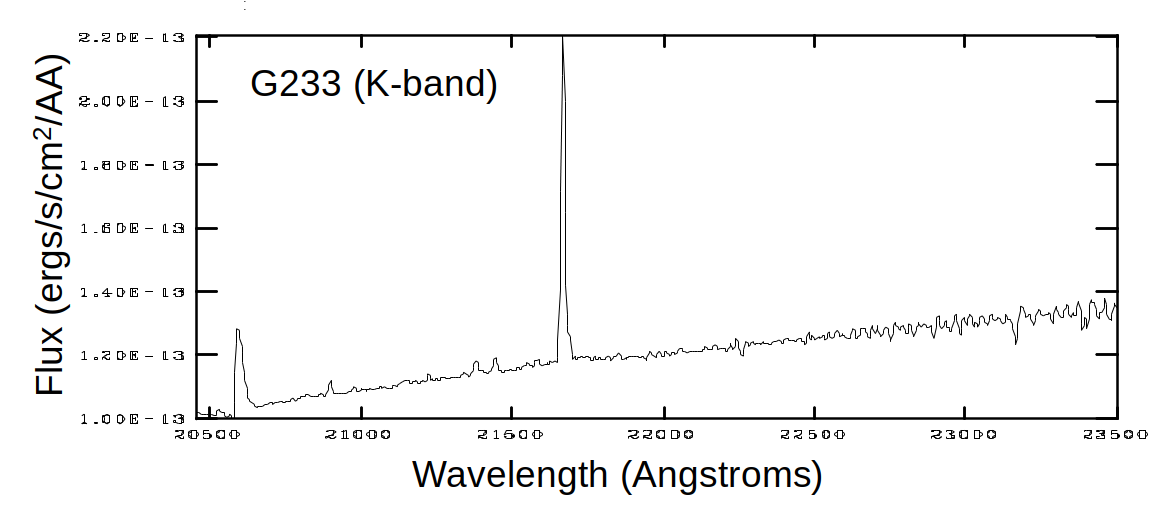}\par 
\end{multicols}
\begin{multicols}{2}
\includegraphics[scale=0.23]{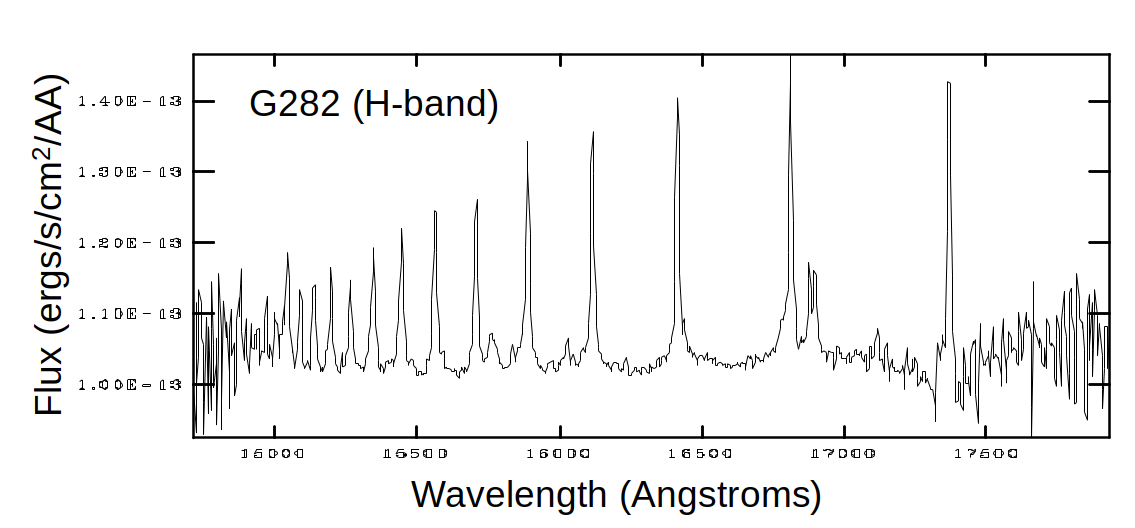}\par
\includegraphics[scale=0.23]{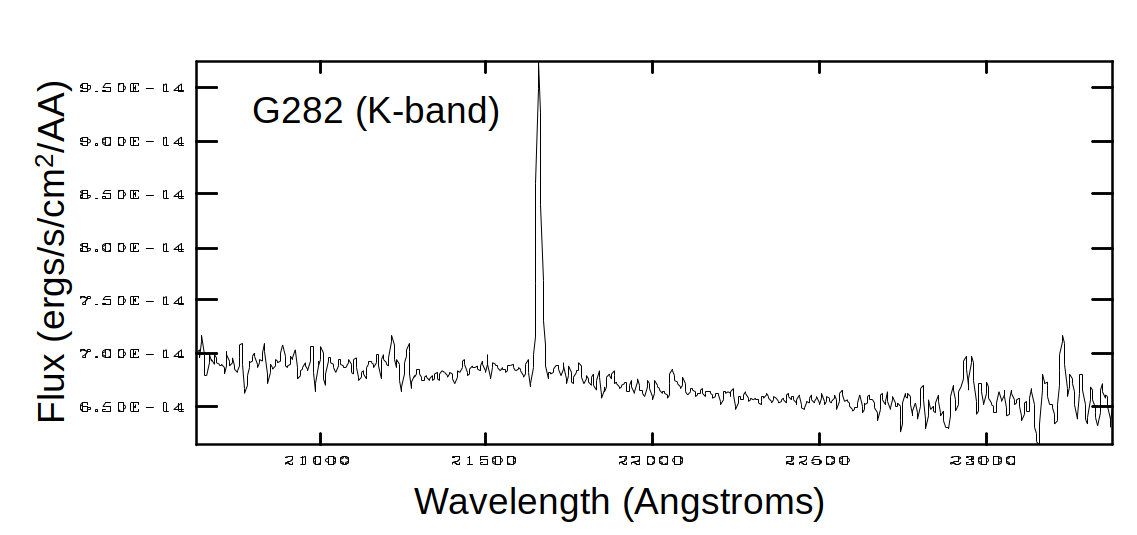}\par 
\end{multicols}
\begin{multicols}{2}
\includegraphics[scale=0.23]{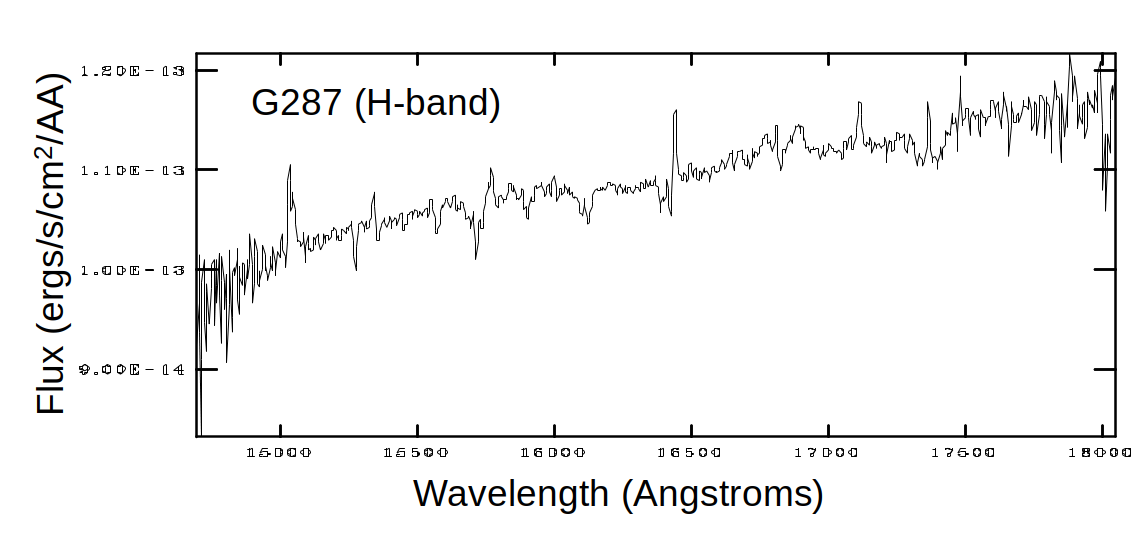}\par
\includegraphics[scale=0.23]{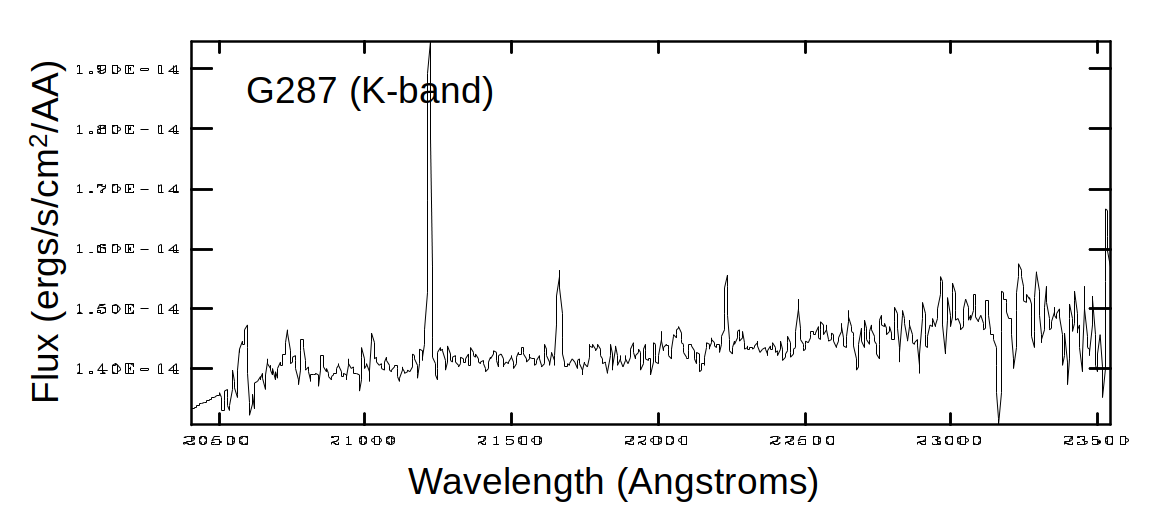}\par 
\end{multicols}
\begin{multicols}{2}
\includegraphics[scale=0.23]{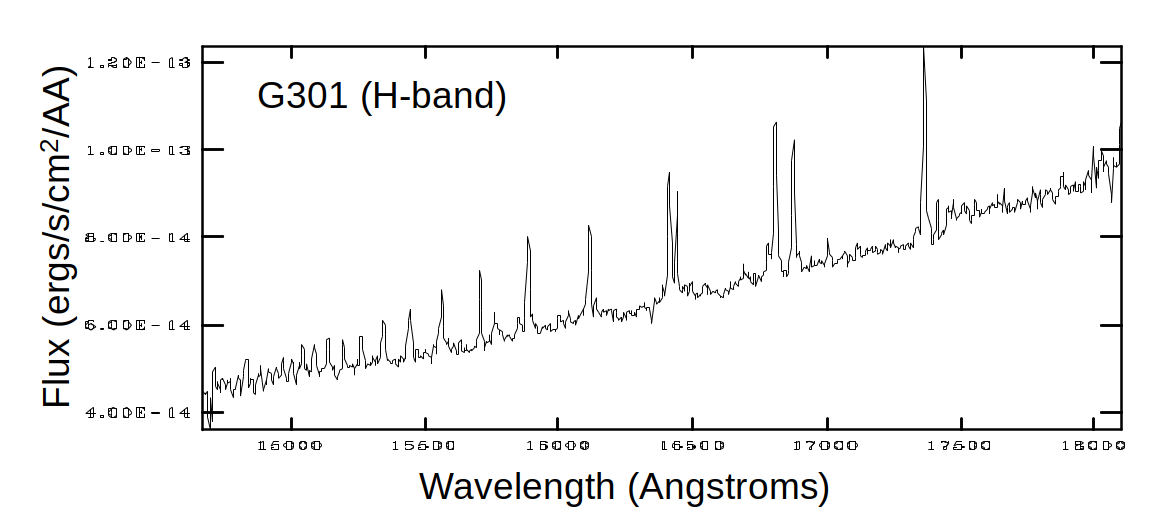}\par
\includegraphics[scale=0.23]{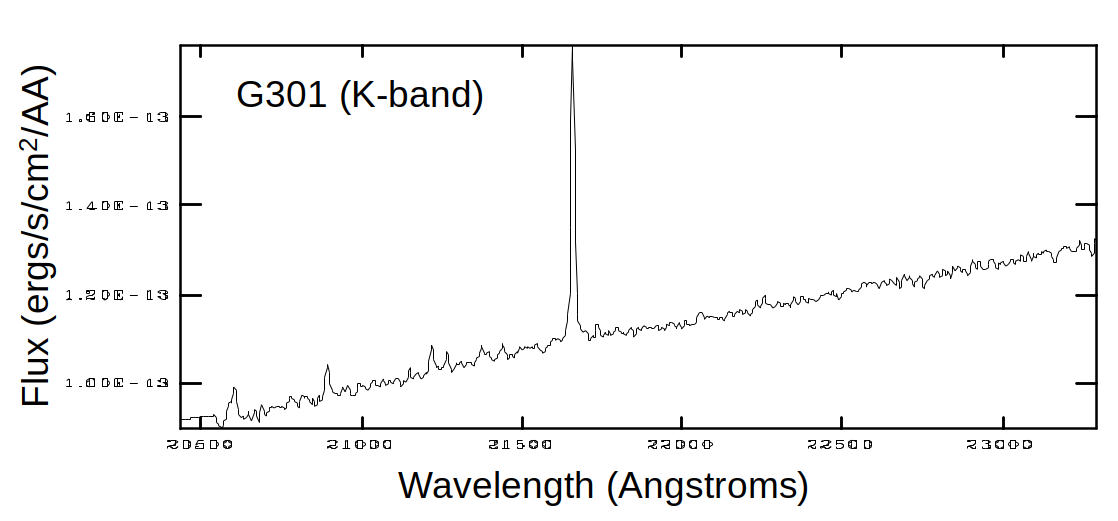}\par 
\end{multicols}
\caption{NIR (left column: H-band, right column: K-band) spectra of G233, G282, G287, and G301 \citep[The spectra of G034 and G231 are published in][]{Cooper2013a, Cooper2013b})}
\label{fig:nir_spectra}
\end{figure*}

\end{appendix}

\end{document}